\documentclass[11pt,pre,fleqn,seceqn,nofootinbib,byrevtex,showpacs]{revtex4}
\usepackage{microtype}
\usepackage{amsmath}
\usepackage{amsfonts}
\usepackage{amssymb}
\usepackage{amsthm}
\usepackage{graphicx}
\usepackage[colorlinks]{hyperref}
\usepackage{times}
\newcommand{\be}{\begin{equation}}
\newcommand{\ee}{\end{equation}}
\newcommand{\om}{\omega}

\newcommand{\ra}{\rightarrow}

\newcommand{\lex}{\left\langle}
\newcommand{\rex}{\right\rangle}

\newcommand{\var}{\textrm{var}}
\newcommand{\reals}{\mathbb{R}}
\newcommand{\mE}{\mathcal{E}}

\newcommand{\rh}{\tilde{h}}
\newcommand{\rs}{\tilde{s}}

\newcommand{\vep}{\epsilon}
\newcommand{\esc}{\textrm{esc}}
\newcommand{\sig}{\sigma}
\newcommand{\D}{\mathrm{d}}
\newcommand{\I}{\mathrm{i}}
\newcommand{\E}{\mathrm{e}}
\newcommand{\Zn}{\mathbb{Z}_n}

\newtheoremstyle{myplain}
{5pt}			%Space above 
{5pt}			%Space below 
{\normalsize}	%Body font 
{}			%Indent amount (empty = no indent, \parindent = para indent) 
{\bfseries}		%Thm head font 
{.}			%Punctuation after thm head 
{.5em}		%Space after thm head: " " = normal interword space; \newline = linebreak 
{\thmname{#1}\thmnumber{ #2}\thmnote{~{(#3)}}}

\theoremstyle{myplain}
\newtheorem{theorem}{Theorem}[section]
\newtheorem{example}[theorem]{Example}

\begin{document}

\title{The large deviation approach to statistical mechanics}
\author{Hugo Touchette}

\address{\mbox{School of Mathematical Sciences, Queen Mary University of London, London E1 4NS, UK}}

\date{\today}

\begin{abstract}
The theory of large deviations is concerned with the exponential decay of probabilities of large fluctuations in random systems. These probabilities are important in many fields of study, including statistics, finance, and engineering, as they often yield valuable information about the large fluctuations of a random system around its most probable state or trajectory. In the context of equilibrium statistical mechanics, the theory of large deviations provides exponential-order estimates of probabilities that refine and generalize Einstein's theory of fluctuations. This review explores this and other connections between large deviation theory and statistical mechanics, in an effort to show that the mathematical language of statistical mechanics is the language of large deviation theory. The first part of the review presents the basics of large deviation theory, and works out many of its classical applications related to sums of random variables and Markov processes. The second part goes through many problems and results of statistical mechanics, and shows how these can be formulated and derived within the context of large deviation theory. The problems and results treated cover a wide range of physical systems, including equilibrium many-particle systems, noise-perturbed dynamics, nonequilibrium systems, as well as multifractals, disordered systems, and chaotic systems. This review also covers many fundamental aspects of statistical mechanics, such as the derivation of variational principles characterizing equilibrium and nonequilibrium states, the breaking of the Legendre transform for nonconcave entropies, and the characterization of nonequilibrium fluctuations through fluctuation relations. 
\end{abstract}

\maketitle

\newpage
\tableofcontents

\newpage
%%%%%%%%%%%%%%%%%%%%%%%%%%%%%%%%%%%%%%%%%%%%%%%
\section{Introduction}

The mathematical theory of large deviations initiated by Cram\'er \cite{cramer1938} in the 1930s, and later developed by Donsker and Varadhan \cite{donsker1975,donsker1975a,donsker1976,donsker1983} and by Freidlin and Wentzell \cite{freidlin1984} in the 1970s, is not a theory commonly studied in physics. Yet it could be argued, without being paradoxical, that physicists have been using this theory for more than a hundred years, and are even responsible for writing down the very first large deviation result \cite{ellis1999}. Whenever physicists calculate an entropy function or a free energy function, large deviation theory is at play. In fact, large deviation theory is almost always involved when one studies the properties of many-particle systems, be they equilibrium or nonequilibrium systems. So what are large deviations, and what is the theory that studies these deviations?

If this question were posed to a mathematician who knows about large deviation theory, he or she might reply with one of the following answers:
\begin{itemize}
\item A theory dealing with the exponential decay of the probabilities of large deviations in stochastic processes;

\item A calculus of exponential-order measures based on the saddle-point approximation or Laplace's method;
\item An extension of Cram\'er's Theorem related to sample means of random variables;
\item An extension or refinement of the Law of Large Numbers and Central Limit Theorem.
\end{itemize}

A physicist, on the other hand, who is minimally acquainted with the concept of large deviations, would probably answer by saying that large deviation theory is
\begin{itemize}
\item A generalization of Einstein's fluctuation theory;
\item A collection of techniques for calculating entropies and free energies;
\item A rigorous expression of saddle-point approximations often used in statistical mechanics;
\item A rigorous formulation of statistical mechanics.
\end{itemize}

These answers do not seem to have much in common, except for the mention of the saddle-point approximation, but they are really all fundamentally related. They differ only in the extent that they refer to two different views of the same theory: one directed at its \emph{mathematical} applications---the other directed at its \emph{physical} applications. 

The aim of this review is to explain this point in detail, and to show, in the end, that large deviation theory and statistical mechanics have much in common. Actually, the message that runs through this review is more ambitious: we shall argue, by accumulating several correspondences between statistical mechanics and large deviation theory, that the mathematics of statistical mechanics, as a whole, is the theory of large deviations, in the same way that differential geometry, say, is the mathematics of general relativity. 

At the core of all the correspondences that will be studied here is Einstein's idea that probabilities can be expressed in terms of entropy functions. The expression of this idea in large deviation theory is contained in the so-called large deviation principle, and an entropy function in this context is called a rate function. This already explains one of the answers given above: large deviation theory is a generalization of Einstein's fluctuation theory. From this first correspondence follows a string of other correspondences that can be used to build and explain, from a clear mathematical perspective, the basis of statistical mechanics. Large deviation theory explains, for example, why the entropy and free energy functions are mutually connected by a Legendre transform, and so provides an explanation of the appearance of this transform  in thermodynamics. Large deviation theory also explains why equilibrium states can be calculated via the extremum principles that are the (canonical) minimum free energy principle and the (microcanonical) maximum entropy principle. In fact, large deviation theory not only justifies these principles, but also provides a prescription for generalizing them to arbitrary macrostates and arbitrary many-particle systems.

These points have already been recognized and ``publicized'' to some extent by a number of people, who see large deviation theory as the proper mathematical framework in which problems of statistical mechanics can be formulated and solved efficiently and, if need be, rigorously. Ellis \cite{ellis1985} is to be credited for providing what is perhaps the most complete expression of this view, in a book that has played a major part in bringing large deviations into physics. The idea that statistical mechanics can be formulated in the language of large deviations has also been expressed in a number of review papers, including one by Oono \cite{oono1989}, two by Ellis \cite{ellis1995,ellis1999}, and the seminal paper of Lanford \cite{lanford1973}, which is considered to be the first work on large deviations and statistical mechanics. Since these works appeared, more applications of large deviations have seen the light, so that the time seems ripe now for a new review. This especially true for the subjects of long-range interaction systems, nonconcave entropies, and nonequilibrium systems, which have all been successfully studied recently using large deviation techniques. 

Our efforts in this review will go towards learning about the many applications of large deviation theory in statistical mechanics, but also, and perhaps more importantly, towards learning about large deviation theory itself. The presentation of this theory covers in fact about half of this review, and is divided into three sections. The first presents a series of simple examples that illustrate the basis of the large deviation principle (Sec.~\ref{secexamples}). There follows a presentation of large deviation theory proper (Sec.~\ref{secldt}), and a section containing many illustrative examples of this theory (Sec.~\ref{secmathapp}). These examples are useful, as they illustrate many important points about large deviations that one must be aware of before studying their applications. 

The content of these three mathematical sections should overall be understandable by most physicists. A great deal of effort has been put into writing an account of large deviation theory which is devoid of the many mathematical details commonly found in textbooks on large deviations. These efforts have concentrated mainly on avoiding the use of measure theory and topology, and on using the level of rigor that prevails in physics for treating limits and approximations. The result is likely to upset mathematicians, but will surely please physicists who are looking for a theory with which to do calculations. Many mathematical elements that are omitted in the presentation are mentioned in the appendices, as well as in various other sections, which also point to many useful references that treat large deviations at the level of rigor demanded by mathematicians.

The physical applications of large deviations are covered in the second part of this review. The list of applications treated in the three sections that make up this part is not exhaustive, but covers most of the important applications related to equilibrium statistical mechanics (Sec.~\ref{secequi}) and nonequilibrium statistical mechanics (Sec.~\ref{secnonequi}). The correspondence between large deviation theory and Einstein's fluctuation theory is fully explained in the section dealing with equilibrium system. Other topics discussed in that section include the interpretation of the entropy as a rate function, the derivation of the Legendre transform connecting the entropy and the free energy, and the derivation of general variational principles that characterize equilibrium states in the microcanonical and canonical ensembles. The topics discussed in the context of nonequilibrium systems are as varied, and include the study of large deviations in stochastic differential equations (Freidlin-Wentzell theory), dynamical models of equilibrium fluctuations (Onsager-Machlup theory), fluctuation relations, and systems of interacting particles. Other applications, related to multifractals, chaotic systems, spin glasses, and quantum systems, are quickly covered in Sec.~\ref{secother}. 

As a warning about the sections covering the physical applications, it should be said that this work is neither a review of statistical mechanics nor a review of large deviation theory---it is a review of the many ways in which large deviation theory can be applied in statistical mechanics. The list of applications treated in this work should be viewed, accordingly, not as a complete list of applications of large deviation theory, but as a selected list or \emph{compendium} of representative examples that should serve as useful points of departure for studying other applications. This is especially true for the examples discussed in the section on nonequilibrium systems (Sec.~\ref{secnonequi}). At the time of writing this review, a complete theory of nonequilibrium systems is still lacking, so it is difficult to provide a unified presentation of these systems based on large deviation theory. The aim of Sec.~\ref{secnonequi} is to give a broad idea of how large deviation techniques can be applied for studying nonequilibrium systems, and to convey a sense that large deviation theory is behind many results related to these systems, just as it is behind many results related to equilibrium systems.
One could go further and argue, following Oono \cite{oono1989} and Eyink \cite{eyink1990} among others, that large deviation theory is not only useful for studying nonequilibrium systems, but provides the proper basis for building a theory of these systems. Section~\ref{secnonequi} was written with this idea in mind. 

%%%%%%%%%%%%%%%%%%%%%%%%%%%%%%%%%%%%%%%%%%%%%%%
\section{Examples of large deviation results}
\label{secexamples} 

Before we immerse ourselves into the theory of large deviations, it is useful to work out a few examples involving random sums to gain a sense of what large deviations are, and a sense of the context in which these deviations arise. The examples are purposely abstract, but are nonetheless simple. The goal in presenting them is to introduce some basic mathematical ideas and notations that will be used throughout this review. Readers who are already familiar with large deviations may skip this section, and start with Sec.~\ref{secldt} or even Sec.~\ref{secequi}.

\begin{figure}[t]
\centering
\includegraphics[scale=1.0]{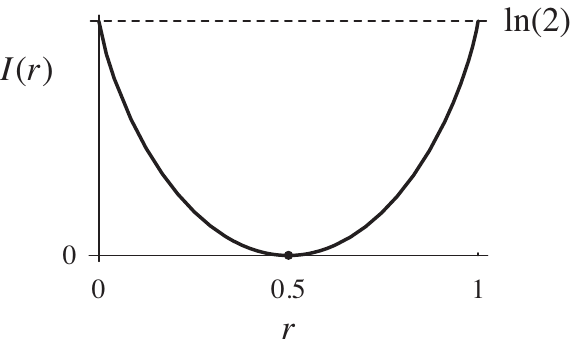}
\caption{Rate function $I(r)$ for Example~\ref{exfrac1}.}
\label{figbinratefct1}
\end{figure}

\begin{example}[Random bits]
\label{exfrac1}
Consider a sequence $b=(b_1,b_2,\ldots,b_n)$ of $n$ independent random bits taking the value $0$ or $1$ with equal probability, and define 
\be
R_n=\frac{1}{n}\sum_{i=1}^n b_i
\ee
to be the fraction of 1's contained in $b$. We are interested to find the probability $P(R_n=r)$ that $R_n$ assumes one of the (rational) values $0,1/n,2/n,\ldots,n/n$. Since the bits are independent and unbiased, we have $P(b)=2^{-n}$ for all $b\in\{0,1\}^n$, so that
\be
P(R_n=r)=\sum_{b:R_n(b)=r} P(b)=\frac{1}{2^n}\frac{n!}{(rn)![(1-r)n]!}.
\ee
Using Stirling's approximation, $n!\approx n^n \E^{-n}$, we can extract from this result a dominant contribution having the form
\be
P(R_n=r)\approx \E^{-nI(r)},\qquad I(r)=\ln 2 +r\ln r +(1-r)\ln (1-r)
\label{eqldp1}
\ee
for $n$ large. The function $I(r)$ entering in the exponential is positive and convex for $r\in[0,1]$, as shown in Fig.~\ref{figbinratefct1}, and has a unique zero is located at $r=1/2$.
\end{example}

The approximation displayed in (\ref{eqldp1}) is an example of large deviation approximation. The exponential-decaying form of this approximation, combined with the expression of the decay or \emph{rate function} $I(r)$, shows that the ``unbalanced'' sequences of $n$ bits that contain more 0's than 1's, or vice versa, are unlikely to be observed as $n$ gets large because $P(R_n)$ decays exponentially with $n$ for $R_n\neq 1/2$. Only the ``balanced'' sequences such that $R_n\approx 1/2$ have a non-negligible probability to be observed as $n$ becomes large. 

The next example discusses a different random sum for which a large deviation approximation also holds.

\begin{figure*}[t]
\centering
\includegraphics[scale=1.0]{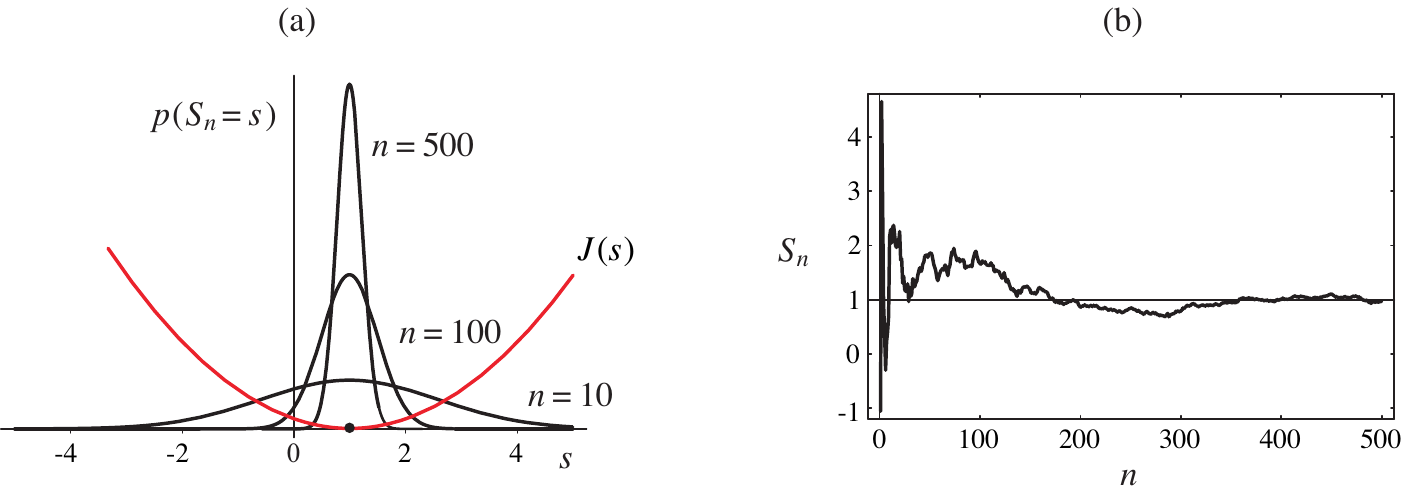}
\caption{Gaussian sample mean with $\mu=\sigma=1$. (a) Probability density $p(S_n=s)$ for increasing values of $n$ together with its corresponding rate function $J(s)$ (red line). (b) Typical realization of $S_n$ converging to its mean.}
\label{figgaussratefct1}
\end{figure*}

\begin{example}[Gaussian sample mean]
\label{exgauss}
The random variable $R_n$, defined in the previous example as a sum of $n$ random variables scaled by $n$, is called in mathematics a \emph{sample mean}. In the present example, we consider a similar sample mean, given by
\be
S_n=\frac{1}{n}\sum_{i=1}^n X_i,
\ee
and assume that the random variables $X_i$ are independent and identically distributed (IID) according to the Gaussian probability density 
\be
p(X_i=x_i)=\frac{1}{\sqrt{2\pi\sigma^2}} \E^{-(x_i-\mu)^2/(2\sigma^2)}.
\label{eqgauss1}
\ee 
The parameters $\mu$ and $\sigma^2$ represent, as usual, the mean and variance, respectively, of the $X_i$'s. 

The probability density of $S_n$ can be written as the integral
\be
p(S_n=s)=\int_{\{x\in\reals^n:S_n(x)=s\} } p(x)\, \D x = \int_{\reals^n} \delta(S_n(x)-s)\, p(x)\, \D x=\lex\delta(S_n-s)\rex,
\label{eqprob1}
\ee
where $x=(x_1,x_2,\ldots,x_n)$ is the vector of random variables, and
\be
p(x)=p(x_1,x_2,\ldots,x_n)=p(x_1)p(x_2)\cdots p(x_n)
\ee
their product density. The solution of this integral is, of course,
\be
p(S_n=s)=\sqrt{\frac{n}{2\pi\sigma^2}} \E^{-n(s-\mu)^2/(2\sigma^2)},
\label{eqgaussexact1}
\ee
since a sum of Gaussian random variables is also exactly Gaussian-distributed. A large deviation approximation is obtained from this exact result by neglecting the term $\sqrt{n}$, which is subdominant with respect to the decaying exponential, thereby obtaining
\be
p(S_n=s)\approx \E^{-n J(s)},\qquad J(s)=\frac{(s-\mu)^2}{2\sigma^2},\quad s\in\reals.
\label{eqratefctgauss}
\ee 
\end{example}

The rate function $J(s)$ that we find here is similar to the rate function $I(r)$ found in the first example---it is convex and possesses a single minimum and zero; see Fig.~\ref{figgaussratefct1}(a). As was the case for $I(r)$, the minimum of $J(s)$ has also for effect that, as $n$ grows, $p(S_n=s)$ gets more and more concentrated around the mean $\mu$ because the mean is the only point for which $J(s)=0$, and thus for which $p(S_n=s)$ does not decay exponentially. In mathematics, this concentration property is expressed by the following limit:
\be
\lim_{n\ra\infty} P(S_n\in [\mu-\delta,\mu+\delta])=1,
\label{llng1}
\ee
where $\delta$ is any positive number. Whenever this limit holds, we say that $S_n$ converges \emph{in probability} to its mean, and that $S_n$ obeys the \emph{Law of Large Numbers}. This point will be studied in more detail in Sec.~\ref{secldt}.

In general, sums of IID random variables involving different probability distributions for the summands have different rate functions. This is illustrated next.

\begin{figure*}[t]
\centering
\includegraphics[scale=1.0]{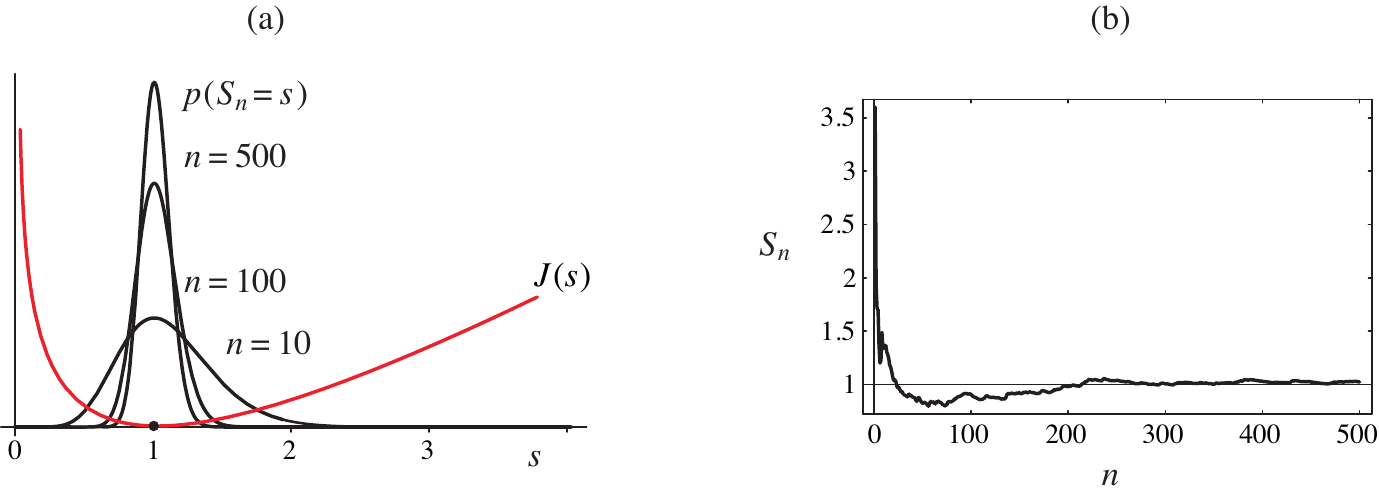}
\caption{Exponential sample mean with $\mu=1$. (a) Probability density $p(S_n=s)$ for increasing values of $n$ together with its corresponding rate function $J(s)$ (red line). (b) Typical realization of $S_n$ converging to its mean.}
\label{figexpratefct1}
\end{figure*}

\begin{example}[Exponential sample mean]
\label{exexp}
Consider the sample mean $S_n$ defined before, but now suppose that the IID random variables $X_1,X_2,\ldots,X_n$ are distributed according to the exponential distribution
\be
p(X_i=x_i)=\frac{1}{\mu} \E^{-x_i/\mu},\quad x_i>0, \mu>0.
\ee
For this distribution, it can be shown that
\be
p(S_n=s)\approx \E^{-n J(s)},\quad J(s)=\frac{s}{\mu}-1-\ln\frac{s}{\mu},\quad s>0.
\label{eqexp1}
\ee
As in the previous examples, the interpretation of the approximation above is that the decaying exponential in $n$ is the dominant term of $p(S_n=s)$ in the limit of large values of $n$. Notice here that the rate function is different from the rate function of the Gaussian sample mean [Fig.~\ref{figexpratefct1}(a)], although it is still positive, convex, and has a single minimum and zero located at $s=\mu$ that yields the most probable or \emph{typical} value of $S_n$ in the limit $n\ra\infty$; see Fig.~\ref{figexpratefct1}(b).
\end{example}

The advantage of expressing $p(S_n=s)$ in a large deviation form is that the rate function $J(s)$ gives a direct and detailed picture of the deviations or \emph{fluctuations} of $S_n$ around its typical value. For the Gaussian sample mean, for example, $J(s)$ is a parabola because the fluctuations of $S_n$ around its typical value (the mean $\mu$) are Gaussian-distributed. For the exponential sample mean, by contrast, $J(s)$ has the form of a parabola only around $\mu$, so that only the \emph{small} fluctuations of $S_n$ near its typical value are Gaussian-distributed. The \emph{large} positive fluctuations of $S_n$ that are away from its typical value are not Gaussian; in fact, the form of $J(s)$ shows that they are exponentially-distributed because $J(s)$ is asymptotically linear as $s\ra\infty$. This distinction between small and large fluctuations explains the ``large'' in ``large deviation theory'', and will be studied in more detail in the next section when discussing the Central Limit Theorem. For now, we turn to another example that shows that large deviation approximations also arise in the context of random vectors.

\begin{example}[Symbol frequencies]
\label{exsanov}
Let $\om=(\om_1,\om_2,\ldots,\om_n)$ be a sequence of IID random variables drawn from the set $\Lambda=\{1,2,\ldots,q\}$ with common probability distribution $P(\om_i=j)=\rho_j>0$. For a given sequence $\om$, we denote by $L_{n,j}(\om)$ the relative frequency with which the number or symbol $j\in\Lambda$ appears in $\om$, that is,
\be
L_{n,j}(\om)=\frac{1}{n}\sum_{i=1}^n \delta_{\om_i,j},
\ee
where $\delta_{i,j}$ is the Kronecker symbol. For example, if $\Lambda=\{1,2,3\}$ and $\omega=(1,3,2,3,1,1)$, then
\be
L_{6,1}(\omega)=\frac{3}{6},\quad L_{6,2}(\omega)=\frac{1}{6},\quad L_{6,3}(\omega)=\frac{2}{6}.
\ee
The normalized vector\footnote{Vectors are not written in boldface. The vector nature of a quantity should be clear from the context in which it appears.}
\be
L_n(\om)=\left(L_{n,1}(\om),L_{n,2}(\om),\ldots,L_{n,q}(\om)\right)
\ee
containing all the symbol frequencies is called the \emph{empirical vector} associated with $\om$ \cite{dembo1998}. It is also called the \emph{type} of $\om$ in information theory \cite{cover1991} or the \emph{statistical distribution} of $\om$ in physics. The name ``distribution'' arises because $L_n(\om)$ has all the properties of a probability distribution, namely, $0\leq L_{n,j}(\om)\leq 1$ for all $j\in\Lambda$, and
\be
\sum_{j\in\Lambda} L_{n,j}(\om)=1
\ee
for all $\om\in\Lambda^n$. It is important to note, however, that $L_n$ is not a probability; it is a random vector associated with each possible sequence or \emph{configuration} $\om$, and distributed according to the multinomial distribution
\be
P(L_n=l)=\frac{n!}{\prod_{j=1}^q (nl_j)!} \prod_{j=1}^q \rho_j^{nl_j}.
\ee
As in Example~\ref{exfrac1}, we can extract from this exact result a large deviation approximation by using Stirling's approximation. The result for large values of $n$ is 
\be
P(L_n=l)\approx \E^{-nI_\rho(l)},\qquad I_\rho(l)=\sum_{j=1}^q l_j\ln\frac{l_j}{\rho_j}.
\ee
\end{example}

The function $I_\rho(l)$ is called the \emph{relative entropy} or \emph{Kullback-Leibler distance} between the probability vectors $l$ and $\rho$ \cite{cover1991}. As a rate function, $I_\rho(l)$ is slightly more complicated than the rate functions encountered so far, although it shares similar properties. It can be shown, in particular, that $I_\rho(l)$ is positive and convex, and has a single minimum and zero located at $l=\rho$, that is, $l_j=\rho_j$ for all $j\in\Lambda$ (see Chap.~2 of~\cite{cover1991}). As before, the zero of the rate function is interpreted as the most probable value of the random variable for which the large deviation result is obtained. This applies for $L_n$ because $P(L_n=l)$ converges to 0 exponentially fast with $n$ for all $l\neq \rho$, since $I_\rho(l)>0$ for all $l\neq \rho$. The only value of $L_n$ for which $P(L_n=l)$ does not converge exponentially to 0 is $l=\rho$. Hence $L_n$ must converge to $\rho$ in probability as $n\ra\infty$.

The next and last example of this section is a simple and classical one in statistical mechanics. It is presented to show that exponential approximations similar to large deviation approximations can be defined for quantities other than probabilities, and that entropy functions are large deviation rate functions in disguise. We will return to these observations, and in particular to the association ``entropy = rate function'', in Sec.~\ref{secequi}.

\begin{example}[Entropy of non-interacting spins]
\label{exentspin}
Consider $n$ spins $\sigma_1,\sigma_2,\ldots,\sigma_n$ taking values in the set $\{-1,1\}$. It is well known that the number $\Omega(m)$ of spin configurations $\sigma=(\sigma_1,\sigma_2,\ldots,\sigma_n)$ having a magnetization per spin
\be
\frac{1}{n}\sum_{i=1}^n \sigma_i
\ee
equal to $m$ is given by the binomial-like formula
\be
\Omega(m)=\frac{n!}{\left[(1-m)n/2\right]! \left[(1+m)n/2\right]!}.
\ee
The similarity of this result with the one found in the example about random bits should be obvious. As in that example, we can use Stirling's approximation to obtain a large deviation approximation for $\Omega(m)$, which we write as
\be
\Omega(m)\approx \E^{n s(m)},\quad s(m)=-\frac{1-m}{2}\ln\frac{1-m}{2}-\frac{1+m}{2}\ln\frac{1+m}{2},\quad m\in[-1,1].
\ee 
The function $s(m)$ is the \emph{entropy} associated with the mean magnetization.

As in the previous example, we can also count the number $\Omega(l)$ of spin configurations containing a relative number $l_+$ of $+1$ spins and a relative number $l_-$ of $-1$ spins. These two relative numbers or frequencies are the components of the two-dimensional empirical vector $l=(l_+,l_-)$, for which we find
\be
\Omega(l)\approx \E^{n\tilde{s}(l)},\quad \tilde{s}(l)=-l_+\ln l_+-l_-\ln l_-
\ee
for $n$ large. The function $\rs(l)$, which plays the role of a rate function, is also called the entropy, although it is now the entropy associated with the empirical vector. Notice that since we can express $m$ as a function of $l$ and vice versa, $s(m)$ can be expressed in terms of $\rs(l)$ and vice versa. 
\end{example}

%%%%%%%%%%%%%%%%%%%%%%%%%%%%%%%%%%%%%%%%%%%%%%%
\section{Large deviation theory}
\label{secldt} 

The cornerstone of large deviation theory is the exponential approximation encountered in the previous examples. This approximation appears so frequently in problems involving many random variables, in particular those studied in statistical mechanics, that we give it a name: \emph{the large deviation principle}. Our goal in this section is to lay down the basis of large deviation theory by first defining  the large deviation principle with more care, and by then deriving a number of important consequences of this principle. In doing so, we will see that the large deviation principle is similar to the laws of thermodynamics, in that a few principles---a single one in this case---can be used to derive many far-reaching results. No attempt will be made in this section to integrate or interpret these results within the framework of statistical mechanics; this will come after Sec.~\ref{secmathapp}.

\subsection{The large deviation principle} 

A basic approximation or scaling law  of the form $P_n\approx \E^{-nI}$, where $P_n$ is some probability, $n$ a parameter assumed to be large, and $I$ some positive constant, is referred to as a \emph{large deviation principle}. Such a definition is, of course, only intuitive; to make it more precise, we need to define what we mean exactly by $P_n$ and by the approximation sign ``$\approx$''. This is done as follows. Let $A_n$ be a random variable indexed by the integer $n$, and let $P(A_n\in B)$ be the probability that $A_n$ takes on a value in a set $B$. We say that $P(A_n\in B)$ satisfies a \emph{large deviation principle} with \emph{rate} $I_B$ if the limit
\be
\lim_{n\ra\infty} -\frac{1}{n}\ln P(A_n\in B)=I_B
\label{eqldp2}
\ee
exists. 

The idea behind this limit should be clear. What we mean when writing $P(A_n\in B)\approx \E^{-nI_B}$ is that the dominant behavior of $P(A_n\in B)$ is a decaying exponential in $n$. Using the small-$o$ notation, this means that
\be
-\ln P(A_n\in B)=n I_B +o(n),
\ee
where $I_B$ is some positive constant. To extract this constant, we divide both sides of the expression above by $n$ to obtain
\be
-\frac{1}{n}\ln P(A_n\in B)=I_B +o(1),
\ee
and pass to the limit $n\ra\infty$, so as to get rid of the $o(1)$ contribution. The end result of these steps is the large deviation limit shown in (\ref{eqldp2}). Hence, if $P(A_n\in B)$ has a dominant exponential behavior in $n$, then that limit should exist with $I_B\neq 0$. If the limit does not exist, then either $P(A_n\in B)$ is too singular to have a limit or else $P(A_n\in B)$ decays with $n$ faster than $\E^{-na}$ with $a>0$. In this case, we say that $P(A_n\in B)$ decays \emph{super-exponentially} and set $I=\infty$. The large deviation limit may also be zero for any set $B$ if $P(A_n\in B)$ is \emph{sub-exponential} in $n$, that is, if $P(A_n\in B)$ decays with $n$ slower than $\E^{-na}$, $a>0$. The cases of interest for large deviation theory are those for which the limit shown in (\ref{eqldp2}) does exist with a non-trivial rate exponent, i.e., different from $0$ or $\infty$.

All the examples studied in the previous section fall under the definition of the large deviation principle, but they are more specific in a way because they refer to particular events of the form $A_n=a$ rather than $A_n\in B$. In the case of the random bits, for example, we found that the probability $P(R_n=r)$ satisfied
\be
\lim_{n\ra\infty} -\frac{1}{n}\ln P(R_n=r)=I(r),
\ee
with $I(r)$ a continuous function that we called in this context a \emph{rate function}. Similar results were obtained for the Gaussian and exponential sample means, although for these we worked with probability densities rather than probability distributions. The ``density'' large deviation principles that we obtained can nevertheless be translated into ``probability'' large deviation principles simply by exploiting the fact that
\be
P(S_n\in [s,s+\D s])=p(S_n=s)\, \D s,
\label{eqdens1}
\ee
where $p(S_n=s)$ is the probability density of $S_n$, in order to write
\be
P(S_n\in [s,s+\D s])\approx \E^{-nJ(s)}\, \D s.
\ee
Proceeding with $P(S_n\in[s,s+\D s])$, the rate function $J(s)$ is then recovered, as in the case of discrete probability distributions, by taking the large deviation limit. Thus
\be
\lim_{n\ra\infty} -\frac{1}{n}\ln P(S_n\in [s,s+\D s])=J(s)+\lim_{n\ra\infty} \frac{1}{n}\ln \D s=J(s),
\ee
where the last equality follows by assuming that $\D s$ is an arbitrary but non-zero infinitesimal element.

\subsection{More on the large deviation principle}

The limit defining the large deviation principle, as most limits appearing in this review, should be understood at a practical rather than rigorous level. Likewise, our definition of the large deviation principle should not be taken as a rigorous definition. In fact, it is not. In dealing with probabilities and limits, there are many mathematical subtleties that need to be taken into account (see Appendix~\ref{appldt}). Most of these subtleties will be ignored in this review, but it may be useful to mention two of them:
\begin{itemize}
\item The limit involved in the definition of the large deviation principle may not exist. In this case, one may still be able to find an upper bound and a lower bound on $P(A_n\in B)$ that are both exponential in $n$:
\be
\E^{-nI_B^-}\leq P(A_n\in B)\leq \E^{-nI_B^+}.
\ee
The two bounds give a precise meaning to the statement that $P(A_n\in B)$ is decaying exponentially with $n$, and give rise to two large deviation principles: one defined in terms of a ``limit inferior'' yielding $I_B^-$, and one defined with a ``limit superior'' yielding $I_B^+$. This approach, which is the one followed by mathematicians, is described in Appendix~\ref{appldt}. For the purposes of this review, we make the simplifying assumption that $I_B^-=I_B^+$ always holds; hence our definition of the large deviation principle involving a simple limit.

\item Discrete random variables are often treated as if they become continuous in the limit $n\ra\infty$. Such a ``discrete to continuous'' limit, or \emph{continuum limit} as it is known in physics, was implicit in many examples of the previous section. In the first example, for instance, we noted that the proportion $R_n$ of $1$'s in a random bit sequence of length $n$ could only assume a rational value. As $n\ra\infty$, the set of values of $R_n$ becomes dense in $[0,1]$, so it is useful in this case to picture $R_n$ as being a continuous random variable taking values in $[0,1]$. Likewise, in Example~\ref{exentspin} we implicitly treated the mean magnetization $m$ as a continuous variable, even though it assumes only rational values for $n<\infty$. In both examples, the large deviation approximations that we derived were continuous approximations involving continuous rate functions.
\end{itemize}

The replacement of discrete random variables by continuous random variables is justified mathematically by the notion of weak convergence. Let $A_n$ be a discrete random variable with probability distribution $P(A_n=a)$ defined on a subset of values $a\in\reals$, and let $\tilde A_n$ be a continuous random variable with probability density $p(\tilde A_n)$ defined on $\reals$. To say that $A_n$ converges weakly to $\tilde A_n$ means, essentially, that any sum involving $A_n$ can be approximated, for $n$ large, by integrals involving $\tilde A_n$, i.e.,
\be
\sum_{a} f(a) P(A_n=a)\stackrel{n\ra\infty}{\approx}\int f(a)\, p(\tilde A_n=a)\, \D a,
\ee
where $f$ is any continuous and bounded function defined over $\reals$. This sort of approximation is common in physics, and suggests the following replacement rule:
\be
P(A_n=a)\longrightarrow p(\tilde A_n=a)\, \D a
\ee
as a formal device for taking the continuum limit of $A_n$. For more information on the notion of weak convergence, the reader is referred to \cite{dupuis1997,capinski2005} and Appendix~\ref{appldt} of this review.

Most of the random variables considered in this review, and indeed in large deviation theory, are either discrete random variables that weakly converge to continuous random variables or are continuous random variables right from the start. To treat these two cases with the same notation, we will try to avoid using probability densities whenever possible, to consider instead probabilities of the form $P(A_n\in [a,a+\D a])$. To further cut in the notations, we will also avoid using a tilde for distinguishing a discrete random variable from its continuous approximation, as done above with $A_n$ and $\tilde A_n$. From now on we thus write
\be
P(A_n\in [a,a+\D a])\approx \E^{-nI(a)}\D a.
\label{eqldp3}
\ee
to mean that $A_n$, whether discrete or continuous, satisfies a large deviation principle. This choice of notation is convenient but arbitrary: readers who prefer probability densities may express a large deviation principle for $A_n$ in the density form $p(A_n=a)\approx \E^{-nI(a)}$ instead of the expression shown in (\ref{eqldp3}). In this way, one need not bother with the infinitesimal element $\D a$ in the statement of the large deviation principle. In this review, we will use the probability notation shown in (\ref{eqldp3}), which has to include the infinitesimal element $\D a$, even though this element is not exponential in $n$. Indeed, without the element $\D a$, the following expectation value would not make sense:
\be
\lex f(A_n)\rex=\int f(a)\, P(A_n\in [a,a+\D a])\asymp \int f(a)\, \E^{-nI(a)}\, \D a.
\ee

There are two final pieces of notation that need to be introduced before we go deeper into the theory of large deviations. First, we will use the more compact expression $P(A_n\in \D a)$ to mean $P(A_n\in [a,a+\D a])$. Next, we will follow Ellis~\cite{ellis1995} and use the sign ``$\asymp$'' instead of ``$\approx$'' whenever we treat large deviation principles. In the end, we thus write 
\be
P(A_n\in \D a)\asymp \E^{-nI(a)}\, \D a
\label{eqldp4}
\ee
to mean that $A_n$ satisfies a large deviation principle, in the sense of (\ref{eqldp2}), with rate function $I(a)$. The sign ``$\asymp$'' is used to stress that, as $n\ra\infty$, the dominant part of $P(A_n\in \D a)$ is the decaying exponential $\E^{-nI(a)}$. We may also interpret the sign ``$\asymp$'' as expressing an equality relationship on a logarithmic scale; that is, we may interpret $a_n\asymp b_n$ as meaning that
\be
\lim_{n\ra\infty} \frac{1}{n}\ln a_n=\lim_{n\ra\infty} \frac{1}{n}\ln b_n .
\ee   
We say in this case that $a_n$ and $b_n$ are equal up to first order in their exponents \cite{cover1991}.

\subsection{Calculating rate functions} 

The theory of large deviations can be described from a practical point of view as a collection of methods that have been developed and gathered together in one toolbox to solve two problems \cite{dembo1998}: \begin{itemize}
\item Establish that a large deviation principle exists for a given random variable; 
\item Derive the expression of the associated rate function.
\end{itemize}

Both of these problems can be addressed, as we have done in the examples of the previous section, by directly calculating the probability distribution of a random variable, and by deriving from this distribution a large deviation approximation using Stirling's approximation or other asymptotic formulae. In general, however, it may be difficult or even impossible to derive large deviation principles through this direct calculation path. Combinatorial methods based on Stirling's approximation cannot be used, for example, for continuous random variables, and become quite involved when dealing with sums of discrete random variables that are non-IID. For these cases, a more general calculation path is provided by a fundamental result of large deviation theory known as the G\"artner-Ellis Theorem \cite{gartner1977,ellis1984}. What we present next is a simplified version of that theorem, which is sufficient for the applications covered in this review; for a more complete presentation, see Sec.~5 of \cite{ellis1995} and Sec.~2.3 of \cite{dembo1998}.

\subsubsection{The G\"artner-Ellis Theorem}

Consider a real random variable $A_n$ parameterized by the positive integer $n$, and define the \emph{scaled cumulant generating function} of $A_n$ by the limit
\be
\lambda(k)=\lim_{n\ra\infty} \frac{1}{n}\ln \left\langle \E^{nkA_n}\right\rangle,
\ee
where $k\in\reals$ and
\be
\left\langle \E^{nkA_n}\right\rangle=\int_\reals \E^{nka}\, P(A_n\in \D a).
\label{eqexpval1}
\ee 
The G\"artner-Ellis Theorem states that, if $\lambda(k)$ exists and is differentiable for all $k\in\reals$, then $A_n$ satisfies a large deviation principle, i.e., 
\be
P(A_n\in \D a)\asymp \E^{-nI(a)}\, \D a,
\ee
with a rate function $I(a)$ given by
\be
I(a)=\sup_{k\in\reals} \{ka-\lambda(k)\}.
\label{eqlf1}
\ee
The symbol ``$\sup$'' above stands for ``supremum of'', which for us can be taken to mean the same as ``maximum of''. The transform defined by the supremum is an extension of the Legendre transform referred to as the \emph{Legendre-Fenchel transform} \cite{rockafellar1970}. The G\"artner-Ellis Theorem thus states in words that, when the scaled cumulant generating function $\lambda(k)$ of $A_n$ is differentiable, then $A_n$ obeys a large deviation principle with a rate function $I(a)$ given by the Legendre-Fenchel transform of $\lambda(k)$.

The next sections will show how useful the G\"artner-Ellis Theorem is for calculating rate functions. It is important to know, however, that not all rate functions can be calculated with this theorem. Some examples of rate functions that cannot be calculated as the Legendre-Fenchel transform of $\lambda(k)$, even though $\lambda(k)$ exists, will be studied in Sec.~\ref{subsecncrf}. The argument presented next is meant to give some insight as to why $I(a)$ can be expressed as the Legendre-Fenchel transform of $\lambda(k)$ when $\lambda(k)$ is differentiable. A full understanding of this argument will also come in Sec.~\ref{subsecncrf}.

\subsubsection{Plausibility argument for the G\"artner-Ellis Theorem}

Two different derivations of the G\"artner-Ellis Theorem are given in Appendix~\ref{appge}. To gain some insight into this theorem, we derive here the second part of this theorem, namely Eq.~(\ref{eqlf1}), by assuming that a large deviation principle holds for $A_n$, and by working out the consequences of this assumption. To start, we thus assume that
\be
P(A_n\in \D a)\asymp \E^{-nI(a)}\, \D a,
\ee
and insert this approximation into the expectation value defined in Eq.~(\ref{eqexpval1}) to obtain
\be
\left\langle \E^{nkA_n}\right\rangle\asymp\int_\reals \E^{n[ka-I(a)]}\, \D a.
\label{eqint1} 
\ee
Next, we approximate the integral by its largest integrand, which is found by locating the maximum of $ka-I(a)$. This approximation, which is known as the \emph{saddle-point approximation} or \emph{Laplace's approximation}\footnote{The saddle-point approximation is used in connection with integrals in the complex plane, whereas Laplace's approximation or Laplace's method is used in connection with real integrals (see Chap.~6 of \cite{bender1978}).}, is a natural approximation to consider here because the error associated with it is of the same order as the error associated with the large deviation approximation itself. Therefore, assuming that the maximum of $ka-I(a)$ exists and is unique, we write
\be
\left\langle \E^{nkA_n}\right\rangle\asymp \exp\left(n \sup_{a\in\reals}\{ka-I(a)\}\right)
\ee
and so
\be
\lambda(k)=\lim_{n\ra\infty}\frac{1}{n}\ln\lex \E^{nkA_n}\rex=\sup_{a\in\reals}\{ka-I(a)\}.
\ee
To obtain $I(a)$ in terms of $\lambda(k)$, we then use the fact that Legendre-Fenchel transforms can be inverted when $\lambda(k)$ is everywhere differentiable (see Sec.~26 of \cite{rockafellar1970}). In this case, the Legendre-Fenchel transform is \emph{self-inverse} (we also say \emph{involutive} or \emph{self-dual}), so that
\be
I(a)=\sup_{k\in\reals} \{ka-\lambda(k)\},
\ee 
which is the result of Eq.~(\ref{eqlf1}).

This heuristic derivation illustrates two important points about large deviation theory. The first is that Legendre-Fenchel transforms appear into this theory as a natural consequence of Laplace's approximation. The second is that the G\"artner-Ellis Theorem is essentially a consequence of the large deviation principle combined with Laplace's approximation. This point is illustrated in Appendix~\ref{appge}, and will be discussed again in the context of another important result of large deviation theory known as Varadhan's Theorem. 

\subsection{Cram\'er's Theorem}

The application of the G\"artner-Ellis Theorem to a sample mean
\be
S_n=\frac{1}{n}\sum_{i=1}^n X_i
\ee
of independent and identically distributed (IID) random variables yields a classical result of probability theory known as \emph{Cram\'er's Theorem} \cite{cramer1938}. In this case, the scaled cumulant generating function has the simple form
\be
\lambda(k)=\lim_{n\ra\infty}\frac{1}{n}\ln \left\langle \E^{k\sum_{i=1}^n X_i}\right\rangle=\lim_{n\ra\infty} \frac{1}{n}\ln\prod_{i=1}^n \left\langle \E^{kX_i}\right\rangle =\ln \left\langle \E^{kX}\right\rangle, 
\ee
where $X$ is any of the summands $X_i$. As a result, one derives a large deviation principle for $S_n$ simply by calculating the \emph{cumulant generating function} $\ln \langle \E^{kX}\rangle$ of a single summand, and by taking the Legendre-Fenchel transform of the result. The next examples illustrate these steps. Note that the differentiability condition of the G\"artner-Ellis Theorem need not be checked for IID sample means because the \emph{generating function} or \emph{Laplace transform} $\langle \E^{kX}\rangle$ of a random variable $X$ is always real analytic when it exists for all $k\in\reals$ (see Theorem~VII.5.1 of \cite{ellis1985}). 

\begin{example}[Gaussian sample mean revisited]
\label{exgaussrev}
Consider again the sample mean $S_n$ of $n$ Gaussian IID random variables considered in Example~\ref{exgauss}. For the Gaussian density of Eq.~(\ref{eqgauss1}), $\lambda(k)$ is easily evaluated to be
\be
\lambda(k)=\ln\lex \E^{kX}\rex=\mu k+\frac{1}{2}\sigma^2 k^2,\quad k\in\reals.
\ee
As expected, $\lambda(k)$ is everywhere differentiable, so that $P(S_n\in \D s)\asymp \E^{-nI(s)}\, \D s$ with
\be
I(s)=\sup_{k} \{ks-\lambda(k)\}.
\ee
This recovers Cram\'er's Theorem. The supremum defining the Legendre-Fenchel transform is solved directly by ordinary calculus. The result is
\be
I(s)=k(s)s-\lambda(k(s))=\frac{(s-\mu)^2}{2\sigma^2},\quad s\in\reals,
\ee 
where $k(s)$ is the unique maximum point of $ks-\lambda(k)$ satisfying $\lambda'(k)=s$. This recovers exactly the result of Example~\ref{exgauss} knowing that $P(S_n\in \D s)=p(S_n=s)\D s$.
\end{example}

\begin{example}[Exponential sample mean revisited]
\label{exexprev}
The calculation of the previous example can be carried out for the exponential sample mean studied in Example~\ref{exexp}. In this case, we find
\be
\lambda(k)=-\ln(1-\mu k),\quad k<1/\mu.
\ee
From Cram\'er's Theorem, we then obtain $P(S_n\in \D s)\asymp \E^{-nI(s)}\, \D s$, where
\be
I(s)=\sup_k\{ks-\lambda(k)\}=k(s)s-\lambda(k(s))=\frac{s}{\mu}-1-\ln\frac{s}{\mu},\quad s>0,
\ee
in agreement with the result announced in (\ref{eqexp1}). It is interesting to note here that the singularity of $\lambda(k)$ at $1/\mu$ translates into a branch of $I(s)$ which is asymptotically linear. This branch of $I(s)$ translates, in turn, into a tail of $P(S_n\in \D s)$ which is asymptotically exponential. If the probability density of the IID random variables is chosen to be a double-sided rather than a single-sided exponential distribution, then both tails of $P(S_n\in \D s)$ become asymptotically exponential.
\end{example}

We will study other examples of IID sums, as well as sums involving non-IID random variables in Sec.~\ref{secmathapp}. It should be clear at this point that the scope of the G\"artner-Ellis Theorem is not limited to IID random variables. In principle, the theorem can be applied to any random variable, provided that one can calculate the limit defining $\lambda(k)$ for that random variable, and that $\lambda(k)$ satisfies the conditions of the theorem. Examples of sample means of random variables for which $\lambda(k)$ fail to meet these conditions will be presented also in Sec.~\ref{secmathapp}.

\subsection{Properties of $\lambda$ and $I$}

We now state and prove a number of properties of scaled cumulant generating functions and rate functions in the case where the latter is obtained via the G\"artner-Ellis Theorem. The properties listed hold for an arbitrary random variable $A_n$ under the conditions stated, not just sample means of IID random variables.

\subsubsection{Properties of $\lambda$ at $k=0$}
\label{propk0}

\begin{figure*}[t]
\centering
\includegraphics[scale=1.0]{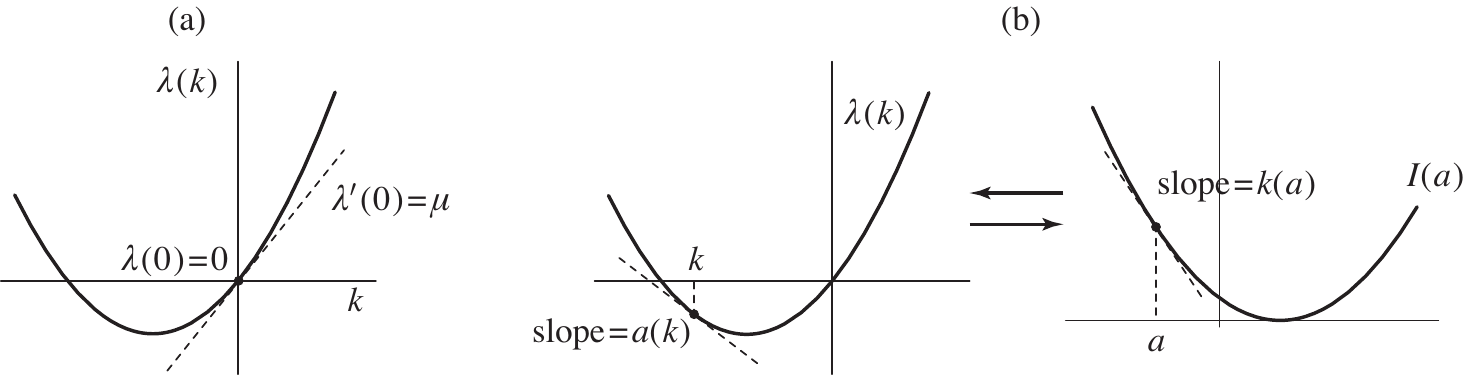}
\caption{(a) Properties of $\lambda(k)$ at $k=0$. (b) Legendre duality: the slope of $\lambda$ at $k$ is the point at which the slope of $I$ is $k$.}
\label{figk0}
\end{figure*}

Since probability measures are normalized, $\lambda(0)=0$. Moreover,
\be
\lambda'(0)=\lim_{n\ra\infty} \left. \frac{\lex A_n \E^{nkA_n}\rex}{\lex \E^{nkA_n}\rex}\right|_{k=0}=\lim_{n\ra\infty} \langle A_n\rangle,
\label{eqm1}
\ee
provided that $\lambda'(0)$ exists. For IID sample means, this reduces to $\lambda'(0)=\langle X\rangle=\mu$; see Fig.~\ref{figk0}(a). Similarly,
\be
\lambda''(0)=\lim_{n\ra\infty} n\left(\langle A_n^2\rangle-\langle A_n\rangle^2\right)=\lim_{n\ra\infty} n\,\var (A_n),
\label{eqvar1}
\ee
which reduces to $\lambda''(0)=\var(X)=\sigma^2$ for IID sample means. 

\subsubsection{Convexity of $\lambda$}
\label{propconv}

The function $\lambda(k)$ is always convex. This comes as a general consequence of H\"older's inequality:
\be
\sum_i |y_i z_i| \leq \left(\sum_i |y_i|^{1/p}\right)^p \left(\sum_i |z_i|^{1/q}\right)^{q},
\ee
where $0\leq p,q \leq 1$, $p+q=1$. Applying this inequality to $\lambda(k)$ yields
\be
\alpha \ln \lex \E^{nk_1 A_n}\rex +(1-\alpha)\ln\lex \E^{nk_2 A_n}\rex\geq \ln\lex \E^{n[\alpha k_1+(1-\alpha)k_2]A_n}\rex
\ee
for $\alpha\in[0,1]$. Hence,
\be
\alpha\lambda(k_1)+(1-\alpha)\lambda(k_2)\geq \lambda\left( \alpha k_1 +(1-\alpha)k_2\right)
\ee 
A particular case of this inequality, which defines a function as being convex \cite{rockafellar1970}, is $\lambda(k)\geq k \lambda'(0)=k\mu$; see Fig.~\ref{figk0}(a). Note that the convexity of $\lambda(k)$ directly implies that $\lambda(k)$ is continuous in the interior of its domain, and is differentiable everywhere except possibly at a denumerable number of points \cite{rockafellar1970,tiel1984}.

\subsubsection{Legendre transform and Legendre duality}

We have seen when calculating the rate functions of the Gaussian and exponential sample means that the Legendre-Fenchel transform involved in the G\"artner-Ellis Theorem reduces to 
\be
I(a)=k(a)a-\lambda(k(a)),
\ee
where $k(a)$ is the unique root of $\lambda'(k)=a$. This equation plays a central role in this review: it defines, as is well known, the \emph{Legendre transform} of $\lambda(k)$, and arises in the examples considered before because $\lambda(k)$ is everywhere differentiable, as required by the G\"artner-Ellis Theorem, and because $\lambda(k)$ is convex, as proved above. These conditions---differentiability and convexity---are the two essential conditions for which the Legendre-Fenchel transform reduces to the better known Legendre transform (see Sec.~26 of \cite{rockafellar1970}).

An important property of Legendre transforms holds when $\lambda(k)$ is differentiable and is \emph{strictly convex}, that is, convex with no linear parts. In this case, $\lambda'(k)$ is monotonically increasing, so that the function $k(a)$ satisfying $\lambda'(k(a))=a$ can be inverted to obtain a function $a(k)$ satisfying $\lambda'(k)=a(k)$. From the equation defining the Legendre transform, we then have $I'(a(k))=k$ and $I'(a)=k(a)$. Therefore, in this case---and this case only---the slopes of $\lambda$ are one-to-one related to the slopes of $I$. This property, which we refer as the \emph{duality property} of the Legendre transform, is illustrated in Fig.~\ref{figk0}(b). 

The next example shows how border points where $\lambda(k)$ diverges translate, by Legendre duality, into branches of $I(a)$ that are linear or asymptotically linear.\footnote{Recall that, because $\lambda(k)$ is a convex function, it cannot have diverging points in the interior of its domain.} A specific random variable for which this duality behavior shows up is the sample mean of exponential random variables studied in Example~\ref{exexprev}. Since we can invert the roles of $\lambda(k)$ and $I(a)$ in the Legendre transform, this example can also be generalized to show that points where $I(a)$ diverges are associated with branches of $\lambda(k)$ that are linear or asymptotically linear; see Example~\ref{exfrac1}. These sorts of diverging points and linear branches arise often in physical applications, for example, in relation to nonequilibrium fluctuations; see Sec.~\ref{subsecfr}.

\begin{figure*}[t]
\centering
\includegraphics[scale=1.0]{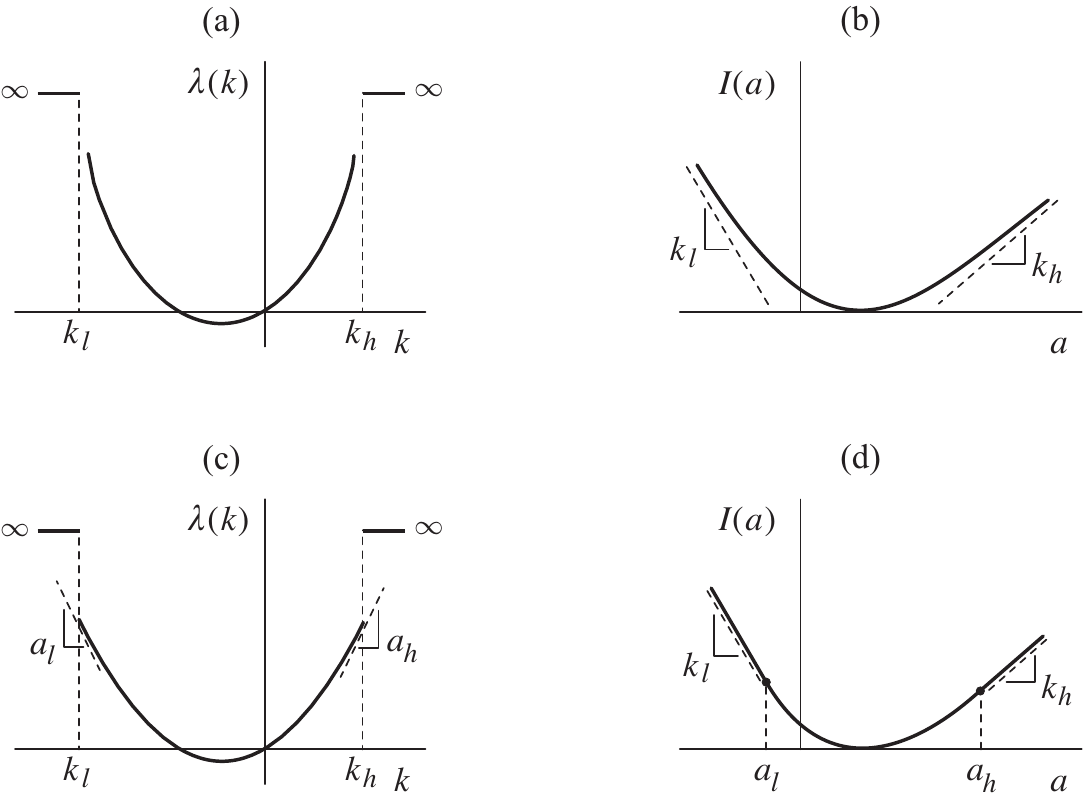}
\caption{(a) Scaled cumulant generating function $\lambda(k)$ defined on the open domain $(k_l,k_h)$, with diverging slopes at the boundaries. (b) The Legendre transform $I(a)$ of $\lambda(k)$ is asymptotically linear as $|k|\ra\infty$. The asymptotic slopes correspond to the boundaries of the region of convergence of $\lambda(k)$. (c) $\lambda(k)$ is defined on $(k_l,k_h)$ as in (a) but has finite slopes at the boundaries. (d) Legendre-Fenchel transform $I(a)$ of the function $\lambda(k)$ shown in (c). The function $I(a)$ has branches that are linear rather than just asymptotically linear, with slopes corresponding to the boundaries of the region of convergence of $\lambda(k)$.}
\label{figbp1}
\end{figure*}

\begin{example}
\label{exbp1}
Consider the scaled cumulant generating function $\lambda(k)$ shown in Fig.~\ref{figbp1}(a). This function has the particularity that it is defined only on a bounded (open) interval $(k_l,k_h)$, and has diverging slopes at the boundaries, that is, $\lambda'(k)\ra\infty$ as $k$ approaches $k_h$ from below and $\lambda'(k)\ra -\infty$ as $k$ approaches $k_l$ from above. To determine the shape of the Legendre transform of $\lambda(k)$, which corresponds to the rate function $I(a)$ associated with $\lambda(k)$ (assume that $\lambda(k)$ is everywhere differentiable), we simply need to use Legendre duality. On the one hand, since the slope of $\lambda(k)$ diverges as $k$ approaches $k_h$, the slope of $I(a)$ must approach the constant $k_h$ as $a\ra\infty$ (remember that slopes of $\lambda$ are abscissas of $I$). On the other hand, since the slope of $\lambda(k)$ goes to $-\infty$ as $k$ approaches $k_l$, the slope of $I(a)$ must approach the constant $k_l$ as $a\ra -\infty$. Overall, $I(a)$ is thus asymptotically linear; see Fig.~\ref{figbp1}(b).

Now suppose that rather than having diverging slopes at the boundaries $k_l$ and $k_h$, $\lambda(k)$ has finite slopes $a_l$ and $a_h$, respectively; see Fig.~\ref{figbp1}(c). What is the rate function $I(a)$ associated with this form of $\lambda(k)$? The answer, surprisingly, is that there is not one but \emph{many} rate functions that may correspond to this $\lambda(k)$. One such rate function is the Legendre-Fenchel transform of $\lambda(k)$ shown in Fig.~\ref{figbp1}(d). This function has the particularity that it has two linear branches which arise, as before, because of the two boundary points of $\lambda(k)$. The difference here is that these branches are really linear, and not just \emph{asymptotically} linear, because the left-derivative of $\lambda(k)$ at $k_h$ is finite, and so is its right-derivative at $k_l$. To understand why these linear branches appear, one must appeal to a generalization of Legendre duality involving the concept of ``supporting lines'' \cite{rockafellar1970}. We will not discuss this concept here; suffice it to say that the value of the left-derivative of $\lambda(k)$ at $k_h$ corresponds to the starting point of the linear branch of $I(a)$ with slope $k_h$. Similarly, the right-derivative of $\lambda(k)$ at $k_l$ corresponds to the endpoint of the linear branch of $I(a)$ with slope $k_l$. 
\end{example}

The reason why the rate function shown in Fig.~\ref{figbp1}(d) is but one candidate rate function associated with the $\lambda(k)$ shown in Fig.~\ref{figbp1}(c) is explained in Sec.~\ref{subsecncrf}. The reason has to do, essentially, with the fact that $\lambda(k)$ is nondifferentiable at its boundaries. In large deviation theory, one says more precisely that $\lambda(k)$ is \emph{non-steep}; see the notes at the end of this section for more information about this concept.

\subsubsection{Varadhan's Theorem}
\label{propvara}
In our heuristic derivation of the G\"artner-Ellis Theorem, we showed that if $A_n$ satisfies a large deviation principle with rate function $I(a)$, then $\lambda(k)$ is the Legendre-fenchel transform of $I(a)$:
\be
\lambda(k)=\lim_{n\ra\infty}\frac{1}{n}\lex \E^{nkA_n}\rex=\sup_a\{ka-I(a)\}.
\label{eqvara1}
\ee
Replacing the product $kA_n$ by an arbitrary continuous function $f$ of $A_n$ yields the more general result
\be
\lambda(f)=\lim_{n\ra\infty} \frac{1}{n}\ln \lex \E^{nf(A_n)}\rex=\sup_{a}\{ f(a)-I(a)\},
\label{eqvara2}
\ee
which is known as \emph{Varadhan's Theorem} \cite{varadhan1966}. The function $\lambda(f)$ thus defined is a \emph{functional} of $f$, as it is a function of the function $f$. 

As we did for the result shown in (\ref{eqvara1}), we can justify (\ref{eqvara2}) as a consequence of the large deviation principle for $A_n$ and Laplace's approximation. It is important to note, however, that Varadhan's Theorem is a consequence of Laplace's approximation only when $A_n$ is a real random variable; for other types of random variables, such as random functions, Varadhan's Theorem still applies, and so \emph{extends} Laplace's approximation to these random variables. Varadhan's Theorem also holds when $f(a)-I(a)$ has more than one maximum, that is, when the integral defining the expected value $\langle \E^{nf(A_n)}\rangle$ has more than one saddle-point. We will come back to this point in Sec.~\ref{secmathapp} when discussing nonconvex rate functions, and again in Sec.~\ref{secequi} when discussing nonconcave entropies.

\subsubsection{Positivity of rate functions}

Rate functions are always positive. This follows by noting that $\lambda(0)=0$ and that $\lambda(k)$ can always be expressed as the Legendre-Fenchel transform of $I(a)$. Hence,
\be
\lambda(0)=\sup_a\{-I(a)\}=-\inf_{a} I(a)=0,
\ee
where ``$\inf$'' denotes the ``infimum of''. A negative rate function would imply that $P(A_n\in \D a)$ diverges as $n\ra\infty$.

\subsubsection{Convexity of rate functions}
\label{propconvex}

Rate functions obtained from the G\"artner-Ellis Theorem are necessarily \emph{strictly convex}, that is, they are convex and have no linear parts.\footnote{This does not mean that all rate functions are strictly convex----only that those obtained from the G\"artner-Ellis Theorem are strictly convex.} That Legendre-Fenchel transforms yield convex functions is easily proved from the definition of these transforms \cite{tiel1984}. To prove that they yield \emph{strictly} convex functions when $\lambda(k)$ is differentiable is another matter; see, e.g., Sec.~26 of \cite{rockafellar1970}. As a special case of interest, let us assume that $\lambda(k)$ is differentiable and has no linear parts, as in our discussion of the Legendre duality property. In this case, the Legendre-Fenchel transform reduces to a Legendre transform, as noted earlier, and the equation defining the Legendre transform then implies
\be
I''(a)=k'(a)=\frac{1}{\lambda''(k)}.
\ee
Since $\lambda(k)$ is convex with no linear parts ($\lambda''(k)> 0$), $I(a)$ must then also be convex with no linear parts ($I''(a)> 0$). This shows, incidentally, that the curvature of $I(a)$ is the inverse curvature of $\lambda(k)$. In the case of IID sample means, in particular, 
\be
I''(a=\mu)=\frac{1}{\lambda''(0)}=\frac{1}{\sigma^2}.
\ee
A similar result holds for non-IID random variables by replacing $\sigma^2$ with the general result of Eq.~(\ref{eqvar1}).

\begin{figure*}[t]
\centering
\includegraphics[scale=1.0]{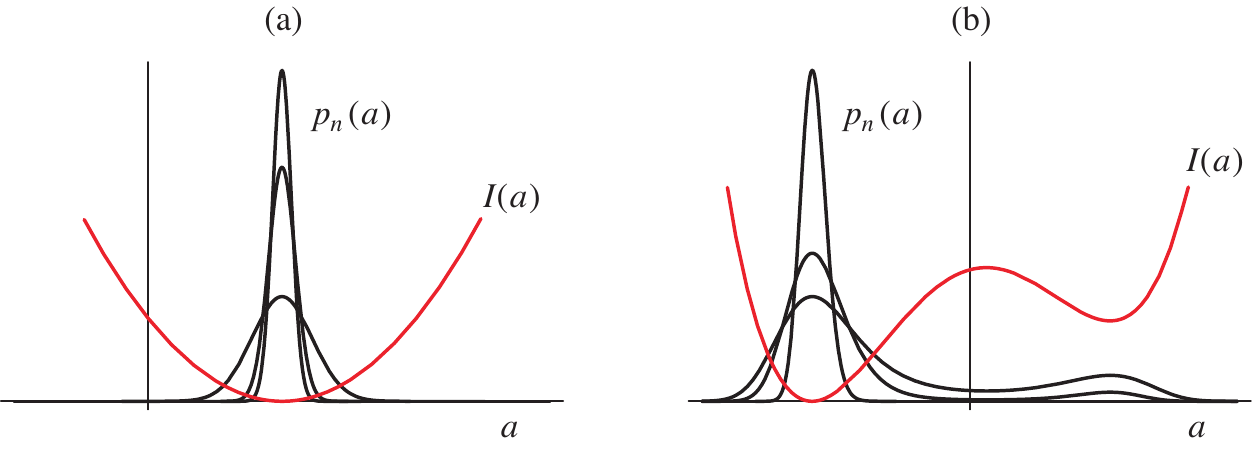}
\caption{(a) Example of a unimodal probability density $p_n(a)$ shown for increasing values of $n$ (black line), and its corresponding convex rate function $I(a)$ (red line). (b) Example of a bimodal probability density $p_n(a)$ shown for increasing values of $n$ characterized by a nonconvex rate function $I(a)$ having a local minimum in addition to its global minimum.}
\label{figratefct1}
\end{figure*}

\subsubsection{Law of Large Numbers}
\label{proplln}

If $I(a)$ has a unique global minimum and zero $a^*$, then
\be
a^*=\lambda'(0)=\lim_{n\ra\infty}\langle A_n\rangle.
\label{eqlln1}
\ee
by Eq.~(\ref{eqm1}). If $I(a)$ is differentiable at $a^*$, we further have $I'(a^*)=k(a^*)=0$. To prove this property, simply apply the Legendre duality property:
\be
I(a^*)=k(a^*) a^*-\lambda(k(a^*))=0\cdot a^*-0=0.
\ee

The global minimum and zero of $I(a)$ has a special property that we noticed already: it corresponds, if it is unique, to the only value at which $P(A_n\in \D a)$ does not decay exponentially, and so around which $P(A_n\in \D a)$ gets more and more concentrated as $n\ra\infty$; see Fig.~\ref{figratefct1}(a). Because of the concentration effect, we have
\be
\lim_{n\ra\infty} P(A_n\in \D a^*)=\lim_{n\ra\infty} P(A_n\in [a^*,a^*+\D a])=1,
\ee
as noted already in Eq.~(\ref{llng1}), and so we call $a^*$ the \emph{most probable} or \emph{typical} value of $A_n$. The existence of this typical value is an expression of the Law of Large Numbers, which states in its weak form that $A_n\ra a^*$ with probability 1. An important observation here is that large deviation theory extends the Law of Large Numbers by providing information as to \emph{how fast} $A_n$ converges in probability to its mean. To be more precise, let $B$ be any set of values of $A_n$. Then
\be
P(A_n\in B)=\int_B P(A_n\in \D a)\asymp \int_B \E^{-nI(a)}\, \D a\asymp \E^{-n\inf_{a\in B}I(a)}
\ee
by applying Laplace's approximation. Therefore, $P(A_n\in B)\ra 0$ exponentially fast with $n$ if $a^*\notin B$, which means that $P(A_n\in B)\ra 1$ exponentially fast with $n$ if $a^*\in B$.

In general, the existence of a Law of Large Numbers for a random variable $A_n$ is a good sign that a large deviation principle holds for $A_n$. In fact, this law can often be used as a point of departure for deriving large deviation principles; see \cite{oconnell1997a,oconnell2000} and Appendix~\ref{appge}. It should be emphasized, however, that $I(a)$ may have more than one global minimum, in which case the Law of Large Numbers may not hold. Rate functions may even have local minima in addition to global ones. The global minima yield typical values of $A_n$ just as in the case of a single minimum, whereas the local minima yield what physicists would call ``metastable'' values of $A_n$ at which $P(A_n\in \D a)$ is locally but not globally maximum; see Fig.~\ref{figratefct1}(b). Physicists would also call a typical value of $A_n$, determined by a global minimum of $I(a)$, an ``equilibrium state''. We will come to this language in Sec.~\ref{secequi}.

\subsubsection{Gaussian fluctuations and the Central Limit Theorem}
\label{propclt}

The Central Limit Theorem arises in large deviation theory when a convex rate function $I(a)$ possesses a single global minimum and zero $a^*$, and is twice differentiable at $a^*$. Approximating $I(a)$ with the first quadratic term,
\be
I(a)\approx\frac{1}{2}I''(a^*)(a-a^*)^2,
\ee
then naturally leads to the Gaussian approximation
\be
P(A_n\in \D a)\approx \E^{- nI''(a^*)(a-a^*)^2/2}\, \D a,
\ee
which can be thought of as a weak form of the Central Limit Theorem. More precise results relating the Central Limit Theorem to the large deviation principle can be found in \cite{martin-lof1982,bryc1993}. We recall that for sample means of IID random variables, $I''(a^*)=1/\lambda''(0)=1/\sigma^2$; see Sec.~\ref{propconvex}.

The Gaussian approximation displayed above can be shown to be accurate for values of $A_n$ around $a^*$ of the order $O(n^{-1/2})$ or, equivalently, for values of $nA_n$ around $a^*$ of the order $O(n^{1/2})$. This explains the meaning of the name ``large deviations''. On the one hand, a \emph{small} deviation of $A_n$ is a value $A_n=a$ for which the quadratic expansion of $I(a)$ is a good approximation of $I(a)$, and for which, therefore, the Central Limit Theorem yields essentially the same information as the large deviation principle. On the other hand, a \emph{large} deviation is a value $A_n=a$ for which $I(a)$ departs sensibly from its quadratic approximation, and for which, therefore, the Central Limit Theorem yields no useful information about the large fluctuations of $A_n$ away from its mean. In this sense, large deviation theory can be seen as a generalization of the Central Limit Theorem characterizing the small as well as the large fluctuations of a random variable. Large deviation theory also generalizes the Central Limit Theorem whenever $I(a)$ exists but has no quadratic Taylor expansion around its minimum; see Examples~\ref{ex2disingm} and \ref{ex2disingc}. Note finally that having a Central Limit Theorem for $A_n$ does not imply that $I(a)$ has a quadratic minimum. A classic counterexample is presented next.

\begin{example}[Sample mean of double-sided Pareto random variables]
\label{expareto1}
Let $S_n$ be a sample mean of $n$ IID random variables $X_1,X_2,\ldots,X_n$ distributed according to the so-called Pareto density
\be
p(x)= \frac{A}{(|x|+c)^{\beta}},
\ee 
with $\beta> 3$, $c$ a real, positive constant, and $A$ a normalization constant. Since the variance of the summands is finite for $\beta>3$, the Central Limit Theorem holds for $n^{1/2} S_n$. Yet it can be verified that the rate function of $S_n$ is everywhere equal to zero because the probability density of $S_n$ has power-law tails similar to those of $p(x)$ \cite{lanford1973}. Note also that the scaled generating function $\lambda(k)$ is diverging for all $k\in\reals$ except $k=0$. 
\end{example}

We will study in the next section another example of sample mean involving a power-law probability density similar to the Pareto density. This time, the power-law density will be one-sided rather than double-sided, and the rate function will be seen to be different from zero for some values of the sample mean.

\subsection{Contraction principle}
\label{subseccp}

We have seen at this point two basic results of large deviation theory. The first is the G\"artner-Ellis Theorem, which can be used to prove that a large deviation principle exists and to calculate the associated rate function from the knowledge of $\lambda(k)$. The second result is Varadhan's Theorem, which can be used to calculate $\lambda(k)$ from the knowledge of a rate function. The last result that we now introduce is a useful calculation device, called the \emph{contraction principle} \cite{donsker1983}, which can be used to calculate a rate function from the knowledge of another rate function. 

The problem addressed by the contraction principle is the following. We have a random variable $A_n$ satisfying a large deviation principle with rate function $I_A(a)$, and we want to find the rate function of another random variable $B_n$ such that $B_n=h(A_n)$, where $h$ is a continuous function. We call $h$ a \emph{contraction} of $A_n$, as this function may be many-to-one. To calculate the rate function of $B_n$ from that of $A_n$, we simply use the large deviation principle for $A_n$ and Laplace's approximation at the level of 
\be
P(B_n\in db)=\int_{\{a:h(a)=b\}} P(A_n\in \D a)
\ee
to obtain 
\be
P(B_n\in db)\asymp \exp{\left(-n \inf_{a:h(a)=b}I_A(a)\right)}\, \D a.
\ee
This shows that if a large deviation principle holds for $A_n$ with rate function $I_A(a)$, then a large deviation principle also holds for $B_n$,
\be
P(B_n\in db)\asymp \E^{-nI_B(b)}\, \D b,
\ee
with a rate function given by
\be
I_B(b)=\inf_{a:h(a)=b} I_A(a).
\label{eqcp1}
\ee
This general reduction of one rate function to another is what is called the contraction principle. If $h$ is a bijective function with inverse $h^{-1}$, then $I_B(b)=I_A(h^{-1}(b))$. Note also that $I_B(b)=\infty$ if there is no value $a$ such that $h(a)=b$, i.e., if the pre-image of $b$ is empty.

The interpretation of the contraction principle should be clear. Since probabilities in large deviation theory are measured on the exponential scale, the probability of any large fluctuation should be approximated, following Laplace's approximation, by the probability of the most probable (although improbable) event leading or giving rise to that fluctuation. We will see many applications of this idea in the next sections, including a derivation of the maximum entropy principle based on the contraction principle. The ``least improbable'' event underlying or leading to a large deviation---be it a ``state'' underlying a large deviation or a ``path'' leading to that deviation---is often referred to as a \emph{dominating} or \emph{optimal} point \cite{ney1983,bucklew1990}.

\subsection{Historical notes and further reading}
\label{subsecnotesldt}

Large deviation theory emerged as a general theory during the 1960s and 1970s from the independent works of Donsker and Varadhan \cite{varadhan1966,donsker1975,donsker1975a,donsker1976,donsker1983}, and Freidlin and Wentzell \cite{freidlin1984}. Prior to that period, large deviation results were known, but there was no unified and general framework that dealt with them. Among these results, it is worth noting Cram\'er's Theorem \cite{cramer1938}, Chebyshev's inequality \cite{dembo1998}, Sanov's Theorem \cite{sanov1961}, which had been anticipated by Boltzmann \cite{boltzmann1877} (see \cite{ellis1999}), as well as extensions of Cram\'er's Theorem obtained by Lanford \cite{lanford1973}, Bahadur and Zabell \cite{bahadur1979}, and by Plachky and Steinebach \cite{plachky1975}. Sanov's Theorem was already encountered in the introductory examples of Sec.~\ref{secexamples}, and will be treated again in the next section. What statisticians call saddle-point approximations (see, e.g., \cite{daniels1954,barndorff1989,butler2007}) are also large deviation results for the probability density of sample means; see Appendix~\ref{appge}. For more information on the development of large deviation theory, see the historical notes found in \cite{dembo1998,bucklew1990,dembo1998} as well as in Sec.~VII.7 of \cite{ellis1985}. 

The G\"artner-Ellis Theorem is the product of a result proved by G\"artner \cite{gartner1977}, which was later generalized by Ellis \cite{ellis1984}. The work of Ellis \cite{ellis1984} explicitly refers to the construction of the large deviation principle currently adopted in large deviation theory (see Appendix~\ref{appldt}), which stems from the work of Varadhan \cite{varadhan1966}. 

As noted before, the statement of the G\"artner-Ellis Theorem given here is a simplification of that theorem. In essence, the result that we have stated and used is that of G\"artner \cite{gartner1977}; it is less general but less technical than the result proved by Ellis \cite{ellis1984}, which can be applied to cases where $\lambda(k)$ exists and is differentiable over some limited interval (so not necessarily the whole line, as in G\"artner's result), provided that a technical condition, known as the \emph{steepness condition}, is verified. For a statement of this condition, see Theorem 5.1 of \cite{ellis1995} or Theorem 2.3.6 of \cite{dembo1998}; for an illustration of it, see Examples~\ref{exskew} and \ref{exnonsteep} of the next section.

The statement of Varadhan's Theorem given here is also a simplification of the original and complete result proved by Varadhan \cite{varadhan1966}; see, e.g., Theorem 4.3.1 in \cite{dembo1998} and Theorem 1.3.4 in \cite{dupuis1997}. An example of rate function for which the full conditions of Varadhan's Theorem are not satisfied is presented in Example~\ref{exnonsteep} of the next section. 

Introductions to the theory of large deviations similar to the one given in this section can be found in review papers by Oono \cite{oono1989}, Amann and Atmanspacher \cite{amann1999}, Ellis \cite{ellis1995,ellis1999}, Lewis and Russell \cite{lewis1996}, and Varadhan \cite{varadhan2003}. Readers who are willing to read mathematical textbooks are encouraged to consult those of Ellis \cite{ellis1985}, Deuschel and Stroock \cite{deuschel1989}, Dembo and Zeitouni \cite{dembo1998}, and den Hollander \cite{hollander2000} for a proper mathematical account of large deviation theory. The main simplifications introduced in this review concern the definition of the large deviation principle, and the fact that we do not state large deviation principles using the abstract language of topological spaces and measure theory. The precise and rigorous definition of the large deviation principle can be found in Appendix~\ref{appldt}.

For an accessible introduction to Legendre-Fenchel transforms and convex functions, see the monograph of van Tiel \cite{tiel1984} and Chap.~VI of \cite{ellis1985}. The definitive reference on convex analysis is the book by Rockafellar \cite{rockafellar1970}.

%%%%%%%%%%%%%%%%%%%%%%%%%%%%%%%%%%%%%%%%%%%%%%%
\section{Mathematical applications}
\label{secmathapp} 

This section is intended to complement the previous section. We review here a number of mathematical problems for which large deviation principles can be formulated. The applications were selected to give an idea of the generality of large deviation theory, to illustrate important points about the G\"artner-Ellis Theorem, and to introduce many ideas and results that will be revisited from a more physical point of view in the next sections. We also discuss here a classification of large deviation results related from top to bottom by the contraction principle.

\subsection{Sums of IID random variables} 
\label{subseciid}

We begin our review of mathematical applications by revisiting the now familiar sample mean
\be
S_n=\frac{1}{n}\sum_{i=1}^n X_i
\ee
involving $n$ IID random variables $X_1,X_2,\ldots,X_n$. The next three examples consider different cases of sample distributions for the $X_i$'s, and derive the corresponding large deviation principle for $S_n$ using the G\"artner-Ellis Theorem or, equivalently in this case, Cram\'er's Theorem. We start with an example closely related to the introductory example of Sec.~\ref{secexamples} which was concerned with spins.

\begin{example}[Binary random variables] 
\label{exbin}
Let the $n$ random variables in $S_n$ be such that $P(X_i=-1)=P(X_i=1)=\frac{1}{2}$. For this distribution,
\be
\lambda(k)=\ln\lex \E^{kX_i}\rex=\ln\cosh k,\quad k\in\reals.
\ee
This function is differentiable for all $k\in\reals$, as expected, so the rate function $I(s)$ of $S_n$ can be calculated as the Legendre transform of $\lambda(k)$. The result is
\be
I(s)=\frac{1+s}{2}\ln(1+s)+\frac{1-s}{2}\ln(1-s),\quad s\in[-1,1].
\ee
The minimum and zero of $I(s)$ is $s=0$.
\end{example}

Surprisingly, not all sample means of IID random variables fall within the framework of Cram\'er's Theorem. Here is an example for which $\lambda(k)$ does not exist, and for which large deviation theory yields in fact no useful information.

\begin{example}[Symmetric L\'evy random variables]
\label{exsymlevy}
The class of \emph{strictly stable} or \emph{strict L\'evy} random variables that are symmetric is defined by the following characteristic function:
\be
\lex \E^{\I\xi X}\rex =\E^{-\gamma |\xi|^{\alpha}},\quad \xi\in\reals, \gamma>0,\alpha\in (0,2).
\label{eqcflevy}
\ee
From this result, it is tempting to make the change of variables $i\xi=k$, often called a Wick rotation, to write $\lambda(k)=-\gamma |k|^\alpha$ for $k\in\reals$, but the correct result for $k$ real is actually 
\be
\lambda(k)=
\left\{
\begin{array}{ll}
0 & \textrm{if } k=0\\
\infty & \textrm{otherwise}.
\end{array}
\right.
\ee
This follows because the probability density $p(x)$ corresponding to the characteristic function of (\ref{eqcflevy}) has power-law tails of the form $p(x)\sim x^{-1-\alpha}$ as $|x|\ra\infty$, which implies that $\langle \E^{kX}\rangle$ does not converge for $k\in\reals\setminus\{0\}$, although it converges when $k$ is purely imaginary, that is, when $k=\I\xi$ with $\xi\in\reals$.

From the point of view of large deviation theory, the divergence of $\lambda(k)$ implies that a large deviation principle cannot be formulated for sums of symmetric L\'evy random variables. This is expected since the probability density of such sums is known to have power-law tails that decay slower than an exponential in $n$ \cite{uchaikin1999}. If we attempt to calculate a rate function in this case, we trivially find $I=0$, as in Example~\ref{expareto1} (see also \cite{lanford1973}). 
\end{example}

In some cases, Cram\'er's Theorem can be applied where $\lambda(k)$ is differentiable to obtain information about the deviations of a random variables for a restricted range of its values. The basis of this \emph{local} or \emph{pointwise} application of Cram\'er's Theorem has to do with Legendre duality. In the case where $\lambda(k)$ is differentiable for all $k\in\reals$, we have seen already that the Legendre-Fenchel transform 
\be
I(s)=\sup_k \{ks-\lambda(k)\}
\label{eqglt1}
\ee
reduces to the Legendre transform 
\be
I(s)=k(s)s-\lambda(k(s)),
\ee
where $k(s)$ is the unique solution of $\lambda'(k)=s$. By Legendre duality, the Legendre transform can also be written as
\be
I(s(k))=ks(k)-\lambda(k),
\label{eqllt1}
\ee
where $s(k)=\lambda'(k)$. Thus we see that if $\lambda(k)$ is differentiable at $k$, then the rate function $I$ at the point $s(k)$ can be expressed through the Legendre transform shown above. By applying this local Legendre transform to all the points $k$ where $\lambda$ is differentiable, we are then able to recover part of $I(s)$ even if $\lambda(k)$ is not everywhere differentiable. This is illustrated next with a variant of the previous example. 

\begin{figure*}[t]
\centering
\includegraphics[scale=1.0]{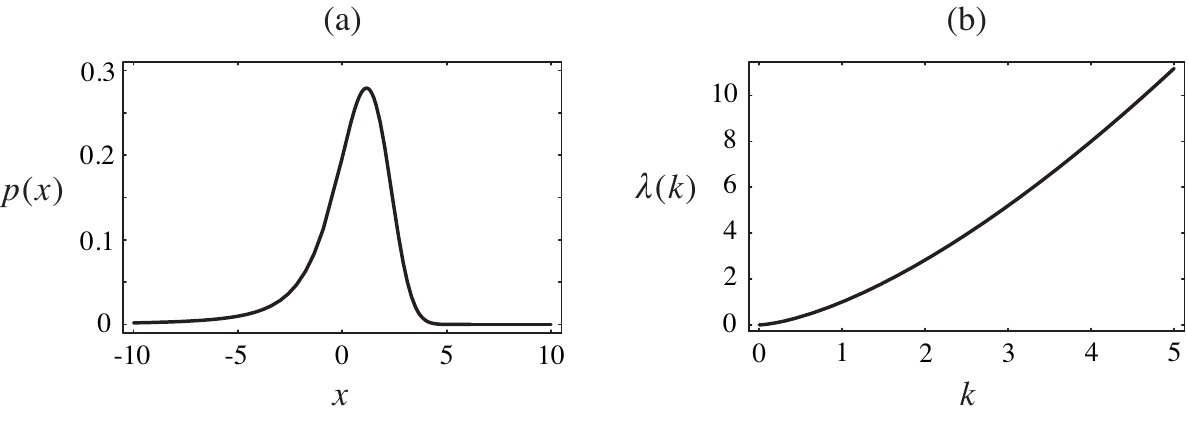}
\caption{(a) Probability density $p(x)$ of a L\'evy random which is totally skewed to the left with $\alpha=1.5$ and $b=1$. The left tail of $p(x)$ decays as $|x|^{-2.5}$, while the right tail decays faster than an exponential; see \cite{nagaev2003} for more details. (b) Corresponding $\lambda(k)$ for $k\geq 0$; $\lambda(k)=\infty$ for $k<0$.}
\label{figskewlevy}
\end{figure*}

\begin{example}[Totally skewed L\'evy random variables]
\label{exskew}
Not all strictly stable random variables have an infinite generating function for $k\neq 0$. A particular subclass of these random variables, known as \emph{totally skewed to the left}, is such that
\be
\lambda(k)=\ln\lex \E^{kX}\rex=
\left\{
\begin{array}{ll}
b k^{\alpha} 	& \textrm{if } k\geq 0\\
\infty			& \textrm{otherwise},
\end{array}
\right.
\label{eqskew1}
\ee
where $b>0$ and $\alpha\in (1,2)$ \cite{uchaikin1999,nagaev2003}. The probability density associated with this log-generating function is shown Fig.~\ref{figskewlevy}. The situation that we face here is that $\lambda(k)$ is not defined for all $k\in\reals$. This prevents us from using Cram\'er's Theorem, and so from using the Legendre-Fenchel transform of Eq.~(\ref{eqglt1}) to obtain the full rate function $I(s)$. However, following the discussion above, we can apply Cram\'er's Theorem locally where $\lambda(k)$ is differentiable to obtain part of the rate function $I(s)$ through the Legendre transform of Eq.~(\ref{eqllt1}). Doing so leads us to obtain $I(s)$ for $s>0$, since $\lambda'(k)>0$ for $k>0$ \cite{nagaev2003}. For $s\leq 0$, it can be proved that the probability density of $S_n$ has a power-law decaying tail \cite{nagaev1999}, so that $I(s)=0$ for $s\leq 0$, as in the previous example. This part of $I(s)$ cannot be obtained from the Legendre transform of Eq.~(\ref{eqllt1}), but yields in any case no useful information about the precise decay of the probability density of $S_n$ for $S_n\leq 0$. 
\end{example}

This trick of locally applying the Legendre transform shown in Eq.~(\ref{eqllt1}) to the differentiable points of $\lambda(k)$ to obtain specific points of $I(s)$ works for any random variables not just sample means of IID random variables. Therefore, although the G\"artner-Ellis Theorem does not rigorously apply when $\lambda(k)$ is not everywhere differentiable, it is possible to obtain part of the rate function associated with $\lambda(k)$ simply by Legendre-transforming the differentiable points of $\lambda(k)$. In a sense, one can therefore say that \emph{the G\"artner-Ellis Theorem holds locally where $\lambda(k)$ is differentiable}. The justification of this statement will come in Sec.~\ref{subsecncrf} when we discuss nonconvex rate functions. 

\subsection{Sanov's Theorem}

Large deviation principles can be formulated for many types of random variable, not just scalar random variables taking values in $\reals$. One particularly important case of large deviation principles is that applying to random vectors taking values in $\reals^d$, $d> 1$. To illustrate this case, let us revisit the problem of determining the probability distribution $P(L_n=l)$ associated with the empirical vector $L_n$ introduced in Example~\ref{exsanov}. Recall that, given a sequence $\om=(\om_1,\om_2,\ldots,\om_n)$ of $n$ IID random variables taking values in a finite set $\Lambda$, the empirical vector $L_n(\om)$ is the vector of empirical frequencies defined by the sample mean
\be
L_{n,j}(\om)=\frac{1}{n}\sum_{i=1}^n \delta_{\om_i,j},\quad j\in\Lambda.
\ee
This vector has $|\Lambda|$ components, and the space of $L_n$, as noted earlier, is the set of probability distributions on $\Lambda$.

To find the large deviations of $L_n$, we consider the vector extension of the G\"artner-Ellis Theorem obtained by replacing the product $kL_n$ in the definition of $\lambda(k)$ by the scalar product $k\cdot L_n$ involving the vector $k\in\reals^\Lambda$. Thus the scaled cumulant generating function that we must now calculate is
\be
\lambda(k)=\lim_{n\ra\infty}\frac{1}{n}\ln\lex \E^{n k\cdot L_n}\rex,\quad k\in\reals^\Lambda.
\ee
Since $L_n$ is a sample mean of IID random variables, the expression of $\lambda(k)$ simplifies to
\be
\lambda(k)=\ln\sum_{j\in\Lambda} \rho_j \E^{k_j},
\label{eqscgf1}
\ee
where $\rho_j=P(\om_i=j)$, $j\in\Lambda$. The expression above is necessarily analytic in $k$ if $\Lambda$ is finite. In this case, we can then use the G\"artner-Ellis Theorem to conclude that a large deviation principle holds for $L_n$ with a rate function $I(l)$ given by
\be
I(l)=\sup_k \{k\cdot l-\lambda(k)\}=k(l)\cdot l-\lambda(k(l)),
\ee
$k(l)$ being the unique root of $\nabla\lambda(k)=l$. Calculating the Legendre transform explicitly yields the rate function
\be
I(l)=\sum_{j\in\Lambda} l_j\ln\frac{l_j}{\rho_j},
\label{eqrelent1}
\ee
which agrees with the rate function calculated by combinatorial means in Example~\ref{exsanov}. 

The complete large deviation principle for $L_n$ is known in large deviation theory as \emph{Sanov's Theorem} \cite{sanov1961}; see \cite{ellis1999} for a discussion of Boltzmann's anticipation of this result. As already noted, $I(l)$ has a unique minimum and zero located at $l=\rho$. Moreover, as most of the rate functions encountered so far, $I(l)$ has the property that it is locally quadratic around it minimum:
\be
I(l)\approx\frac{1}{2}\sum_{i,j\in\Lambda} (l_j-\rho_j)
\left.
\frac{\partial^2 I}{\partial l_j\partial l_i} 
\right|_{l=\rho}
(l_i-\rho_i)=\frac{1}{2}\sum_{i\in\Lambda} \frac{(l_i-\rho_i)^2}{\rho_i}.
\ee

Extensions of Sanov's Theorem exist when $\Lambda$ is infinite or even continuous. The mathematical tools needed to treat these cases are quite involved, but the essence of these extensions is easily explained at a heuristic level. For definiteness, consider the case where the IID random variables $\om_1,\om_2,\ldots,\om_n$ take values in $\reals$ according to a probability density $\rho(x)$. For this sequence, the continuous extension of the empirical vector $L_n$ is the \emph{empirical density}
\be
L_n(x)=\frac{1}{n}\sum_{i=1}^n \delta(\om_i-x),\quad x\in\reals,
\label{eqed1}
\ee
involving Dirac's delta function $\delta$. This is a normalized density in the sense that
\be
\int_{-\infty}^\infty L_n(x)\, \D x=1
\ee
for all $\om\in\reals^n$. Since $L_n$ is now a function (it is a random function to be more precise), the vector $k$ used in the discrete version of Sanov's Theorem must be replaced by a function $k(x)$, so that
\be
k\cdot L_n=\int_{-\infty}^\infty k(x)L_n(x)\, \D x.
\ee 
Similarly, the analog of $\lambda(k)$ found in Eq.~(\ref{eqscgf1}) is now a functional of $k(x)$ having the form
\be
\lambda(k)=\ln\int_{-\infty}^\infty \rho(x)\, \E^{k(x)}\, \D x=\ln\lex \E^{k(X)}\rex_\rho.
\ee
To apply the G\"artner-Ellis Theorem to this functional, we note that $\lambda(k)$ is differentiable in the sense of functional derivatives:
\be
\frac{\delta \lambda(k)}{\delta k(y)}=\frac{\rho(y)\E^{k(y)}}{\lex \E^{k(X)}\rex_\rho}.
\ee
By analogy with the discrete case, $L_n$ must then satisfy a large deviation principle with a rate function $I(\mu)$ equal to the (functional) Legendre transform of $\lambda(k)$. The result of that transform, as should be expected, is the continuous version of the relative entropy:
\be
I(\mu)=\int_{-\infty}^\infty \D x\, \mu(x)\ln\frac{\mu(x)}{\rho(x)}.
\ee
To complete this result, it must be added that $I(\mu)=\infty$ if $\mu$ has a larger support than $\rho$, that is mathematically, if $\mu$ is not continuous relative to $\rho$. This makes sense: the realizations of $L_n$ cannot have a support larger than that of $\rho$. 

\subsection{Markov processes}
\label{subsecmp}

Sample means of IID random variables constitute the simplest example of stochastic processes for which large deviation principles can be derived. The natural application to consider next concerns the class of Markov processes. Large deviation results have been formulated for this class of processes mainly by Donsker and Varadhan \cite{donsker1975,donsker1975a,donsker1976,donsker1983}, who established through their work much of the basis of large deviation theory as we know it today. Our treatment of these processes will follow the path of the G\"artner-Ellis Theorem, and will be presented, for simplicity, for finite Markov chains. The case of continuous-time Markov processes will be discussed in Sec.~\ref{secnonequi} when dealing with nonequilibrium systems. Some subtleties of infinite-state Markov chains will also be discussed in Sec.~\ref{secnonequi}.

The study of Markov chains is similar to the study of IID sample means, in that we consider a sequence $\om=(\om_1,\om_2,\ldots,\om_n)$ of $n$ random variables taking values in some finite set $\Lambda$, and study the sample mean
\be
S_n=\frac{1}{n}\sum_{i=1}^n f(\om_i)
\label{eqap1}
\ee
involving an arbitrary function $f:\Lambda\ra\reals^d$, $d\geq1$. The difference with the IID case, apart from the added function $f$, is that we now assume that the $\om_i$'s form a \emph{Markov chain} defined by
\be
P(\om)=P(\om_1,\om_2,\ldots,\om_n)=\rho(\om_1)\prod_{i=2}^n \pi(\om_i|\om_{i-1}).
\label{eqmc1}
\ee
In this expression, $\rho(\om_1)$ denotes the probability distribution of the initial state $\om_1$, while $\pi(\om_i|\om_{i-1})$ is the conditional probability of $\om_{i}$ given $\om_{i-1}$.  We consider here the case where $\pi$ is a fixed function of $\om_i$ and $\om_{i-1}$, in which case the Markov chain is said to to be \emph{homogeneous}. The sample mean $S_n$ thus defined on the Markov chain $\om$ is often referred to as a \emph{Markov additive process} \cite{ney1987,ney1987a,bucklew1990,dembo1998}. 

To derive a large deviation principle for $S_n$, we proceed as before to calculate $\lambda(k)$. The generating function of this random variable can be written as
\begin{eqnarray}
\lex \E^{nk\cdot S_n}\rex  	&=&\sum_{\om_1,\om_2,\ldots,\om_n} \rho(\om_1)\E^{k\cdot f(\om_1)} \pi(\om_2|\om_1)\E^{k\cdot f(\om_2)}\cdots\pi(\om_n|\om_{n-1})\E^{k\cdot f(\om_n)}\nonumber\\
				&=&\sum_{\om_1,\om_2,\ldots,\om_n} \pi_k(\om_n|\om_{n-1})\cdots\pi_k(\om_2|\om_1)\rho_k(\om_1),
\end{eqnarray}
by defining $\rho_k(\om_1)=\rho(\om_1)\E^{k\cdot f(\om_1)}$ and $\pi_k(\om_{i}|\om_{i-1})=\pi(\om_i|\om_{i-1})\E^{k\cdot f(\om_i)}$. We recognize in the second equation a sequence of matrix products involving the vector of values $\rho_k(\om_1)$ and the \emph{transition matrix} $\pi_k(\om_i|\om_{i-1})$. To be more explicit, let us denote by $\rho_k$ the vector of probabilities $\rho_k(\om_1=i)$, that is, $(\rho_k)_i=\rho_k(\om_1=i)$, and let $\Pi_k$ denote the matrix formed by the elements of $\pi_k(\om_i|\om_{i-1})$, that is, $(\Pi_k)_{ji}=\pi_k(j|i)$. In terms of $\rho_k$ and $\Pi_k$, we then write
\be
\lex \E^{nk\cdot S_n}\rex =\sum_{j\in\Lambda} \left( \Pi_k^{n-1}\rho_k\right)_j,
\ee
The function $\lambda(k)$ is extracted from this expression by determining the asymptotic behavior of the product $\Pi_k^{n-1}\rho_k$ using the Perron-Frobenius theory of positive matrices. Depending on the form of $\Pi$, one of three cases arises:

\begin{description}
\item[Case A:] $\Pi$ is ergodic (irreducible and aperiodic), and has therefore a unique stationary probability distribution $\rho^*$ such that $\Pi\rho^*=\rho^*$. In this case, $\Pi_k$ has a unique \emph{principal} or \emph{dominant} eigenvalue $\zeta(\Pi_k)$ from which it follows that $\lex \E^{nk\cdot S_n}\rex\asymp {\zeta(\Pi_k)}^n$, and thus that $\lambda(k)=\ln\zeta(\Pi_k)$. Given that $\Pi$ is assumed to be finite, $\zeta(\Pi_k)$ must be analytic in $k$. From the G\"artner-Ellis Theorem, we therefore conclude that $S_n$ satisfies a large deviation principle with rate function
\be
I(s)=\sup_k\{k\cdot s-\ln\zeta(\Pi_k)\}.
\label{eqmcldp1}
\ee 
 
\item[Case B:] $\Pi$ is not irreducible, which means that it has two or more stationary distributions (broken ergodicity). In this case, $\lambda(k)$ exists but depends generally on the initial distribution $\rho(\om_1)$. Furthermore, $\lambda(k)$ may be nondifferentiable, in which case the G\"artner-Ellis Theorem does not apply. This arises, for example, when two of more eigenvalues of $\Pi_k$ compete to be the dominant eigenvalue for different initial distribution $\rho(\om_1)$ and different $k$. 

\item[Case C:] $\Pi$ has no stationary distributions (e.g., $\Pi$ is periodic). In this case, no large deviation principle can generally be found for $S_n$. In fact, in this case, the Law of Large Numbers does not even hold in general.
\end{description}

The next two examples study Markov chains falling in Case A. The first example is a variation of Example~\ref{exfrac1} on random bits, whereas the second generalizes Sanov's Theorem to Markov chains. For examples of Markov chains falling in Case B, see \cite{dinwoodie1992,dinwoodie1993}.

\begin{figure*}[t]
\centering
\includegraphics[scale=1.0]{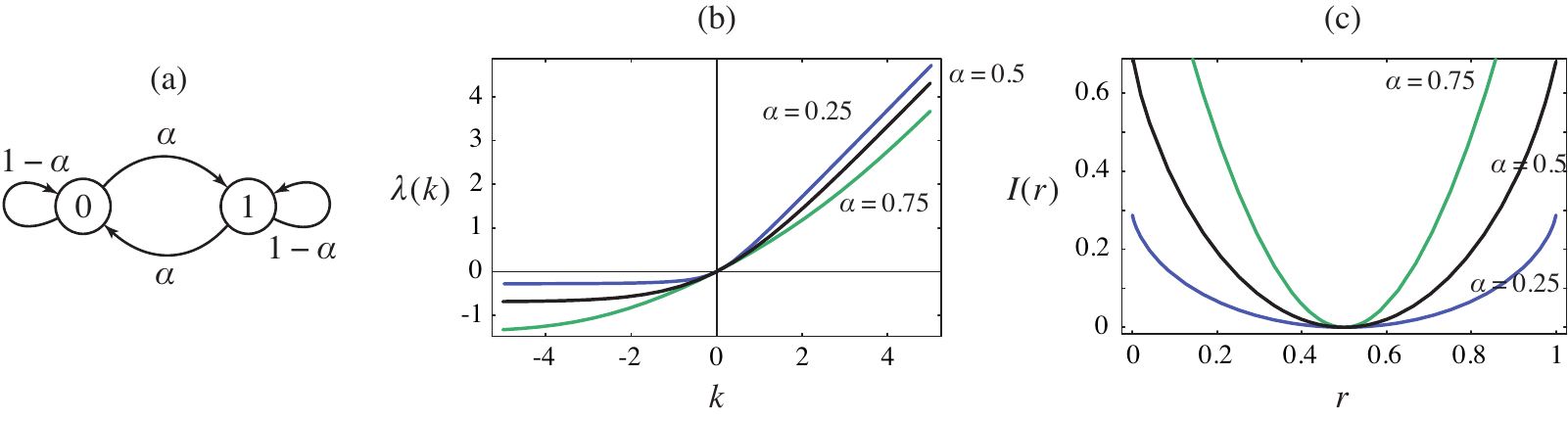}
\caption{(a) Transition probabilities between the two states of the symmetric binary Markov chain of Example~\ref{exsymmcbit}. (b) Corresponding scaled cumulant generating function $\lambda(k)$ and (c) rate function $I(r)$ for $\alpha=0.25$, $0.5$, and $0.75$.}
\label{figsymmc}
\end{figure*}

\begin{example}[Balanced Markov bits \cite{oono1989}]
\label{exsymmcbit}
Consider again the bit sequence $b=(b_1,b_2,\ldots,b_n)$ of Example~\ref{exfrac1}, but now assume that the bits have the Markov dependence shown in Fig.~\ref{figsymmc}(a) with $\alpha\in(0,1)$. The symmetric and irreducible transition matrix associated with this Markov chain is
\be
\Pi=
\left(
\begin{array}{cc}
\pi(0|0) & \pi(0|1) \\
\pi(1|0) & \pi(1|1)
\end{array}
\right)
=
\left(
\begin{array}{cc}
1-\alpha & \alpha \\
\alpha & 1-\alpha
\end{array}
\right).
\ee
The largest eigenvalue of 
\be
\Pi_k=
\left(
\begin{array}{cc}
1-\alpha & \alpha \\
\alpha \E^k & (1-\alpha)\E^k
\end{array}
\right)
\ee
can be calculated explicitly to obtain $\lambda(k)$. The result is shown in Fig.~\ref{figsymmc} for various values of $\alpha$. Also shown in this figure is the corresponding rate function $I(r)$ obtained by calculating the Legendre transform of $\lambda(k)$. The rate function clearly differs from the rate function found for independent bits. In fact, to second order in $\varepsilon=1/2-\alpha$ we have
\be
I(r)\approx I_0(r)+2(1-2r)^2\varepsilon +(2-32r^2+64r^3-32r^4)\varepsilon^2,
\ee
where $I_0(r)$ is the rate function of the independent bits obtained here for $\alpha=1/2$ or, equivalently, for $\varepsilon=0$; see Eq.~(\ref{eqldp1}). Note that the zero of $I(s)$ does not change with $\alpha$ because the stationary distribution of the Markov chain is uniform for all $\alpha\in(0,1)$, which means that the most probable sequences are the balanced sequences such that $R_n= 1/2$. What changes with $\alpha$ is the propensity of generating repeated strings of $0$'s or $1$'s in a given bit sequence. For $\alpha<1/2$, a bit is more likely to be followed by the same bit, while for $\alpha>1/2$, a bit is more likely to be followed by its opposite. The effect of this correlation, as can be seen from Fig.~\ref{figsymmc}(c), is that empirical frequencies of $1$'s close to $0$ or $1$ are exponentially more probable for $\alpha<1/2$ than for $\alpha>1/2$.
\end{example}

Oono \cite{oono1989} discusses an interesting variant of the example above having absorbing states and a corresponding linear rate function. General quadratic approximations of rate functions of Markov chains are also discussed in that paper.

\begin{example}[Sanov's Theorem for Markov chains]
\label{exsanovmc}
The extension of Sanov's Theorem to irreducible Markov chains can be derived from the general Legendre-Fenchel transform shown in (\ref{eqmcldp1}) by choosing $f(\om_i)=\delta_{\om_i,j}$, $j\in\Lambda$, in which case $\pi_k(j|i)=\pi(j|i)\E^{k_j}$. Ellis shows in \cite{ellis1984} (see Theorem III.1) that the supremum over all vectors $k\in\reals^\Lambda$ involved in that transform can be simplified to the following supremum:
\be
I(l)=\sup_{u>0}\sum_{j\in\Lambda} l_j\ln\frac{u_j}{(\Pi u)_j}=\sup_{u>0}\lex \ln\frac{u}{\Pi u}\rex_l,
\ee
which involves only the strictly positive vectors $u$ in $\reals^\Lambda$. This result was first obtained by Donsker and Varadhan \cite{donsker1975}; for a proof of it, see Sec.~3.1.2 of \cite{dembo1998} or Sec.~V.B of \cite{bucklew1990}. The minimum and zero of this rate function is the stationary distribution $\rho^*$ of $\Pi$.
\end{example}

The expression of the rate function for the empirical vector is obviously more complicated for Markov chains because of the correlations introduced between the random variables $\om_1,\om_2,\ldots,\om_n$. Since these random variables ``interact'' only between pairs, it may be expected that a rate function similar in structure to the relative entropy is obtained if we replace the single-site empirical vector by a double-site empirical vector or \emph{empirical matrix}, that is, if we look at the frequencies of occurrences of pair values in a Markov chain. Mathematically, this pair empirical matrix should be defined as
\be
Q_n(x,y)=\frac{1}{n}\sum_{i=1}^{n} \delta_{\om_i,x}\delta_{\om_{i+1},y},\quad x,y\in\Lambda
\ee
by requiring that $\om_{n+1}=\om_1$ because $Q_n(x,y)$ then has the nice property that
\be
\sum_{x\in\Lambda} Q_n(x,y)=L_n(y),\quad\text{and}\quad \sum_{y\in\Lambda} Q_n(x,y)=L_n(x),
\label{eqtr1}
\ee
where $L_n$ is the usual empirical vector of the random sequence $\om=(\om_1,\om_2,\dots,\om_n)$. In this case, $Q_n$ is said to be \emph{balanced} or to have \emph{shift-invariant} marginals. 

The rate function of $Q_n$ can be derived in many different ways. One which is particularly elegant focuses on the sequence $\zeta=(\zeta_1,\zeta_2,\ldots,\zeta_n)$ which is built from the contiguous pairs $\zeta_i=(\om_i,\om_{i+1})$ appearing in the sequence $\om=(\om_1,\om_2,\ldots,\om_n)$. The empirical vector of $\zeta$ is the pair empirical distribution of $\om$, and the probability distribution of $\zeta$ factorizes in a way that partially mimics the IID case. Combining these two observations in Sanov's Theorem, it then follows that $Q_n$ satisfies a large deviation principle with rate function
\be
I_3(q)=\sum_{(x,y)\in\Lambda^2} q(x,y) \ln\frac{q(x,y)}{\pi(y|x)l(x)},
\label{eqi3}
\ee
where $l(x)$ is the marginal of $q(x,y)$. The complete derivation of this large deviation result can be found in Sec.~3.1.3 of \cite{dembo1998}. Note that the zero of $I_3(q)$ is reached when $q(x,y)/l(x)=\pi(y|x)$, in which case $l(x)=\rho^*(x)$, where $\rho^*$ is again the unique stationary distribution of $\Pi$. 

\subsection{Nonconvex rate functions}
\label{subsecncrf}

Since Legendre-Fenchel transforms yield functions that are necessarily convex (see Sec.~\ref{propconvex}), one obvious limitation of the G\"artner-Ellis Theorem is that it cannot be used to calculate \emph{nonconvex} rate functions and, in particular, rate functions that have two or more local or global minima. The breakdown of this theorem for this class of rate functions is related to the differentiability condition on $\lambda(k)$. This is illustrated and explained next using a combination of examples and results about convex functions.

\begin{example}[Multi-atomic distribution \cite{ellis1995}] A nonconvex rate function is easily constructed by considering a continuous random variable having a Dirac-like probability density supported on two or more points. The rate function associated with $p(Y_n=y)=\frac{1}{2}\delta({y\pm1})$, for example, is 
\be
I(y)=\left\{
\begin{array}{ll}
0 & \textrm{if } y=\pm 1\\
\infty & \textrm{otherwise},
\end{array}
\right.
\ee
and is obviously nonconvex as it has two minima corresponding to its two non-singular values (a convex function always has only one minimum). Therefore, it cannot be expressed as the Legendre-Fenchel transform of the scaled cumulant generating function $\lambda(k)$ of $Y_n$. To be sure, calculate $\lambda(k)$:
\be
\lambda(k)=\lim_{n\ra\infty}\frac{1}{n}\ln\frac{\E^{-nk}+\E^{nk}}{2}=|k|
\ee
and its Legendre-Fenchel transform:
\be
I^{**}(y)=
\sup_k \{ky-\lambda(k)\}=
\left\{
\begin{array}{ll}
0 & \textrm{if } y\in [-1,1]\\
\infty & \textrm{otherwise}.
\end{array}
\right.
\ee
The result does indeed differ from $I(y)$; in fact, $I(y)\neq I^{**}(y)$ for $y\in (-1,1)$. 

The G\"artner-Ellis Theorem is obviously not applicable here because $\lambda(k)$ is not differentiable at $k=0$. However, as in the example of the skewed L\'evy random variables (Example~\ref{exskew}), we could apply the Legendre transform of Eq.~(\ref{eqllt1}) locally where $\lambda(k)$ is differentiable to obtain some part of $I(y)$. In this case, we obtain only two points of this function, namely, $I(-1)=0$ and $I(1)=0$, since $\lambda'(k)=-1$ for $k<0$ and $\lambda'(k)=1$ for $k>0$.
\end{example} 

The previous example raises a number of important questions related to the G\"artner-Ellis Theorem and the way rate functions are calculated. The most obvious has to do with the differentiability of $\lambda(k)$: Is there a general connection between the differentiability of this function and the convexity of rate functions? Indeed, why is $\lambda(k)$ required to be differentiable in the G\"artner-Ellis Theorem? Moreover, what is the result of the Legendre-Fenchel transform of $\lambda(k)$ in general? To answer these questions, we list and discuss next four results of convex analysis that characterize the Legendre-Fenchel transform. All of these results can be found in \cite{rockafellar1970} (see also \cite{tiel1984} and Chap.~VI of \cite{ellis1985}).

\begin{figure*}[t]
\centering
\includegraphics[scale=1.0]{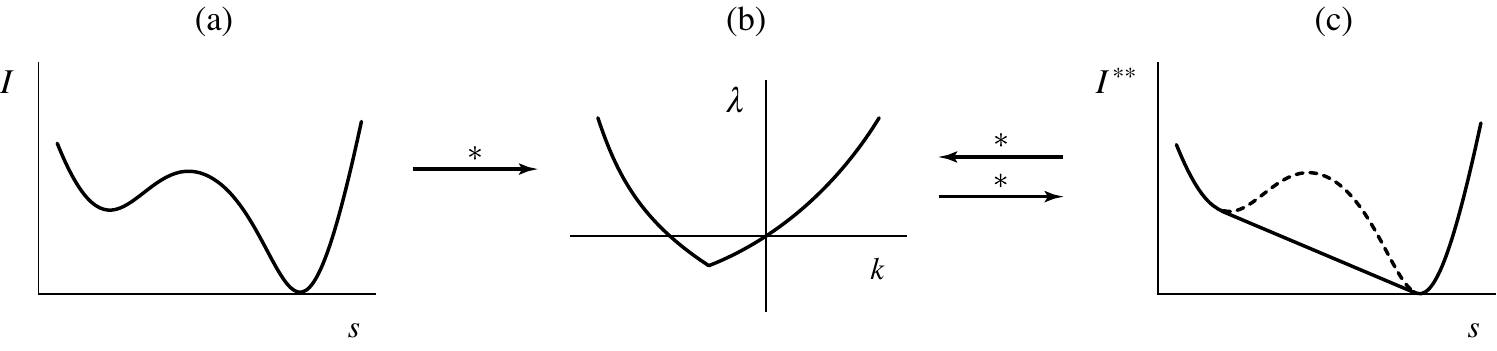}
\caption{Legendre-Fenchel transforms connecting (a) a nonconvex rate function $I(s)$, (b) its associated scaled cumulant generating function $\lambda(k)$, and (c) the convex envelope $I^{**}(s)$ of $I(s)$. The arrows illustrate the relations $I^*=\lambda$, $\lambda^*=I^{**}$ and $(I^{**})^*=\lambda$.}
\label{fignondual}
\end{figure*}

\begin{description}
\item[Result 1:] The Legendre-Fenchel transform of $I$ yields $\lambda$ whether $I$ is convex or not. 
\end{description}

This result follows essentially because $\lambda(k)$ is always convex. In convex analysis, the Legendre-Fenchel transform of $I$ is denoted by $I^*$. Thus $I^*=\lambda$ for all $\lambda$, in accordance with Varadhan's Theorem.

\begin{description}
\item[Result 2:] If $I$ is nonconvex, then the Legendre-Fenchel transform of $\lambda$, denoted by $\lambda^*$, does not yield $I$; rather, it yields the \emph{convex envelope} of $I$. 
\end{description}

This result is illustrated in Fig.~\ref{fignondual}. The convex envelope is usually denoted by $I^{**}$, since it is given by the double Legendre-Fenchel transform of $I$, and is such that $I^{**}\leq I$. With this notation, we then have $\lambda^*=I^{**}\neq I$ if $I$ is nonconvex, and $I=\lambda^*=I^{**}$ if $I$ is convex. Accordingly, when a rate function $I$ is convex, it can be calculated as the Legendre-Fenchel transform of $\lambda$.

\begin{description}
\item[Result 3:] The convex envelope $I^{**}$ of $I$ has the same Legendre-Fenchel transform as $I$, that is, $(I^{**})^*=I^*=\lambda$; see Fig.~\ref{fignondual}. In general, functions having the same convex envelope have the same Legendre-Fenchel transform.
\end{description}

This property explains why nonconvex rate functions cannot be obtained from $\lambda$. Put simply, the Legendre-Fenchel transform is a many-to-one transformation for the class of nonconvex functions. We also say that the Legendre-Fenchel transform is \emph{non-self-dual} or \emph{non-involutive} for nonconvex functions.

\begin{figure*}[t]
\centering
\includegraphics[scale=1.0]{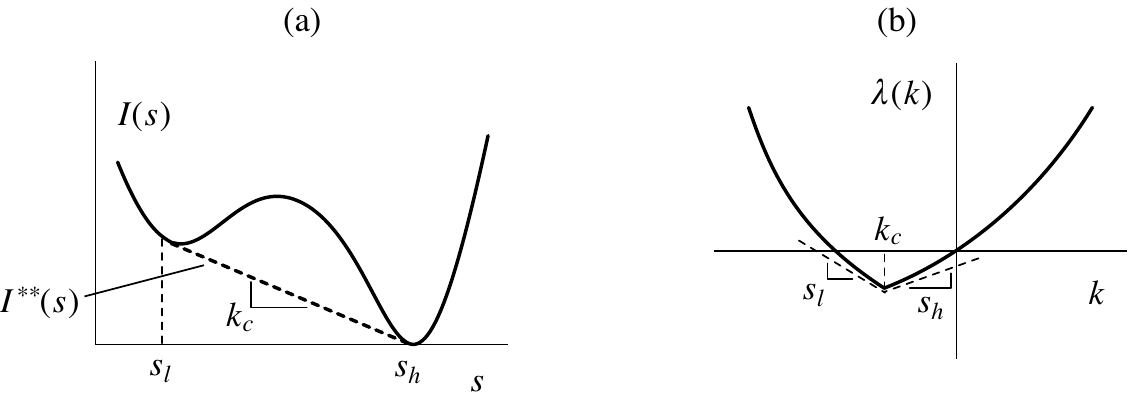}
\caption{(a) Nonconvex rate function $I$ and its convex envelope $I^{**}$. (b) Associated scaled cumulant generating function $\lambda(k)$ having a nondifferentiable point at $k_c$.}
\label{fignonduallegprop}
\end{figure*}

\begin{description}
\item[Result 4:] $\lambda$ is nondifferentiable if $I$ is nonconvex. To be more precise, suppose that $I(s)$ differs from its convex envelope $I^{**}(s)$ over some open interval $(s_l,s_h)$, as in Fig.~\ref{fignonduallegprop}(a). Then its Legendre-Fenchel transform $I^*=\lambda$ is nondifferentiable at some value $k_c$ corresponding to the slope of $I^{**}(s)$ over the interval $(s_l,s_h)$; see Fig.~\ref{fignonduallegprop}(b). Moreover, the left- and right-derivatives of $\lambda$ at $k_c$ equal $s_l$ and $s_h$, respectively. The same results hold when $I(s)$ is linear (we also say \emph{affine}) over $(s_l,s_h)$.
\end{description}

The condition of differentiability of $\lambda(k)$ entering in the G\"artner-Ellis Theorem can be understood from these results as follows. From the results 2 and 3, we have that a rate function $I$ can be obtained as the Legendre-Fenchel transform of $\lambda$ only if $I$ is convex. This leaves us with two possibilities: either $I$ is strictly convex, that is, it is convex with no linear parts, or else $I$ is convex but has one or more linear parts. The second possibility leads to a nondifferentiable $\lambda(k)$, as is the case for a nonconvex $I$ according to the results 3 and 4, so these two cases cannot be distinguished from the point of view of $\lambda$; see Fig.~\ref{fignondual}. Hence, the only possibility for which the sole knowledge of $\lambda$ enables us to write $I=\lambda^*$ is when $I$ is strictly convex. In this case, $\lambda$ is differentiable by the result 4, as required by the G\"artner-Ellis Theorem. 

This reasoning shows, incidentally, that the differentiability of $\lambda(k)$ is a \emph{sufficient but not a necessary} condition for having $I=\lambda^*$. Simply consider the case where $I$ is convex but has one or more linear parts. Then $I=\lambda^*$, since $I$ is convex, but $\lambda=I^*$ is nondifferentiable by the result 4. The problem with rate functions having linear parts, as pointed out, is that they cannot be distinguished from nonconvex rate functions if we only know $\lambda$. That is, without any a priori knowledge of $I$, we know for sure that $I=\lambda^*$ only when $\lambda(k)$ is differentiable. The next example puts these observations into practice.

\begin{figure*}[t]
\centering
\includegraphics[scale=1.0]{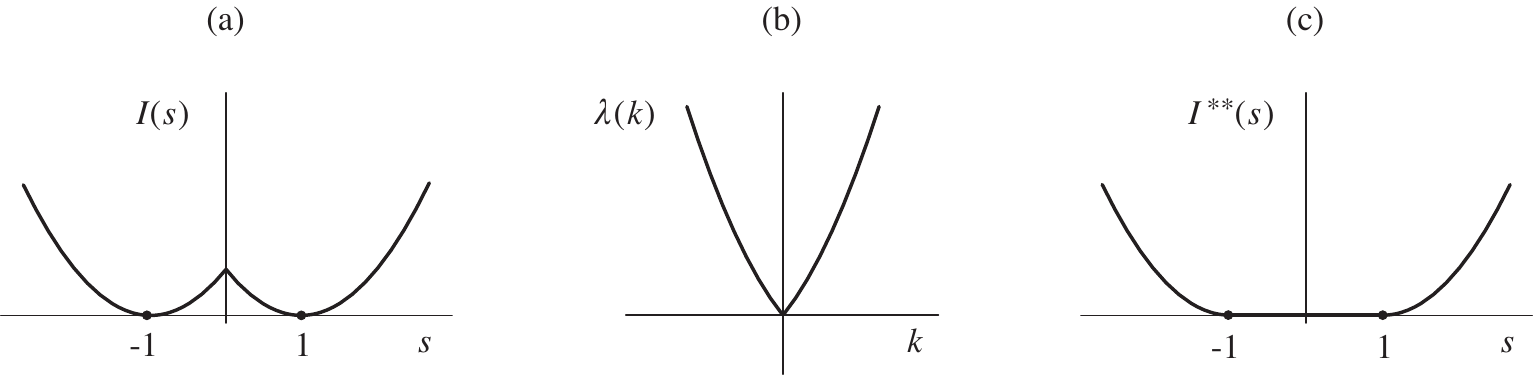}
\caption{(a) Nonconvex rate function $I(s)$ for Example~\ref{exmixedsum}. (b) Corresponding $\lambda(k)$. (c) Convex envelope $I^{**}(s)$ of $I(s)$.}
\label{figmixedsum}
\end{figure*}

\begin{example}[Mixed Gaussian sum]
\label{exmixedsum}
This example is due to Ioffe \cite{ioffe1993}. Consider the sum
\be
S_n=Y+\frac{1}{n}\sum_{i=1}^n X_i
\label{eqmixsum}
\ee
where the $X_i$'s are IID random variables distributed according to the normal distribution with unit mean and unit variance, and where $Y$ is a discrete random variable, taken to be independent of the $X_i$'s and such that $P(Y=-1)=P(Y=1)=\frac{1}{2}$. To find the rate function of $S_n$, we use our knowledge of  the Gaussian sample mean (see Examples~\ref{exgauss} and \ref{exgaussrev}) to write
\be
P(S_n\in \D s| Y=\pm 1)\asymp \E^{-nI_{\pm}(s)}\, \D s,
\ee
where $I_{\pm}(s)=(s\mp 1)^2/2$. As a result,
\be
P(S_n\in \D s)=\sum_{y=\pm 1} P(S_n\in \D s|Y=y)P(Y=y)\asymp \E^{-n I_-(s)}\, \D s +\E^{-n I_+(s)}\, \D s\asymp \E^{-nI(s)}\, \D s,
\ee
where
\be
I(s)=\min\{ I_-(s),I_+(s)\}=
\left\{
\begin{array}{ll}
I_-(s) & \textrm{if } s<0 \\
I_+(s) & \textrm{if } s\geq 0.
\end{array}
\right.
\ee
This rate function is nonconvex, as seen in Fig.~\ref{figmixedsum}(a). 

We can verify that the Legendre-Fenchel transform of $\lambda(k)$ yields the convex envelope of $I(s)$ rather than $I(s)$ itself. Following the previous example, we find here 
\be
\lambda(k)=\frac{1}{2}k^2 + |k|,\quad k\in\reals.
\ee
Notice that $\lambda(k)$ is nondifferentiable at $k=0$; see Fig.~\ref{figmixedsum}(b). Taking the Legendre-Fenchel transform yields
\be
I^{**}(s)=
\sup_k \{ks-\lambda(k)\}=
\left\{
\begin{array}{ll}
I_-(s) & \textrm{if } s<-1\\
0 	 & \textrm{if } s\in [-1,1]\\
I_+(s) & \textrm{if } s>1.
\end{array}
\right.
\ee
Figure~\ref{figmixedsum}(c) shows that $I^{**}(s)$ is the convex envelope of $I(s)$, which differs from $I(s)$ for $s\in (-1,1)$. The part of $I(s)$ that can be obtained by applying the local Legendre transform of Eq.~(\ref{eqllt1}) to the differentiable branches of $\lambda(k)$ is the part of $I(s)$ that coincides with its convex envelope. We leave it to the reader in the end to show that $(I^{**})^*=I^*=\lambda$. The calculation of these Legendre-Fenchel transforms involves an interplay of local and global maximizers which accounts for the nondifferentiable point of $\lambda(k)$.
\end{example}

The last example of this section is there to show that boundary points of $\lambda(k)$ can also be thought of as nondifferentiable points for the purpose of the G\"artner-Ellis Theorem.

\begin{figure*}[t]
\centering
\includegraphics[scale=1.0]{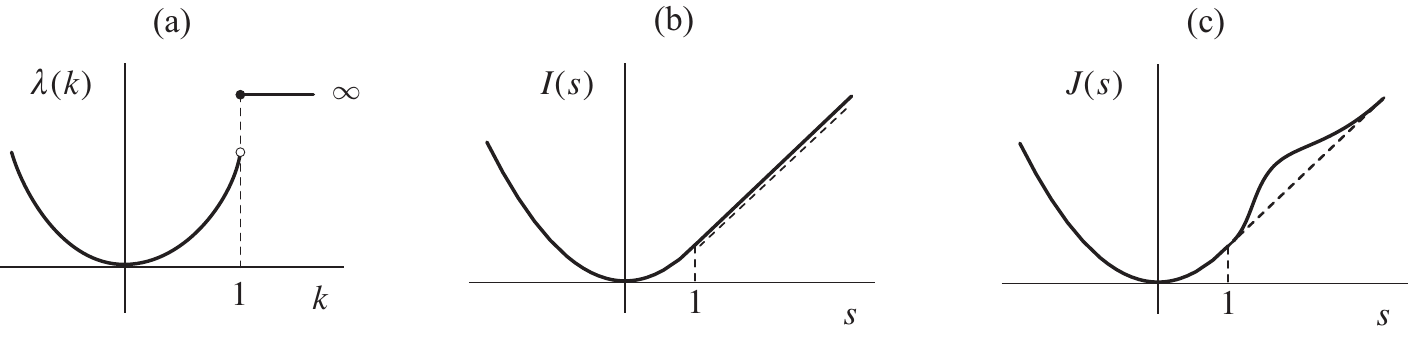}
\caption{(a) $\lambda(k)$ for Example~\ref{exnonsteep}. (b) Rate function $I(s)$ for that example. (c) Nonconvex rate function $J(s)$ having the same Legendre-Fenchel transform as $I(s)$.}
\label{figborder}
\end{figure*}

\begin{example}[Non-steep $\lambda$]
\label{exnonsteep}
Consider again the sum $S_n$ shown in (\ref{eqmixsum}), but let $Y=Z/n$, where $Z$ is an exponentially-distributed random variable with unit mean, that is, $p(Z=z)=\E^{-z}$, $z\geq 0$. The calculation of $\lambda(k)$ for $S_n$ yields
\be
\lambda(k)=\frac{k^2}{2}+\lim_{n\ra\infty}\frac{1}{n}\ln\lex \E^{kZ}\rex
=
\left\{
\begin{array}{ll}
k^2/2 & \textrm{if } k<1\\
\infty & \textrm{if } k\geq 1.
\end{array}
\right.
\label{eql2}
\ee 
Applying the Legendre transform of Eq.~(\ref{eqllt1}) to the differentiable branch of this function leads to $I(s)=s^2/2$ for $s<1$, since $\lambda'(k)<1$ for $k\in(-\infty,1)$; see Fig.~\ref{figborder}(a). As in the previous examples, the Legendre transform of $\lambda(k)$ yields here only part of $I(s)$ because the image of $\lambda'(k)$ does not cover the whole range of $S_n$. In the present example, we say that $\lambda(k)$ is \emph{non-steep} because its derivative is upper bounded.

To obtain the full rate function of $S_n$, we can follow the previous example by noting that, conditionally on $Z=z$, $S_n$ must be Gaussian with mean $z/n$ and unit variance. Therefore,
\be
p(S_n=s)=\int_0^\infty p(S_n=s|Z=z)\, p(Z=z)\, \D z\asymp \int_0^\infty \E^{-n(s-z/n)^2/2}\, \E^{-z}\, \D z.
\ee
A large deviation principle is extracted from the last integral by performing the integral exactly or by using Laplace's approximation. In both cases, we obtain $p(S_n=s)\asymp \E^{-nI(s)}$, where
\be
I(s)=
\left\{
\begin{array}{ll}
s-1/2	 & \textrm{if } s>1\\
s^2/2	& \textrm{if } s\leq 1.
\end{array}
\right.
\ee
As seen in Fig.~\ref{figborder}(b), $I(s)$ is not strictly convex, since it is linear for $s>1$; hence the fact that this part cannot be obtained from the G\"artner-Ellis Theorem. The nondifferentiable point of $\lambda(k)$ associated with this part of $I(s)$ is non-trivial: it is the boundary point of $\lambda(k)$ located at $k=1$. That nondifferentiable point would also arise if $I(s)$ were nonconvex for $s>1$ instead of being linear; see Fig.~\ref{figborder}(c). 

There is an extra mathematical subtlety related to the boundary point of $\lambda(k)$, namely, that the Legendre-Fenchel transform of $I(s)$ does not yield $\lambda(k)$ at $k=1$. This may seem to contradict Varadhan's Theorem, but there is in fact no contradiction here. Simply, there is a technical condition associated with this theorem which we have not mentioned (see, e.g., Theorem 5.1 of \cite{ellis1995} or Theorem 4.3.1 of \cite{dembo1998}), and which happens to be violated in the present example. Mathematically, the problem arises because Legendre-Fenchel transforms yield functions that are \emph{lower semi-continuous} in addition to being convex \cite{rockafellar1970,tiel1984}. Here $\lambda(k)$ is convex but not lower semi-continuous, since its domain is open; hence the fact that $\lambda\neq I$ at $k=1$. In practice, boundary points of $\lambda(k)$ appear to be the only points for which we may have $\lambda\neq I^*$, and thus for which Varadhan's Theorem must be applied with care.
\end{example}

Other examples of nonconvex rate functions related to non-irreducible Markov chains having more that one stationary distributions are discussed by Dinwoodie \cite{dinwoodie1992,dinwoodie1993}. These examples relate to the Case B of Markov chains mentioned before. In the end, it should be kept in mind that nonconvex rate functions pose no limitations for large deviation theory; they pose only a limitation for the G\"artner-Ellis Theorem because of the way that theorem relies on Legendre-Fenchel transforms. We will see in the next section other methods that can be used to calculate rate functions, be they convex or not. 

\subsection{Self-processes}
\label{subsecsp}

There are many random variables, apart from sample means, that can be studied from the point of view of large deviation theory. One which is often studied in information theory and in nonequilibrium statistical mechanics is
\be
A_n(\om)=-\frac{1}{n}\ln P_n(\om),
\ee
where $P_n(\om)$ denotes, as usual, the probability distribution of the sequence $\om=(\om_1,\om_2,\ldots,\om_n)$. We call $A_n$ a \emph{self-process}, since it is a transformed version of $P_n(\om)$. Therefore, studying the large deviations of $A_n$ with respect to $P_n$ is, in a way, the same as studying the large deviations of $P_n$ with respect to itself.

The rate function of $A_n$ can be calculated using the G\"artner-Ellis Theorem once the nature of $\om$ is specified. The random variables $\om_1,\om_2,\ldots,\om_n$ can form, for example, a Markov chain or can be IID, in which case $A_n$ reduces to a simple IID sample mean. Note in this case the form of $\lambda(k)$:
\be
\lambda(k)=\lim_{n\ra\infty}\frac{1}{n}\ln \lex \E^{-k\ln P_n}\rex=\lim_{n\ra\infty}\frac{1}{n}\ln\sum_{\om\in\Lambda^n} P_n(\om)^{1-k}=\ln \sum_{j\in\Lambda} P(\om_i=j)^{1-k}.
\ee
Other types of sequences of random variables or stochastic processes can be dealt with by calculating $\lambda(k)$ from its definition; see in particular \cite{lecomte2005,lecomte2007} for the treatment of continuous-time Markov processes.

An important point to note here is that, if the rate function $I(a)$ of $A_n$ has a unique global minimum and zero, as is the case when $I(a)$ is convex, then
\be
\lim_{n\ra\infty} \lex A_n\rex=a^*
\label{eqlim1}
\ee
and
\be
\lim_{n\ra\infty} A_n=a^*
\label{eqlim2}
\ee
with probability 1. The first limit involving $\lex A_n\rex$ implies, on the one hand, that the \emph{mean Boltzmann-Gibbs-Shannon entropy}  
\be
H_n= -\frac{1}{n}\sum_{\om\in\Lambda^n} P_n(\om)\ln P_n(\om) 
\label{eqer1}
\ee
converges to the constant $a^*$, which is called the \emph{entropy rate} \cite{cover1991} or \emph{Kolmogorov-Sinai entropy} \cite{beck1993,gaspard1998}. On the other hand, the limit (\ref{eqlim2}) involving $A_n$ alone implies
\be
-\frac{1}{n}\ln P_n(\om_1,\om_2,\ldots,\om_n)\ra a^*
\ee
with probability 1 as $n\ra\infty$. The latter limit is known as the \emph{Asymptotic Equipartition Theorem} or the \emph{Shannon-McMillan-Breiman Theorem} \cite{cover1991}. What this result says concretely is that most of the probability $P_n(\om)$ is concentrated on sequences in $\Lambda^n$ such that $P_n(\om)\asymp \E^{-na^*}$. The set $\mathcal{T}_n$ containing these sequences is commonly called the \emph{typical set} of $\Lambda^n$ \cite{cover1991}. Thus $P(\mathcal{T}_n)\ra 1$ and $P_n(\om)\asymp \E^{-na^*}$ for all $\om\in\mathcal{T}_n$ in the limit $n\ra\infty$, which implies that $\mathcal{T}_n$ must contain about $\E^{na^*}$ typical sequences. These results are fundamental in information theory; for a more detailed discussion of this point, see Chap.~X of \cite{bucklew1990}, Sec.~3.6 of \cite{dembo1998} or Chap.~12 of~\cite{cover1991}. For an application of the self-process in the context of nonequilibrium systems, see Example~\ref{exentprod1}.

\begin{example}[Entropy rate of a Markov source] 
\label{exmarkovsource}
For an ergodic Markov chain with transition matrix $\pi(j|i)$, the entropy rate is
\be
a^*=\lambda'(0)=-\sum_{i,j\in\Lambda} \pi(j|i)\rho^*(i)\ln \pi(j|i),
\ee
$\rho^*$ being as usual the unique stationary distribution of the Markov chain.
\end{example}
 
\subsection{Level 1, 2 and 3 of large deviations}

It is customary since the work of Donsker and Varadhan to define three levels of large deviation results referred to as the Level-1, 2 and 3 of large deviations \cite{ellis1985}. \emph{Level-1} is the level of sample means, whereas \emph{Level-2} is the level of Sanov's Theorem, that is, the level of the large deviations of the empirical vector $L_n$. The reason for ordering the large deviations of the empirical vector above those of sample means is that the latter can be derived from the former using the contraction principle. To see this, let $\om=(\om_1,\om_2,\ldots,\om_n)\in\Lambda^n$ be a sequence of random variables, which are not necessarily independent, and let 
\be
S_n=\frac{1}{n}\sum_{i=1}^n f(\om_i).
\ee 
In terms of the empirical vector $L_n$ of $\om$, $S_n$ can always be written as
\be
S_n=\int_\Lambda f(x) L_n(x)\, \D x=f\cdot L_n
\ee
Thus, given the rate function $I_2(\mu)$ of $L_n$, we can use the contraction formula (\ref{eqcp1}) with $h(\mu)=f\cdot\mu$ to express the rate function $I_1(s)$ of $S_n$ as
\be
I_1(s)=\inf_{\mu: h(\mu)=s} I_2(\mu).
\label{eqctr1}
\ee

The contraction function $h(\mu)$ is often written as $\mu(f)$ or $\lex f\rex_\mu$ to emphasize that it is an average of $f$ taken with respect to the random density $\mu$. The empirical vectors solving the constrained minimization problem of Eq.~(\ref{eqctr1}) have an interesting probabilistic interpretation: they are, in the limit $n\ra\infty$, the most probable vectors $L_n$ such that $h(L_n)=s$. Consequently, these vectors must maximize the conditional probability
\be
P(L_n\in d\mu| h(L_n)\in \D s)=\frac{P(L_n\in d\mu,h(L_n)\in \D s)}{P(h(L_n)\in \D s)},
\ee
which implies that they must also globally minimize the rate function
\be
I_2^s(\mu)=
\left\{
\begin{array}{ll}
I_2(\mu)-I_1(s) & \textrm{if } h(\mu)=s\\
\infty & \textrm{otherwise}.
\end{array}
\right.
\label{eqcrf1}
\ee
Consequently, $I_1(s)=\inf_\mu I_2^s(\mu)$.

\begin{example}[Sample means via Sanov's Theorem]
\label{exsmst1}
The constrained minimization arising from the contraction of Level-2 to Level-1 can be solved explicitly for IID sample means. For this case, $I_2(\mu)$ is the relative entropy and the contraction formula (\ref{eqctr1}) is referred to as the \emph{minimum relative entropy principle} \cite{oono1989}. To solve the constrained minimization, we use Lagrange's multipliers method and search for the unconstrained critical points of
\be
F_\alpha(\mu)=\alpha h(\mu)-I_2(\mu),\quad \alpha\in\reals.
\ee
Since $I_2(\mu)$ is strictly convex and $h(\mu)=\langle f\rangle_\mu$ is a linear and differentiable functional of $\mu$, $F_\alpha(\mu)$ has a unique maximum $\mu_\alpha$ for all $\alpha\in\reals$ satisfying $\delta F_\alpha(\mu_\alpha)=0$. Given the expression of the relative entropy, the expression of $\mu_\alpha$ is found to be
\be
\mu_\alpha(x)=\frac{\rho(x)\E^{\alpha f(x)}}{W(\alpha)},\quad W(\alpha)=\int_\Lambda \rho(x)\E^{\alpha f(x)}\, \D x=\lex \E^{\alpha f(X)}\rex_\rho
\label{eqmua1}
\ee
with the value of $\alpha$ implicitly determined by the constraint $h(\mu_\alpha)=\langle f\rangle_{\mu_\alpha}=s$ or, equivalently, $W'(\alpha)/W(\alpha)=s$. The expression for $\mu_\alpha$ makes sense, obviously, provided that $W(\alpha)<\infty$. If this is the case, then $I_1(s)=I_2(\mu_\alpha)$. Equivalently,
\be
I_1(s)=\alpha s-\ln W(\alpha),
\label{eqct2}
\ee
since
\be
F_\alpha(\mu_\alpha)=\alpha h(\mu_\alpha)-I_2(\mu_\alpha)=\alpha s-I_2(\mu_\alpha)=\ln W(\alpha).
\ee
We recognize in Eq.~({\ref{eqct2}}) the result of Cram\'er's Theorem. 
\end{example}

\begin{example}[Symmetric L\'evy random variable revisited]
The contraction of Level-2 to Level-1 does not work for sample means of symmetric L\'evy random variables because $W(\alpha)=\infty$ for all $\alpha\neq 0$. This case is discussed by Lanford \cite{lanford1973}, who proves the result reached in  Example~\ref{exsymlevy}, namely, $I(s)=0$.
\end{example}

The \emph{Level-3} of large deviations, from which Level-2 is obtained by contraction, is the level of the pair empirical distribution $Q_n(x,y)$, which is commonly completed by including all the $m$-tuple empirical distributions defined on $\om=(\om_1,\om_2,\ldots,\om_n)$, $2\leq m\leq n$. In defining $m$-tuple empirical distributions, we require, as we did for the pair empirical distribution, that $\om_{n+1}=\om_1$, $\om_{n+2}=\om_2$, and so forth until $\om_{n+m}=\om_m$, so as to guarantee that all the $(m-1)$-tuple distributions obtained by contraction of an $m$-tuple distribution are the same. The $n$-tuple distribution of an $n$-tuple sequence is the ultimate empirical distribution that can be defined. In the limit $n\ra\infty$, such a distribution becomes an infinite joint empirical distribution called the \emph{empirical process} \cite{donsker1983}. The construction of this abstract process is explained in Sec.~6.5.3 of \cite{dembo1998} or Chap.~IX of \cite{ellis1985}. We will limit ourselves here to noting that, for sequences of IID random variables, the empirical process possesses a convex rate function, and that the zero of this rate function is the infinite product of $\rho$, the common probability distribution of the IID random variables. 

We close this section by noting an alternative characterization of the Level-2 rate function of Markov chains, derived by contracting the large deviations of the pair empirical matrix.

\begin{example}[Sanov's Theorem for Markov chains revisited] 
Let $L_n$ and $Q_n$ denote, respectively, the empirical vector and empirical matrix of an irreducible Markov chain. The contraction $h(Q_n)=L_n$ that takes $Q_n$ to $L_n$ is the usual ``tracing-out'' operation expressed in Eq.~(\ref{eqtr1}). Given the rate function $I_3(q)$ of $Q_n$ found in Eq.~(\ref{eqi3}), we then have
\be
I_2(\mu)=\inf_{q:h(q)=\mu} I_3(q)
\ee
for the rate function of $L_n$. 
\end{example}

%%%%%%%%%%%%%%%%%%%%%%%%%%%%%%%%%%%%%%%%%%%%%%%
\section{Large deviations in equilibrium statistical mechanics}
\label{secequi}

The previous sections introduced all the large deviation results that will now be applied to study the properties of physical systems composed of many particles. We start in this section with the equilibrium properties of many-particle systems described at a probabilistic level by statistical-mechanical ensembles, such as the microcanonical or canonical ensembles. The use of large deviation techniques for studying these systems has its roots in the work of Ruelle \cite{ruelle1969}, Lanford \cite{lanford1973}, and especially Ellis \cite{ellis1985,ellis1995,ellis1999}. Of these sources, Ellis \cite{ellis1985} is the first that explicitly referred to the mathematical theory of large deviations, as developed by Donsker and Varadhan \cite{donsker1975,donsker1975a,donsker1976,donsker1983}, among others. The many links that exist between large deviations and equilibrium statistical mechanics have also been discussed by Lewis, Pfister, and Sullivan \cite{lewis1988,lewis1988a,lewis1989,lewis1994a,lewis1995,lewis1995a}, as well as by Oono \cite{oono1989}. A basic overview of some of these links can be found in \cite{amann1999}.

The material presented in this section borrows from all these sources. By defining statistical-mechanical ensembles in a way that is explicitly focused on large deviation theory, we aim to show here that the study of equilibrium states and their fluctuations in a given ensemble can be reduced to the study of properly-defined rate functions. In the process, many connections between equilibrium statistical mechanics and large deviation theory will be established and discussed. We will see, in particular, that entropy functions are special rate functions, and that variational principles, such as the maximum entropy principle or the minimum free energy principle, follow from the contraction principle. The last observation is especially useful because it provides us with a clear explanation of why variational principles arise in equilibrium statistical mechanics. It also provides us with a systematic method or \emph{scheme} for deriving such variational principles in general.

\subsection{Basic principles}

The following list of common definitions and postulates, inspired from \cite{ruelle1969,lanford1973,ellis1985}, establishes the basis of equilibrium statistical mechanics on which we will work:

\begin{itemize}
\item We consider a collection of $n$ particles (atoms, spins, molecules, etc.) that interact with one another through some forces or potentials.\footnote{In keeping with the notations of the previous sections, we use $n$ to denote the number of particles rather that the more common $N$ used in physics.} 

\item The \emph{collective} or \emph{joint} state of the $n$ particles is denoted by a sequence $\om=(\om_1,\om_2,\ldots,\om_n)$ of $n$ variables, with $\om_i$ denoting the state of the $i$th particle. 

\item A sequence $\om$ is called a \emph{microstate}, as it gives a complete description of the $n$-particle system at the microscopic level. The set or space $\Lambda_n$ of all microstates is the $n$-fold product $\Lambda^n$ of the one-particle state space $\Lambda$.  

\item The physical interactions or dependencies between the $n$ particles are determined by a \emph{Hamiltonian} or \emph{energy function} $H_n(\om)$. Given $H_n(\om)$, we define the \emph{mean energy} or \emph{energy per particle} by $h_n(\om)=H_n(\om)/n$.

\item The microstate $\om$ of the $n$-particle system is modeled abstractly as a random variable, which is distributed according to a reference or \emph{prior} probability measure $P(\D\om)$ on $\Lambda_n$. The form of $P(\D\om)$ is determined by physical considerations. For most models, Liouville's Theorem dictates that $P(\D\om)$ be the uniform measure $P(\D\om)=\D\om/|\Lambda_n|$, where $|\Lambda_n|=|\Lambda|^n$ is the \emph{volume} of $\Lambda_n$. Since $|\Lambda_n|$ is a constant, one can work equivalently with the unnormalized (Lebesgue) measure $P(\D\om)=\D\om$.

\item The probabilistic description of the $n$-particle system is completed by specifying the external conditions or constraints under which that system is prepared or studied. The specification of these conditions is tantamount to selecting a given \emph{statistical-mechanical ensemble}, which corresponds mathematically to a probability distribution on $\Lambda_n$ involving the constraints and the prior distribution $P(\D\om)$.

\item The interactions between the particles give rise to a \emph{macroscopic} or \emph{thermodynamic} behavior of the whole system that can be described by having recourse to a few macroscopic or ``coarse-grained'' variables called \emph{macrostates}. Mathematically, a macrostate is just a function $M_n(\om)$ of the microstates.

\item The thermodynamic behavior of the whole system is characterized by one or more \emph{equilibrium states}, defined as the most probable values of a set of macrostates in a chosen ensemble. 

\item When calculating equilibrium states, the limit $n\ra\infty$ is assumed to obtain states that are representative of macroscopic systems. This limit is called the \emph{thermodynamic limit}, and entails in many cases the continuum limit.
\end{itemize}

The mathematical basis for the notion of thermodynamic behavior is the Law of Large Numbers \cite{lanford1973}. The idea, in a nutshell, is that the outcomes of a macrostate, say $M_n(\om)$, involving $n$ particles should concentrate in probability around certain stable or equilibrium values  (macroscopic determinism) despite the fact that the particles' state is modeled by a random variable $\om$ (microscopic chaos). Large deviation theory enters this picture by noting that, in many cases, the outcomes of $M_n$ are ruled by a large deviation principle, and that, in these cases, the concentration of $M_n$ around equilibrium values is ``exponentially effective'' in the limit $n\ra\infty$, as the probability of observing a departure from these equilibrium values is exponentially small with the number $n$ of particles. Consequently, all that is needed to describe the state of a large many-particle system at the macroscopic level is to know the equilibrium values of $M_n$ which correspond to the global minima of the rate function governing the fluctuations of $M_n$. 

These considerations summarize the application of large deviation techniques in equilibrium statistical mechanics. What remains to be done at this point is to show how the probabilities of microstates and macrostates are to be constructed depending on the nature of the many-particle system studied, and to show how  rate functions are extracted from these probabilities. For simplicity, we will review here only two types of many-particle systems, namely, closed systems at constant energy, and open systems exchanging energy with a heat bath at constant temperature.  The first type of system is modeled, as is well known, by the \emph{microcanonical ensemble}, whereas the second type is modeled by the \emph{canonical ensemble}. The treatment of other ensembles follows the treatment of these two ensembles. 

\subsection{Large deviations of the mean energy}

Before exploring the large deviations of general macrostates, it is useful to study the large deviations of the mean energy $h_n(\om)$ with respect to the prior distribution $P(\D\om)$ defined on $\Lambda_n$. The rate function turns out in this case to be the microcanonical entropy function up to an additive constant, whereas the scaled cumulant generating function of $h_n$ turns out to be the canonical free energy function, again up to a constant. These associations and their consequences for thermodynamics are explained next. 

\subsubsection{Entropy as a rate function}

Using the notation developed in the previous sections, we write the probability distribution of $h_n$ with respect to the prior $P(\D \om)$ on $\Lambda_n$ as
\be
P(h_n\in \D u)=\int_{\{\om\in\Lambda_n:h_n(\om)\in \D u\}} P(\D\om),
\label{eqh1}
\ee
where $\D u=[u,u+\D u]$ is an infinitesimal interval of mean energy values. For the uniform prior measure $P(\D\om)=\D\om/|\Lambda|^n$, $P(h_n\in \D u)$ is proportional to the volume 
\be
\Omega(h_n\in \D u)=\int_{\{\om\in\Lambda_n:h_n(\om)\in \D u\}} \D\om
\ee
of microstates $\om$ such that $h_n(\om)\in \D u$. Therefore, if $P(h_n\in \D u)$ scales exponentially with $n$, then so must $\Omega(h_n\in \D u)$. Defining the rate function $I(u)$ of $P(h_n\in \D u)$ by the usual limit
\be
I(u)=\lim_{n\ra\infty} -\frac{1}{n}\ln P(h_n\in \D u),
\ee
we must then have
\be
I(u)=\ln|\Lambda|-s(u),
\label{eqis1}
\ee
where
\be
s(u)= \lim_{n\ra\infty} \frac{1}{n}\ln\Omega(h_n\in \D u)
\label{eqs1}
\ee
is the \emph{microcanonical entropy} or \emph{entropy density}. Equation~(\ref{eqis1}) proves our first claim, namely, that the rate function $I(u)$, if it exists, is the negative of the entropy $s(u)$ up to the additive constant $\ln |\Lambda|$. 

In the following, we shall absorb the constant $\ln|\Lambda|$ by re-defining the entropy using the limit
\be
s(u)=\lim_{n\ra\infty}\frac{1}{n}\ln P(h_n\in \D u)
\label{eqnews1}
\ee
rather than the limit displayed in Eq.~(\ref{eqs1}). This re-definition simply amounts to replacing the Lebesgue measure $\D\om$ in the integral of $\Omega(h_n\in u)$ by the uniform prior measure $P(\D\om)$, in which case $I(u)=-s(u)$. This minor re-definition of the entropy complies with the definition used in works on large deviations and statistical mechanics (see, e.g., \cite{ellis1985,ellis1995,oono1989,pfister1991,lewis1995a}). It brings, for one thing, the notion of entropy closer to large deviation theory, and allows one to use prior measures that are not uniform. Note that throughout this review, we also use $k_B=1$. 

\subsubsection{Free energy as a scaled cumulant generating function}

The proportionality of $P(h_n\in \D u)$ and $\Omega(h_n\in \D u)$ noted above implies that
\be
\lambda(k)=\lim_{n\ra\infty}\frac{1}{n}\ln \lex \E^{nkh_n}\rex
\ee
satisfies 
\be
\lambda(k)=-\left.\varphi(\beta)\right|_{\beta=-k}-\ln |\Lambda|. 
\label{eqlp1}
\ee
where
\be
\varphi(\beta)=\lim_{n\ra\infty} -\frac{1}{n}\ln Z_n(\beta),
\label{eqf1}
\ee
and 
\be
Z_n(\beta)=\int_{\Lambda_n} \E^{-n\beta h_n(\om)}\, \D\om=\int_{\Lambda_n} \E^{-\beta H_n(\om)}\, \D\om.
\label{eqpartf1}
\ee
The latter function is the well-known $n$-particle \emph{partition function} associated with $H_n$; accordingly, $\varphi(\beta)$ is the \emph{canonical free energy function}. With these associations, we see, as announced, that $\lambda(k)$ is the free energy function of the canonical ensemble up to a constant and a change of variable ($\beta=-k$). As we did for the entropy, we shall absorb the constant $\ln|\Lambda|$ in $\varphi(\beta)$ by re-defining this function as
\be
\varphi(\beta)=\lim_{n\ra\infty} -\frac{1}{n}\ln \int_{\Lambda_n} \E^{-n\beta h_n(\om)}\, P(\D\om)=\lim_{n\ra\infty} -\frac{1}{n}\ln\lex \E^{-n\beta h_n(\om)} \rex
\label{eqnewf1}
\ee
using $P(\D\om)$ instead of $\D\om$ as the measure entering in the integral of the partition function. We this new definition, we then have $\lambda(k)=-\left.\varphi(\beta)\right|_{\beta=-k}$. This form of free energy function will be used from now on, since it also complies with the form used in works on large deviations and statistical mechanics.

It should be noted for correctness that what is commonly referred to as the free energy in thermodynamics is not the function $\varphi(\beta)$ but the function $f(\beta)=\varphi(\beta)/\beta$. Here we use $\varphi(\beta)$ as the free energy because this function has the convenient property of always being concave in $\beta$. The function $f(\beta)$, by contrast, is concave or convex (negative concave) depending on the sign of $\beta$. In textbooks of statistical mechanics, $\varphi(\beta)$ is sometimes called the \emph{Massieu potential} \cite{balian1991}.

\subsubsection{Legendre transforms in thermodynamics}

The relationships that we have established between $I(u)$ and $s(u)$, on the one hand, and $\lambda(k)$ and $\varphi(\beta)$, on the other, are important because they imply that the G\"artner-Ellis Theorem and Varadhan's Theorem can be applied to express $s(u)$ as the Legendre-Fenchel transform of $\varphi(\beta)$, and vice versa. By transposing Varadhan's Theorem at the level of $s(u)$ and $\varphi(\beta)$, we indeed obtain
\be
\varphi(\beta)=\inf_{u} \{ \beta u -s(u)\},
\label{eqfree1}
\ee
whereas for the the G\"artner-Ellis Theorem, we obtain
\be
s(u)=\inf_{\beta} \{\beta u-\varphi(\beta)\},
\label{eqs2}
\ee
provided that $\varphi(\beta)$ exists and is differentiable. In both expressions, ``$\inf$'' stands as before for the ``infimum of''. The reason why the Legendre-Fenchel transform is now expressed with an ``$\inf$'' instead of a ``$\sup$'' is, of course, because the entropy is defined as the negative of a rate function; see Eq.~(\ref{eqis1}).

The two Legendre-Fenchel transforms shown above are fundamental in statistical mechanics. The first one shown in Eq.~(\ref{eqfree1}) provides a precise formulation of the basic thermodynamic principle that states that the free energy is the Legendre transform of the entropy. Varadhan's Theorem refines this result by establishing that $\varphi(\beta)$ is, in general, the \emph{Legendre-Fenchel} transform of $s(u)$, not simply the \emph{Legendre} transform, and that this Legendre-Fenchel transform is valid for essentially any $s(u)$. Legendre-Fenchel transforms rather than Legendre transforms must be used in particular when $s(u)$ is not concave. 

The second Legendre-Fenchel transform shown in Eq.~(\ref{eqs2}) is the converse of the first, expressing the entropy as the Legendre-Fenchel transform of the free energy. This result is also well known in thermodynamics, but in a form that usually also involves the Legendre transform rather than the Legendre-Fenchel transform, and without reference to any conditions about the validity of that transform. These conditions are the conditions of the G\"artner-Ellis Theorem; they are important, and will be studied in detail later when discussing nonconcave entropies. 

For convenience, we will often refer thereafter to the two Legendre-Fenchel transforms shown above using the ``star'' notation introduced in Sec.~\ref{subsecncrf}. With this notation, Eq.~(\ref{eqfree1}) is expressed as $\varphi=s^*$, while Eq.~(\ref{eqs2}) is expressed as $s=\varphi^*$.

\subsection{Microcanonical ensemble}

We now come to the problem that we set ourselves to solve in this section: we consider an $n$-particle system represented by a Hamiltonian function $H_n(\om)$, and attempt to derive a large deviation principle for a macrostate $M_n(\om)$ of that system. We consider first the case of a closed system constrained to have a fixed energy $H_n(\om)=U$. Other constraints can also be included (see, e.g., \cite{ellis2000}). The statistical-mechanical ensemble that models the stationary properties of such a system is, as is well known, the microcanonical ensemble, which we define mathematically next.

\subsubsection{Definition of the ensemble}

The microcanonical ensemble is based on the assumption that all the microstates $\om\in\Lambda_n$ such that $H_n(\om)=U$ or, equivalently, such that $h_n(\om)=U/n=u$ are equally probable (equiprobability postulate). Therefore, what we call the microcanonical ensemble at the level of microstates is the conditional probability measure 
\be
P^u(\D\om)=P(\D\om|h_n\in \D u)=
\left\{
\begin{array}{ll}
\displaystyle\frac{P(\D\om)}{P(h_n\in \D u)} & \textrm{if } h_n(\om)\in \D u\\
0 & \textrm{otherwise},
\end{array}
\right.
\ee
which assigns a non-zero and constant probability only to those microstates having a mean energy lying in the interval $\D u$. The probability $P(h_n\in \D u)$ was introduced earlier, and is there to make $P^u(\D\om)$ a normalized measure:
\be
\int_{\Lambda_n} P^u(\D\om)=\frac{1}{P(h_n\in \D u}\int_{\{\om\in\Lambda_n: h_n(\om)\in \D u\}} P(\D\om)=1.
\ee

The extension of the microcanonical measure $P^u(\D\om)$ to macrostates follows the standard rules of probability theory. Given a macrostate $M_n(\om)$, we define $P^u(M_n\in \D m)$ to be the conditional or constrained probability measure given by
\be
P^u(M_n\in \D m)=P(M_n\in \D m| h_n\in \D u)= \frac{P(h_n\in \D u,M_n\in \D m)}{P(h_n\in \D u)},
\label{eqmicro1}
\ee
where
\be
P(h_n\in \D u,M_n\in \D m)=\int_{\{\om\in\Lambda_n:h_n(\om)\in \D u,M_n(\om)\in \D m \}} P(\D\om)
\ee
is the joint probability of $h_n$ and $M_n$. It is this probability measure that we have to use to find the most probable values of $M_n$ given that the system represented by the Hamiltonian $H_n$ has a fixed energy $H_n=U$ or, equivalently, a fixed mean energy $h_n=U/n=u$. The latter expression of the energy constraint involving $h_n$ is generally preferred over the former involving $H_n$, since we are interested in finding the most probable values of $M_n$ in the large-$n$ or thermodynamic limit. Thermodynamic limits involving a different rescaling of the energy are also conceivable, depending on the form of the Hamiltonian.

\subsubsection{Microcanonical large deviations}

The theory of large deviations enters in the description of the microcanonical ensemble as a basic tool for finding the the values of $M_n$ that maximize the microcanonical probability measure $P^u(M_n\in \D m)$. From our knowledge of sample means, we should expect at this point to be able to prove a large deviation principle for $P^u(M_n\in \D m)$, and to find the most probable values of $M_n$ by locating the global minima of the corresponding rate function. As shown next, this is possible if a large deviation principle holds for the \emph{unconstrained} measure $P(M_n\in \D m)$, and if there exists a contraction of $M_n$ to $h_n$. In this case, the equilibrium values of $M_n$ that globally minimize the rate function of the \emph{constrained} measure $P^u(M_n\in \D m)$ can be calculated as the global minima of the rate function of the \emph{unconstrained} measure $P(M_n\in \D m)$ subject to the constraint $h_n(\om)=u$ \cite{ellis1995,ellis1999}.

To prove this result, consider a macrostate $M_n(\om)$, and suppose that a large deviation principle holds for this macrostate with respect to the unconstrained prior measure $P(\D\om)$, that is,
\be
P(M_n\in \D m)=\int_{\{\om\in\Lambda_n:M_n(\om)\in \D m\}} P(\D\om)\asymp \E^{n\rs(m)}\, \D m.
\label{eqsldp1}
\ee
The rate function of this large deviation principle is written without the usual minus sign to conform with the notation used in physics. The negative ``rate function'' $\rs(m)$ is called the \emph{macrostate entropy} of $M_n$, since it effectively corresponds to the entropy of $M_n$ defined with the volume measure $\Omega(M_n\in \D m)$ up to the constant $\ln|\Lambda|$. 

Suppose now that the mean energy or energy per particle $h_n(\om)$ can be rewritten as a function of the macrostate $M_n(\om)$. That is to say, suppose that there exists a bounded, continuous function $\rh(m)$ of $M_n$, called the \emph{energy representation function}, such that $h_n(\om)=\rh(M_n(\om))$ for all $\om\in\Lambda_n$ or, more generally, such that 
\be
|h_n(\om)-\rh(M_n(\om))|\ra 0 
\ee
uniformly over all $\om\in\Lambda_n$ as $n\ra\infty$. Given that this function exists, it is readily seen that the most probable values $m$ of $M_n(\om)$ with respect to $P^u(M_n\in \D m)$ are those that maximize the macrostate entropy $\rs (m)$ subject to the constraint $\rh(m)=u$. To be sure, construct the explicit large deviation principle for $P^u(M_n\in \D m)$. Assuming that $P(M_n\in \D m)$ satisfies the large deviation principle shown in (\ref{eqsldp1}), it follows by contraction that $P(h_n\in \D u)$ also satisfies a large deviation principle which we write, as before, as
\be
P(h_n\in \D u)\asymp \E^{ns(u)}\, \D u.
\ee
Combining these large deviations in the expression of $P^u(M_n\in \D m)$ shown in Eq.~(\ref{eqmicro1}), we then obtain
\be
P^u(M_n\in \D m)\asymp \E^{-nI^u(m)}\, \D m,
\label{eqmldp1}
\ee
where
\be
I^u(m)=
\left\{
\begin{array}{ll}
s(u)-\rs(m) & \textrm{if } \rh(m)=u \\
\infty & \textrm{otherwise}.
\end{array}
\right.
\label{eqmrf1}
\ee
The rate function $I^u(m)$ is similar to the rate function $I^s_2(\mu)$ discussed in connection with the contraction of the level-2 large deviations to the level-1 large deviations; see Eq.~(\ref{eqcrf1}). The main point to observe here is that the global minimizers of $I^u(m)$, which correspond to the equilibrium values of $M_n$ in the microcanonical ensemble with $h_n=u$, are the global maximizers of the macrostate entropy $\rs (m)$ subject to the constraint $\rh(m)=u$. Denoting by $\mE^u$ the set of all such \emph{equilibrium values} or \emph{equilibrium states}, we then write
\be
\mE^u=\{m:I^u(m)=0\}=\{m:m \textrm{ globally maximizes } \rs(m) \textrm{ with } \rh(m)=u\}.
\label{eqeu1}
\ee
The class of macrostates for which $\mE^u$ can be calculated using the formula above depends on the model studied and, more precisely, on the form of its Hamiltonian. This point will be discussed in more detail later. 

It useful to know that $\mE^u$ can be calculated, at least in theory, without the macrostate entropy $\rs(m)$ and the energy representation function $\rh(m)$. If we can prove, for instance, that $P(h_n\in \D u,M_n(\om)\in \D m)$ satisfies a joint large deviation principle of the form
\be
P(h_n\in \D u,M_n(\om)\in \D m)\asymp \E^{n\rs(u,m)}\, \D u\, \D m,
\label{eqjldp1}
\ee
then we obtain
\be
P^u(M_n\in \D m)\asymp \E^{-nJ^u(m)}\, \D m,
\label{eqmldp2}
\ee 
where $J^u(m)=s(u)-\rs(u,m)$. In this case,
\be
\mE^u=\{m: J^u(m)=0\}=\{m: m \textrm{ globally maximizes } \rs(u,m)\}.
\label{eqeu2}
\ee

One may also attempt to obtain $I^u(m)$ directly using the G\"artner-Ellis Theorem. This method, however, is of limited use, since the calculation of the scaled cumulant generating function of $M_n$ in the microcanonical ensemble involves a constrained integral on $\Lambda_n$ which can be evaluated explicitly only for certain combinations of macrostates and Hamiltonians (e.g., non-interacting particles). The G\"artner-Ellis Theorem is also limited in that it cannot be used, as we have seen, to calculate nonconvex rate functions, which implies that it cannot be used to calculate nonconcave entropy functions. In this case, the representation of $\mE^u$ based on $\rs(m)$ and $\rh(m)$ or $\rs(u,m)$ alone should be used. The former representation based on $\rs(m)$ and $\rh(m)$ is generally more practical.

\subsubsection{Einstein's fluctuation theory and the maximum entropy principle}

The microcanonical large deviation principles displayed in (\ref{eqmldp1}) and (\ref{eqmldp2}) provide a precise formulation of Einstein's theory of microcanonical fluctuations \cite{einstein1987}. They embody the main result of that theory, which is that probabilities in the microcanonical ensemble can be expressed in terms of entropies. But they also refine that result, in that
\begin{itemize}
\item They provide a precise expression of the exponential scaling of $P^u(M_n\in \D m)$ with $n$, which need not be satisfied by all macrostates. That exponential scaling is somewhat hidden in Einstein's theory in the implicit assumption that the entropy is an extensive quantity.\footnote{Of course, one can always put a probability $P$ in the form $P=\E^S$ simply by defining $S=\ln P$. The non-trivial result stated in Einstein's theory, and more precisely in large deviation theory, is that $S=\ln P$ is extensive with the number of particles, which means that $P$ decays exponentially fast with the number of particles. This statement is the essence of the large deviation principle.}

\item They lead us to identify the equilibrium values of $M_n$ not just as those values $m$ maximizing $P^u(M_n\in \D m)$, but, more precisely, as the zeros of $I^u(m)$ or $J^u(m)$, thereby bringing the study of equilibrium states in direct contact with the Law of Large Numbers.\footnote{This shows, incidentally, that thermodynamics and statistical mechanics are not really concerned about \emph{average} values so much as about \emph{most probable} values. Equilibrium states are, first and foremost, most probable states.}

\item They suggest a procedure---a scheme---for deriving general \emph{maximum entropy principles} that can be used to find the equilibrium values of $M_n$, and to calculate $s(u)$ in terms of a maximization involving a macrostate entropy. 
\end{itemize}

The last point simply follows by examining the two representations of the set $\mE^u$ of microcanonical equilibrium states, defined in Eqs.~(\ref{eqeu1}) and (\ref{eqeu2}). The representation of Eq.~(\ref{eqeu1}), which involves the rate function $I^u(m)$, implies on the one hand that
\be
s(u)=\sup_{m: \rh(m)=u} \rs(m).
\label{eqme1}
\ee
On the other hand, the representation of Eq.~(\ref{eqeu2}), which involves $J^u(m)$ instead of $I^u(m)$, implies that
\be
s(u)=\sup_{m}\ \rs(u,m).
\label{eqmer2}
\ee
These variational formulae can also be derived from the contraction principle. In the first formula, the contraction is the energy representation function, whereas in the second, the contraction is the map $(u,m)\ra u$. Each formula provides, in the end, a general \emph{maximum entropy principle} that can be used to calculate the microcanonical entropy $s(u)$ from the knowledge of a macrostate entropy. The well-known maximum entropy principle of Jaynes \cite{jaynes1957a,jaynes2003} is a particular application of these formulae, obtained by considering the Level-2 large deviations of systems of independent particles. This is explained in the next example. 

\begin{example}[Jaynes's Maximum Entropy Principle] 
\label{exjaynes1}
Consider a system of $n$ particles with individual energies $\varepsilon_1, \varepsilon_2,\ldots, \varepsilon_n$, $\varepsilon_i\in\Lambda$. Assuming that the particles do not interact with each other, we write the mean energy $h_n$ of the $n$ particles as 
\be
h_n=\frac{1}{n}\sum_{i=1}^n \varepsilon_i.
\ee
An obvious choice of macrostate for this model, apart from the mean energy itself, is the empirical vector or \emph{one-particle energy distribution}
\be
L_n(\varepsilon)=\frac{1}{n}\sum_{i=1}^n \delta(\varepsilon_i-\varepsilon)
\ee
which counts the relative number of particles having an energy $\varepsilon$. The energy representation for this choice of macrostate is
\be
\rh(L_n)=\int_\Lambda \varepsilon L_n(\varepsilon)\, \D\varepsilon.
\ee
From Sanov's Theorem, the large deviations of $L_n$ are ruled by the relative entropy $I_\rho(\mu)$. For the uniform prior $\rho=|\Lambda|^{-1}$, $I_\rho(\mu)$ is related to the Boltzmann-Gibbs-Shannon entropy
\be
\rs(\mu)=-\int_\Lambda \D\varepsilon\, \mu(\varepsilon)\ln \mu(\varepsilon)
\ee
through $I_\rho(\mu)=-\rs(\mu)+\ln|\Lambda|$. Therefore,
\be
s(u)=\sup_{l:\rh(\mu)=u} \rs(\mu)-\ln|\Lambda|
\ee
by the general maximum entropy principle derived in (\ref{eqme1}). For independent particles, the microcanonical entropy $s(u)$ is thus obtained by maximizing the Boltzmann-Gibbs-Shannon entropy $\rs(\mu)$ subject to the energy constraint $\rh(\mu)=u$. It is this version of the maximum entropy principle that we refer to as Jaynes's maximum entropy principle \cite{jaynes1957a,jaynes2003}. 

A variational problem similar to the one displayed above was solved in Example~\ref{exsmst1} when treating the contraction of Level-2 to Level-1. Its explicit solution, re-written in a more thermodynamic form, is
\be
\mu_\beta(\varepsilon)=\frac{\E^{-\beta\varepsilon}}{Z(\beta)},\quad Z(\beta)=\int_\Lambda \E^{-\beta\varepsilon}\, \D\varepsilon
\ee
with $\beta$ implicitly determined by $\rh(\mu_\beta)=u$ or, equivalently, by $Z'(\beta)/Z(\beta)=u$. Similarly as in Example~\ref{exsmst1}, we therefore obtain
\be
s(u)=\rs(\mu_\beta)-\ln|\Lambda|=\beta u-\varphi(\beta),
\ee
which is nothing but Cram\'er's Theorem written in terms of $s(u)$ and $\varphi(\beta)$.\footnote{Cram\'er's Theorem appears here because the mean energy $h_n$ of non-interacting particles is a sample mean of IID random variables under the uniform prior measure $P(\D\om)$.} Since the linear form of $\rh(L_n)$ is directly related to the additive form of $h_n$, the explicit expression of $\mu_\beta$ does not carry over to the case where there is some interaction between the particles. Thus, strictly speaking, Jaynes's maximum entropy principle is only applicable to non-interacting particles.\footnote{The form of the entropy function also depends on the observable considered. To be more precise, one should therefore say that Jaynes's maximum entropy principle applies only to the \emph{one-particle distribution} of \emph{non-interacting} particle systems. Other maximum principles that are applicable to other observables and other systems can be derived, following this section, by obtaining an explicit large deviation expression for the probability of a given observable, defined in the context of a given system and ensemble. In the end, it is probability that defines the form of a maximum entropy principle, not the choice of an arbitrary entropy function or some \emph{ad hoc} information-based argument.}
\end{example}

\subsubsection{Treatment of particular models}

The microcanonical equilibrium properties of systems of non-interacting particles can always be treated, as in the previous example, at the Level-2 of large deviations using Sanov's Theorem, or directly at the Level-1 using Cram\'er's Theorem (see, e.g., \cite{lehtonen1990}). These two levels of large deviations can also be used in general to study the equilibrium properties of mean-field models of particles involving an all-to-all coupling between particles. Examples of such models, for which the general maximum entropy principles mentioned before have been applied successfully, include the mean-field versions of the Curie-Weiss model \cite{ellis1985,eisele1988,ellis1995} and its parent model, the Potts model \cite{orey1988,ellis1990,ellis1995,costeniuc2005a}, the Blume-Emery-Griffiths model \cite{barre2001,ellis2004,ellis2005}, the mean-field Hamiltonian model \cite{barre2005}, as well as mean-field versions of the spherical model \cite{kastner2006,casetti2007}, and the $\phi^4$ model \cite{hahn2005,hahn2006,campa2007}. In all of these models, the energy representation function is either a nonlinear function of the empirical vector (Level-2) or a function of properly-chosen Level-1 macrostates, commonly referred to as \emph{mean fields} or \emph{order parameters}. This is illustrated in the next two examples.

\begin{example}[Mean-field Potts model]
\label{exmfp}
The mean-field Potts model with $q$ states is defined by the Hamiltonian
\be
H_n(\om)=-\frac{1}{2n}\sum_{i,j} \delta_{\om_i,\om_j},
\ee
where $\om_i\in\Lambda=\{1,2,\ldots,q\}$. The factor $n$ in front of the sum is there to make the energy an extensive variable or, equivalently, to make the mean energy an intensive variable. In terms of the empirical vector
\be
L_n(\om)=(L_{n,1}(\om),L_{n,2}(\om),\ldots,L_{n,q}(\om)),\quad L_{n,j}(\om)=\frac{1}{n}\sum_{i=1}^n \delta_{i,j}
\ee 
we obviously have $h_n(\om)=\rh(L_n(\om))$, where
\be
\rh(\mu)=-\frac{1}{2} \mu\cdot\mu=-\frac{1}{2}\sum_{j=1}^q \mu_j^2.
\ee
The reader is referred to \cite{costeniuc2005a} for the complete calculation of $s(u)$ based on this energy representation function, and for the calculation of the equilibrium values of $L_n$ in the microcanonical ensemble. Note that the macrostate entropy that needs to be maximized here is the Boltzmann-Gibbs-Shannon entropy, or relative entropy, as in the case of non-interacting particles; however, now the energy representation function is a nonlinear function of the empirical vector.
\end{example}

\begin{example}[Mean-field $\phi^4$ model~\cite{campa2007,hahn2005,hahn2006}]
\label{exmfp4}
The mean-field $\phi^4$ model is defined by the Hamiltonian
\be
H_n=\sum_{i=1}^n \left( \frac{p_i^2}{2}-\frac{q_i^2}{2}+\frac{q_i^4}{4}\right)-\frac{1}{4n}\sum_{i,j=1}^n q_iq_j,
\ee
where $p_i,q_i\in\reals$. We can re-write the mean energy of this model using the following energy representation function:
\be
\rh(k,v,m)=k+v-\frac{m^2}{4},
\ee
where 
\be
k=\frac{1}{2n}\sum_{i=1}^n p_i^2,\quad v=\frac{1}{n}\sum_{i=1}^n \left(\frac{q_i^4}{4}-\frac{q_i^2}{2}\right),\quad m=\frac{1}{n}\sum_{i=1}^n q_i
\ee
are, respectively, the mean kinetic energy, the mean potential energy, and the mean magnetization of the model. The entropy functions of these macrostates can be derived using the G\"artner-Ellis Theorem, since they are all strictly convex. The entropy $\rs(m)$ of $m$, for example, is the magnetization entropy found in Example~\ref{exentspin}, which is also the negative of the rate function calculated for the binary sample mean of Example~\ref{exbin}. The calculation of the two other macrostate entropies $\rs(k)$ and $\rs(v)$ is reported in \cite{campa2007}. In the end, we obtain $s(u)$ by solving
\be
s(u)=\sup_{(k,v,m):\rh(k,v,m)=u} \rs(k)+\rs(v)+\rs(m).
\ee
The details of this calculation can be found in \cite{campa2007}.
\end{example}

When going beyond non-interacting and mean-field systems, two different classes of systems must be distinguished: those involving \emph{long-range} interactions, such as systems of gravitating particles, and those involving \emph{short-range} interactions, such as the nearest-neighbor Ising model. From the point of view of the formalism developed here, long-range systems (see \cite{draw2002} for a definition of long-range interactions) are similar to mean-field systems, in that their mean energy often admits a representation function involving Level-1 or Level-2 macrostates \cite{bouchet2005}. This is the case, for example, for systems of gravitating particles and plasmas, which can be investigated in the mean-field or Vlasov limit using the empirical distribution (see \cite{draw2002,campa2008,campa2009} for recent reviews). Some statistical models of two-dimensional (2D) turbulence can also be treated with an energy representation function involving a relatively simple macrostate (see, e.g., \cite{ellis2000}). A particularity of these models is that the prior distribution $P(\D\om)$ is not always chosen to be the uniform measure~\cite{ellis2002}. Another model worth mentioning, finally, is the so-called $\alpha$-Ising model in one dimension, which admits the local magnetization function as a mean-field \cite{barre2005}. Other models of long-range systems are discussed in \cite{campa2008,campa2009}.

Systems involving short-range interactions are much more complicated to study  due to the fact that their large deviation analysis must be based on the Level-3 empirical process mentioned in Sec.~\ref{secmathapp}. The empirical process can be used in principle to study non-interacting and mean-field models, but this is never done in practice, as there are simpler macrostates to work with. The problem with short-range models is that the empirical process is, in general, the \emph{only} macrostate admitting an energy representation function. This is the case, for example, for the nearest-neighbor Ising model, which has been studied extensively in one and two dimensions from the point of view of the empirical process (see, e.g.,~\cite{ellis1985,pfister1991,ellis1995,pfister2002}). We summarize in the next example the equilibrium properties of the mean magnetization of the 2D version of this model, obtained by contracting the large deviations of the empirical process down to the mean energy and mean magnetization. The main sources for this example are Ellis \cite{ellis1995}, Pfister \cite{pfister2002}, and Kastner~\cite{kastner2002}. For a discussion of the Ising model in one dimension, see \cite{ellis1985,lewis1994}.

\begin{figure*}[t]
\centering
\includegraphics[scale=1.0]{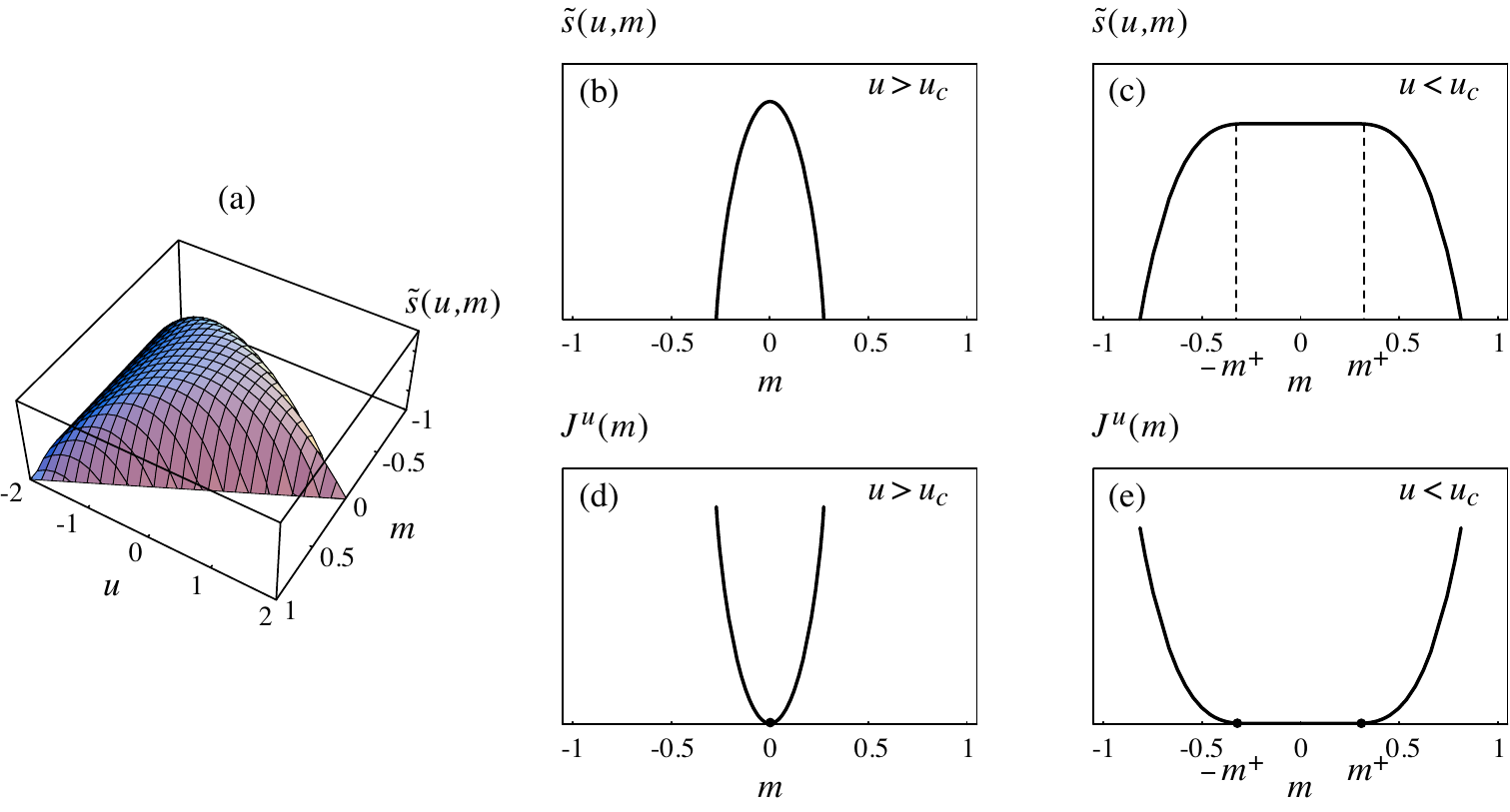}
\caption{(a) Sketch of the joint entropy $\rs(u,m)$ of the 2D Ising model (after \cite{kastner2002}). (b) Projection of $\rs(u,m)$ as a function of $m$ for $u<u_c$. (c) Projection of $\rs(u,m)$ as a function of $m$ for $u>u_c$. (d)-(e) Microcanonical rate function $J^u(m)$ above and below the critical mean energy $u_c$.}
\label{fig2dising}
\end{figure*}

\begin{example}[2D Ising model]
\label{ex2disingm}
Consider the 2D nearest-neighbor Ising model, defined by the usual Hamiltonian
\be
H_n=-\frac{1}{2}\sum_{\langle i,j\rangle} \sigma_i\sigma_j,\quad \sigma_i\in\{-1,1\},
\ee
where $\langle i,j\rangle$ denotes first-neighbor sites on the finite 2D square lattice containing $n$ spins. The entropy of this model is sketched, following \cite{kastner2002}, in Fig.~\ref{fig2dising} as a function of the mean energy $u$ and mean magnetization $m$. There are many properties of this entropy worth noting from the point of large deviations:
\begin{itemize}
\item $\rs(u,m)$ is strictly concave in $m$ for $u\in [u_c,2]$, where $u_c=-\sqrt{2}$, with a maximum located at $m=0$; see Fig.~\ref{fig2dising}(b). 

\item $\rs(u,m)$ is concave in $m$ for $u\in [-2,u_c)$, but not strictly concave. In fact, for this range of mean energies, $\rs(u,m)$ is constant in $m$ in the interval $[-m^+(u),m^+(u)]$; see Fig.~\ref{fig2dising}(c). The boundary point $m^+(u)$ is called the \emph{spontaneous magnetization}, and is such that $m^+(u_c)=0$ and $m^+(u)\ra 1$ as $u\ra -2$. 

\item By contraction, $s(u)=\rs(u,0)$, since $m=0$ is always a maximum of $\rs(u,m)$ for all $u\in [-2,2]$. Although not plotted, $s(u)$ is known to be concave and differentiable, which means that it can be calculated in principle as the Legendre transform of the canonical free energy calculated by Onsager \cite{onsager1944}. 

\item The rate function $J^u(m)=s(u)-\rs(u,m)$ has a unique minimum and zero for $[u_c,2]$, corresponding to the unique equilibrium value of the mean magnetization in the microcanonical ensemble for all $u\in [u_c,2]$; see Fig.~\ref{fig2dising}(d).

\item For $u\in [-2,u_c)$, $J^u(m)$ is zero for all $m$ in the interval $[-m^+(u),m^+(u)]$, which is called the \emph{phase transition interval} or \emph{phase coexistence region} \cite{ellis1985,ellis1995}; see Fig.~\ref{fig2dising}(e). 

\item In the coexistence region, $P^u(M_n\in \D m)$ decays as $\E^{- a\sqrt{n}}$, where $a$ is some positive constant, instead of the anticipated (bulk) decay $\E^{-bn}$, $b>0$. This ``slower'' large deviation principle describes a surface effect related to a change of phase, and depends on boundary conditions imposed on the model (see, e.g., \cite{ioffe1994,pfister2002}).
\end{itemize}

The fact that $J^u(m)$ is zero over the whole coexistence region when $u<u_c$ means that the large deviations of the mean magnetization are undetermined in that region \cite{lanford1973,ellis1995}. In particular, we cannot conclude from the shape of $J^u(m)$ that there is a whole interval of equilibrium values for the mean magnetization. The actual equilibrium values are determined by the refined large deviation principle for $P^u(M_n\in \D m)$ mentioned in the last point above. The same remark applies when studying the 2D Ising model in the canonical ensemble as a function of the temperature; see Example~\ref{ex2disingc}.
\end{example}

\subsection{Canonical ensemble}

The canonical ensemble differs from the microcanonical ensemble in the way microstates are weighted. In the microcanonical ensemble, the control parameter is the energy $U$ or the mean energy $u=U/n$, and the microstates $\om$ are taken to be distributed according to the constrained prior distribution $P^u(\D\om)$, which assigns the same probabilistic weight to all the microstates with the same energy $U$ or mean energy $u$. In the canonical ensemble, the control parameter is the \emph{temperature} $T$ or, equivalently, the \emph{inverse temperature} $\beta=(k_B T)^{-1}$, and the relevant probability measure that one considers on $\Lambda_n$ is the so-called \emph{canonical} or \emph{Gibbs measure}
\be
P_\beta(\D\om)=\frac{\E^{-\beta H_n(\om)}}{Z_n(\beta)}P(\D\om).
\label{eqcano1}
\ee
In this expression, $Z_n(\beta)$ is the $n$-particle partition function defined earlier in Eq.~(\ref{eqpartf1}). We will not discuss the physical interpretation of $P_\beta(\D\om)$, apart from mentioning that it arises as the probability distribution of a sample system with Hamiltonian $H_n(\om)$ placed in thermal contact with a heat bath at inverse temperature $\beta$ (see, e.g., Sec.~28 of \cite{landau1991}). It should be mentioned also that, although the expression of $P_\beta(\D\om)$ involves the exponential function, that expression is not a large deviation principle---it is just the definition of a measure on $\Lambda_n$.

The derivation of a large deviation principle for a general macrostate $M_n(\om)$ in the canonical ensemble proceeds similarly as in the microcanonical ensemble. On the one hand, if an energy representation $\rh(m)$ and a macrostate entropy $\rs(m)$ exist for $M_n$, it is relatively easy to show that
\be
P_\beta(M_n\in \D m)=\int_{\{\om\in\Lambda_n:M_n(\om)\in \D m\}} P_\beta(\D\om)\asymp \E^{-nI_\beta(m)}\, \D m
\ee
where
\be
I_\beta(m)=\beta \rh(m)-\rs(m)-\varphi(\beta)
\label{eqcldp1}
\ee
and $\varphi(\beta)$ is the free energy defined in Eq.~(\ref{eqnewf1}) \cite{ellis2000}. On the other hand, if one knows that a joint large deviation holds for $P(h_n\in \D u,M_n\in \D m)$ with macrostate entropy $\rs(u,m)$, then 
\be
P_\beta(M_n\in \D m)\asymp \E^{-nJ_\beta(m)}\, \D m,
\ee
where
\be
J_\beta(m)=\inf_u \{\beta u-\rs(u,m)\}-\varphi(\beta).
\label{eqcldp2}
\ee
These two large deviation principles generalize Einstein's theory of fluctuations to the canonical ensemble. The rate functions $I_\beta(m)$ and $J_\beta(m)$ that we obtain in this ensemble are the macrostate free energies that form the basis of the Ginzburg-Landau theory of phase transitions \cite{landau1991}.

As in the microcanonical ensemble, the global minima of $I_\beta(m)$ or $J_\beta(m)$ define the most probable or \emph{equilibrium} values of the macrostate $M_n$ which now appear in the canonical ensemble with inverse temperature $\beta$. The set $\mE_\beta$ containing these canonical equilibrium values is thus defined as
\be
\mE_\beta=\{m:I_\beta(m)=0\}\quad\textrm{or}\quad\mE_\beta=\{m:J_\beta(m)=0\},
\ee
depending on the rate function ($I_\beta$ or $J_\beta$, respectively) used for analyzing the canonical large deviations of $M_n$. Equivalently, we have
\be
\mE_\beta=\{m: m\textrm{ is a global minimum of }I_\beta(m)\}
\ee
or
\be
\mE_\beta=\{m:m\textrm{ is a global minimum of }J_\beta(m)\},
\ee
The canonical analog of the maximum entropy principle that we obtain from these definitions of $\mE_\beta$ is called the \emph{minimum free energy principle}, and is expressed as
\be
\varphi(\beta)=\inf_m \{\beta\rh(m)-\rs(m)\}
\label{eqmfep1}
\ee
or
\be
\varphi(\beta)=\inf_m \inf_u\{\beta u-\rs(u,m)\},
\label{eqmfep2}
\ee
depending again on the rate function ($I_\beta$ or $J_\beta$, respectively) used. These formulae play the same role as the maximum entropy principle, in that they enable us to obtain the thermodynamic free energy $\varphi(\beta)$ as the solution of a variational problem involving a function that we call the macrostate free energy. The solutions of this variational principle are the canonical equilibrium values of $M_n$. 

There is connection with the maximum entropy principle of the microcanonical ensemble that can be made here. Since the two infima in Eq.~(\ref{eqmfep2}) can be interchanged, we can used the maximum entropy principle of (\ref{eqmer2}) to write
\be
\varphi(\beta)=\inf_u\inf_m\{\beta u-\rs(u,m)\}=\inf_u\left\{\beta u -\sup_m \rs(u,m)\right\}=\inf_u\{\beta u -s(u)\}.
\ee
We thus recover the basic Legendre-Fenchel transform found in Eq.~(\ref{eqfree1}). The formulae (\ref{eqmfep1}) and (\ref{eqmfep2}) can also be derived by recasting the integral defining $Z_n(\beta)$ as an integral over $M_n$, and by applying Laplace's Method to the latter integral \cite{ellis2004}. In the case where $\rh(m)$ and $\rs(m)$ exist, for example, we can write
\be
Z_n(\beta)\asymp \int \E^{-n\beta \rh(m)}\, P(M_n\in \D m)\asymp\int \E^{-n\{\beta\rh(m)-\rs(m)\}}\, \D m\asymp \E^{-n\inf\{\beta\rh(m)-\rs(m) \}}.
\ee

The class of macrostates for which large deviation principles can be derived in the canonical ensemble is exactly the same as in the microcanonical ensemble, since the rate functions of these two ensembles are built from the same energy representation function and macrostate entropies. Hence, if a large deviation principle holds for some macrostate $M_n$ in the microcanonical ensemble, then a large deviation principle also holds for $M_n$ in the canonical ensemble, and vice versa. This is not to say that the two ensembles yield the same sets of equilibrium states. There are models for which the microcanonical equilibrium set $\mE^u$ and the canonical equilibrium set $\mE_\beta$ are \emph{equivalent}, in the sense that they can be put in a one-to-one correspondence. But there are also models for which the two sets are not equivalent. This problem of ensemble equivalence is the subject of the next subsection.

To close our discussion of canonical large deviations, we discuss next two examples of canonical large deviations: one involving the mean energy $h_n$, the other the mean magnetization of the 2D Ising model. The first example is important for understanding the content of the next section. For a discussion of other large deviation results derived in the canonical ensemble, the reader is referred to \cite{ellis1985,ellis1995,ellis1999,barre2005}.

\begin{example}[Equilibrium mean energy]
\label{exmec1}
The probability distribution of the mean energy $h_n$ in the canonical ensemble is
\be
P_\beta(h_n\in \D u)=\int_{\{\om\in\Lambda_n:h_n(\om)\in \D u\}}P_\beta(\D\om)=\frac{\E^{-n\beta u}}{Z_n(\beta)} P(h_n\in \D u).
\ee
Assuming that the microcanonical entropy $s(u)$ exists, that is, assuming that $P(h_n\in \D u)\asymp \E^{ns(u)}\, \D u$, then
\be
P_\beta(h_n\in \D u)\asymp \E^{-nI_\beta(u)}\, \D u,\quad I_\beta(u)=\beta u-s(u)-\varphi(\beta).
\label{eqegme1}
\ee
The mean energy $u_\beta$ realized at equilibrium in the canonical ensemble at inverse temperature $\beta$ is determined from this large deviation principle by requiring that $I_\beta(u_\beta)=0$. Thus,
\be
\varphi(\beta)=\inf_u \{\beta u-s(u)\}=\beta u_\beta -s(u_\beta).
\ee
By Legendre duality, $u_\beta$ must be such that $\varphi'(\beta)=u_\beta$ if $\varphi(\beta)$ is differentiable. If $s(u)$ is differentiable, then we also have $s'(u_\beta)=\beta$, thereby recovering the standard thermodynamic definition of the inverse temperature. Observe, however, that $I_\beta(u)$ may have many critical points satisfying $s'(u)=\beta$; the global minimizer $u_\beta$ is only one of them.
\end{example}

\begin{figure*}[t]
\centering
\includegraphics[scale=1.0]{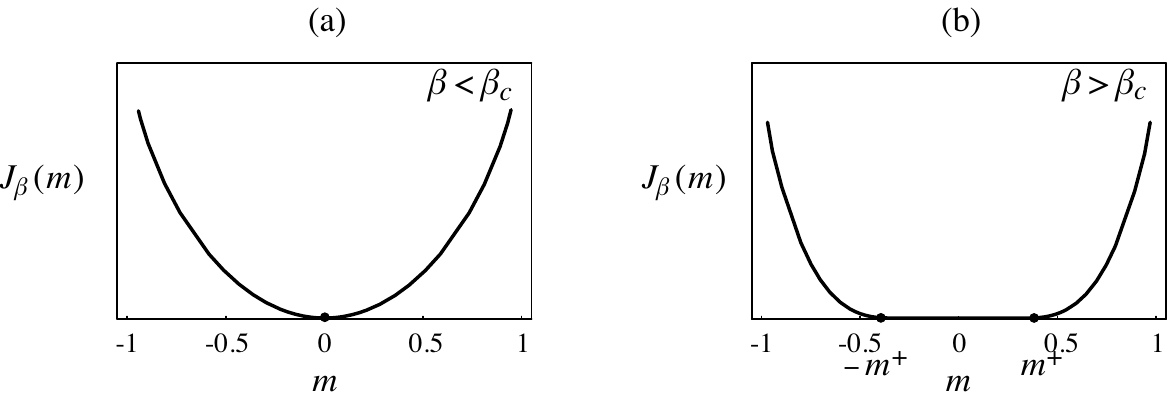}
\caption{Canonical rate function $J_\beta(m)$ for (a) $\beta< \beta_c$ and (b) $\beta>\beta_c$.}
\label{fig2disingc}
\end{figure*}

\begin{example}[2D Ising model]
\label{ex2disingc}
The rate function $J_\beta(m)$ associated with the mean magnetization of the 2D Ising model in the canonical ensemble is sketched in Fig.~\ref{fig2disingc}. This rate function is directly obtained from the macrostate entropy $\rs(u,m)$ discussed in Example~\ref{ex2disingm} via Eq.~(\ref{eqcldp2}). The properties of $J_\beta(m)$ are similar to its microcanonical counterpart $J^u(m)$. In particular, 
\begin{itemize}
\item $J_\beta(m)$ is a symmetric function of $m$ and is convex for all $\beta\in\reals$. One difference with $J^u(m)$ is that $J_\beta(m)$ is finite for all $m\in (-1,1)$.
\item For $\beta \leq \beta_c$, $J_\beta(m)$ is \emph{strictly} convex, which implies that it has a unique global minimum located at $m=0$; see Fig.~\ref{fig2disingc}(a). In other words, for $\beta\geq\beta_c$, $m_\beta=0$ is the unique equilibrium mean magnetization.
\item For $\beta>\beta_c$,  $J_\beta(m)$ is convex but achieves, as in the case of $J^u(m)$, its zero on a whole interval of mean magnetizations denoted by $[-m^+(\beta),m^+(\beta)]$; see Fig.~\ref{fig2disingc}(b). This canonical phase transition interval is such that $m^+(\beta)\ra 0$ when $\beta\ra\beta_c$ and $m^+(\beta)\ra 1$ when $\beta\ra\infty$.
\item As in the microcanonical case, $P(M_n\in \D m)\asymp \E^{-b\sqrt{n}}$ with $b>0$ inside the phase transition interval. This ``slower'' large deviation principle is also a surface effect.
\item The phase transition in the canonical ensemble is second-order, as in the microcanonical ensemble, with critical inverse temperature $\beta_c=s'(u_c)$. The critical exponent associated with the phase transition is non-trivial, as is well known, because $J_\beta(m)$ has no Taylor expansion around $m=0$ above the critical $\beta_c$.
\end{itemize}

The 2D Ising model is interesting from the point of view of large deviation theory because it shows that the interactions or correlations between the components of a system (here the spins) can change the scaling of a large deviation principle, and can lead to a breakdown of the Central Limit Theorem and the Law of Large Numbers. The breakdown of the Central Limit Theorem is related here to the fact that $J_\beta(m)$ is not locally quadratic around $m=0$ when $\beta>\beta_c$. The breakdown of the Law of Large Numbers, on the other hand, is related to the fact that $J_\beta(m)$ does not have a unique concentration point (i.e., global minimum or equilibrium state) for $\beta>\beta_c$. 
\end{example}

\subsection{Equivalence of ensembles}
\label{secequival}

\begin{figure*}[t]
\centering
\includegraphics[scale=1.0]{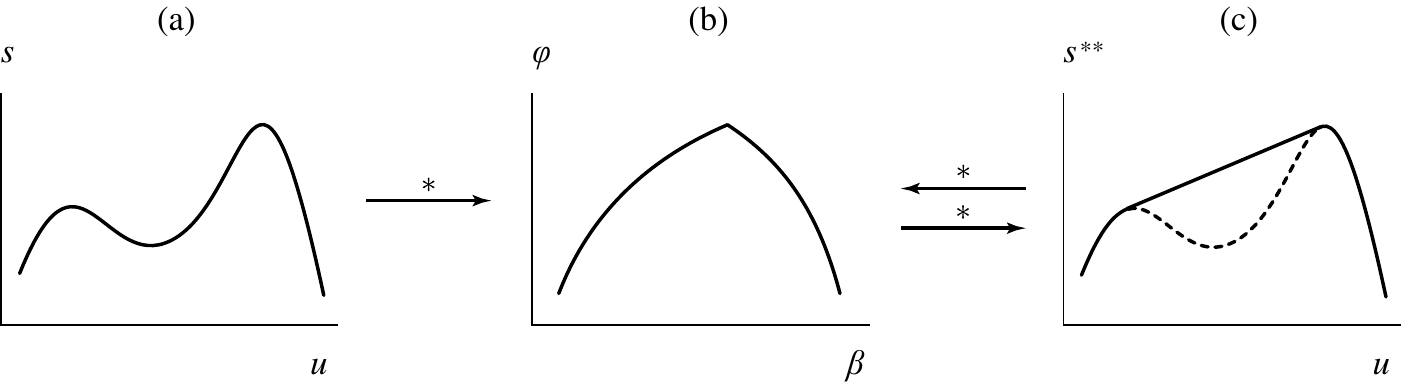}
\caption{Thermodynamic nonequivalence of the microcanonical and canonical ensembles as related to the nonconcavity of the entropy (see also Fig.~\ref{fignondual}). The Legendre-Fenchel transform of $\varphi(\beta)$ yields the concave envelope $s^{**}(u)$ of $s(u)$ rather than $s(u)$ itself. The Legendre-Fenchel transform of both $s(u)$ and $s^{**}(u)$ yield $\varphi(\beta)$.}
\label{fignoneq1}
\end{figure*}

We have seen before that not all rate functions $I$ can be obtained as the Legendre-Fenchel transform of a scaled generating cumulant function $\lambda$; only those that are convex are such that $I= \lambda^*$. When applied to entropy functions, this observation directly implies that \emph{entropy functions that are nonconcave cannot be calculated as the Legendre-Fenchel transform of free energies}. Consider, for instance, the entropy $s(u)$ as a function of the mean energy. Then we have $s=\varphi^*$ if $s$ is concave in $u$, but $s\neq\varphi^*$ if $s$ is nonconcave in $u$; see Fig.~\ref{fignoneq1}. In the first case, namely when $s(u)$ is concave, we say that the microcanonical and canonical ensembles are \emph{equivalent at the thermodynamic level} because $s(u)$ and $\varphi(\beta)$ are then one-to-one related by Legendre-Fenchel transform. In the second case, namely when $s(u)$ is nonconcave, we say that the two ensembles are \emph{nonequivalent at the thermodynamic level}, since part of $s(u)$ cannot be obtained from $\varphi(\beta)$ \cite{ellis2000}. What is obtained by taking the Legendre-Fenchel transform of $\varphi(\beta)$ is the \emph{concave envelope} $s^{**}(u)$ rather than $s(u)$ itself; see Fig.~\ref{fignoneq1}. Recall that $\varphi=s^*$ always holds, as noted after Eq.~(\ref{eqfree1}), so that the nonequivalence of the microcanonical and canonical ensembles only goes in one direction: the free energy can always be obtained as the Legendre-Fenchel transform of the entropy, but the entropy can be obtained as the Legendre-Fenchel transform of the free energy only when the entropy is concave.  

\begin{figure*}[t]
\centering
\includegraphics[scale=1.0]{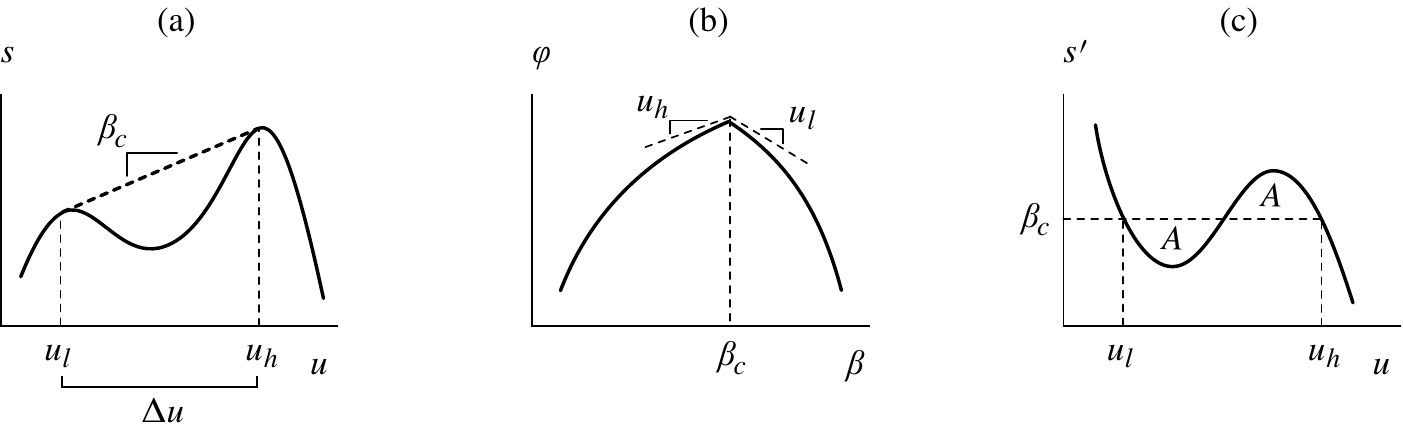}
\caption{(a) Nonconcave entropy $s(u)$ and its concave envelope $s^{**}(u)$. (b) Associated free energy $\varphi(\beta)$ characterized by a nondifferentiable point. (c) Maxwell's construction: The two areas $A$ defined by the intersection of $s'(u)$ and $s^{**}{'}(u)$ are equal.}
\label{fignoneq2}
\end{figure*}

These results and definitions are simple applications of the results that we have discussed before in the context of nonconvex rate functions. Further applications arise from what we know about Legendre-Fenchel transforms. In particular, the result relating the nonconvexity or affinity of rate functions with the nondifferentiability of the scaled cumulant generating function implies at the level of $s(u)$ and $\varphi(\beta)$ that, if $s(u)$ is nonconcave or is affine, then $\varphi(\beta)$ is nondifferentiable. Physically, this means that a first-order phase transition in the canonical ensemble can arise in two ways from the point of view of the microcanonical ensemble: either $s(u)$ is nonconcave or $s(u)$ is affine \cite{touchette2006}. The \emph{latent heat} $\Delta u$ of the phase transition corresponds in both cases to the length of the affine portion of the concave envelope $s^{**}(u)$ of $s(u)$. Indeed, if $s^{**}(u)$ if affine over some open interval $(u_l,u_h)$, then $\varphi(\beta)$ is nondifferentiable at a critical inverse temperature $\beta_c$ corresponding to the slope of $s^{**}(u)$ over $(u_l,u_h)$; see Figs.~\ref{fignoneq1} and \ref{fignoneq2}. Moreover, $\varphi'(\beta_c+0)=u_l$ and $\varphi'(\beta_c-0)=u_h$, so that 
\be
\Delta u=\varphi'(\beta_c-0)-\varphi'(\beta_c+0)=u_h-u_l.
\ee
The Maxwell or equal-area construction \cite{maxwell1875,huang1987} used in physics to calculate the latent heat is nothing but the construction of the concave envelope $s^{**}(u)$ \cite{ellis2004}; see Fig.~\ref{fignoneq2}(c).

These relationships between entropies, free energies, and phase transitions lead us to one last ``physical'' reformulation of a mathematical result that we have discussed before, namely, the G\"artner-Ellis Theorem. Put simply: \textit{If there is no first-order phase transition in the canonical ensemble, then the microcanonical entropy is the Legendre transform of the canonical free energy}. This follows by noting that if there is no phase transition at the level of the free energy or only a second-order phase transition, then the free energy is once-differentiable, which implies by the G\"artner-Ellis Theorem that the entropy can be calculated as the Legendre transform of the free energy. Of course, a concave yet affine entropy, such as the entropy $\tilde s(u,m)$ of the 2D Ising model, can also be calculated as the Legendre(-Fenchel) transform of the free energy, even though the latter has a nondifferentiable point. But, as in the case of nonconvex rate functions, it is impossible to distinguish from the sole knowledge of the free energy an affine entropy from a nonconcave entropy. 

Examples of models with nonconcave entropies include the mean-field Blume-Emery-Griffiths model \cite{barre2001,ellis2004,ellis2005}, the mean-field Potts model \cite{ispolatov2000,costeniuc2005a}, some models of plasmas \cite{kiessling2003} and 2D turbulence \cite{kiessling1997,ellis2002} mentioned before, as well as models of gravitational systems (see \cite{chavanis2006} for a recent review). The latter systems were historically the first to be discovered as having nonconcave entropies or, equivalently, as having negative heat capacities due to the long-range nature of the gravitational force (see \cite{lynden1999,draw2002}). In general, the long-range nature of the interaction in a many-particle system is a necessary, but not sufficient, condition for having nonconcave entropies. 

In some cases, the entropy may be concave as a function of $u$ alone but nonconcave as a function of some other macrostate. The mean-field $\phi^4$ model, for example, has a concave $s(u)$ but a nonconcave macrostate entropy $\rs(u,m)$ involving the mean energy $u$ and mean magnetization $m$ \cite{campa2007,hahn2005,hahn2006}. Thus, although $s(u)$ for this model can be calculated in the spirit of the G\"artner-Ellis Theorem as the Legendre-Fenchel transform of $\varphi(\beta)$, $\rs(u,m)$ cannot be obtained as the Legendre-Fenchel transform of a free energy function because that Legendre-Fenchel transform yields a concave function. The nonconcave $s(u,m)$ can be obtained, however, by other methods. One suggested by large deviation theory is to obtain $s(u,m)$ by contraction of another macrostate entropy that can hopefully be calculated using the G\"artner-Ellis Theorem; see Example~\ref{exmfp4}.

The next example discusses the calculation of $\rs(u,m)$ in the case where this function is concave.

\begin{example}[Concave entropy involving two macrostates]
If the macrostate entropy $\rs(u,m)$ is concave, as in the case of the 2D Ising model (see Example~\ref{ex2disingm}), then it can be expressed as the Legendre-Fenchel transform of a free energy function $\varphi(\beta,\eta)$ involving two variables or \emph{conjugated fields}: one for $h_n$, the other for $M_n$. By analogy with the definition $\varphi(\beta)$, $\varphi(\beta,\eta)$ is constructed as
\be
\varphi(\beta,\eta)=\lim_{n\ra\infty} -\frac{1}{n}\ln\int_{\Lambda_n} \E^{-n\beta h_n-n\eta M_n}\, P(\D\om).
\ee
The concavity of $\rs(u,m)$ then implies
\be
\rs(u,m)=\inf_{\beta,\eta} \{\beta u+\eta m-\varphi(\beta,\eta)\}.
\ee
The free energy $\varphi(\beta,\eta)$ can be put in the more familiar form
\be
\varphi(\beta,\eta')=\lim_{n\ra\infty} -\frac{1}{n}\ln\int_{\Lambda_n} \E^{-n\beta [h_n- \eta' M_n]}\, P(\D\om)
\ee
by defining $\eta'=-\eta/\beta$. In this form, $\varphi(\beta,\eta')$ is nothing but the standard canonical free energy $\varphi(\beta)$ of the modified mean Hamiltonian $h'_n=h_n-\eta' M_n$ involving the magnetic field $\eta'$. This field is the parameter of the canonical ensemble which is conjugated to the mean magnetization constraint $M_n=m$ of the microcanonical ensemble, whereas $\beta$ is the usual canonical parameter conjugated to the mean energy constraint $h_n=u$.
\end{example}

The thermodynamic equivalence of the microcanonical and canonical ensemble is also the basis for the equivalence of Gibbs's entropy and Boltzmann's entropy \cite{lebowitz1993}. This is the subject of the next example.  

\begin{example}[Boltzmann versus Gibbs entropy]
Gibbs's canonical entropy is defined for a discrete set of microstates $\om$ as
\be
S_{G}(\beta)=-\sum_{\om\in\Lambda_n} P_\beta(\om)\ln P_\beta(\om),
\ee
where $P_\beta(\om)$ is the canonical probability distribution. From the definition of this distribution, given in Eq.~(\ref{eqcano1}), we can write
\be
S_G(\beta)=n\beta\lex h_n\rex_\beta+\ln Z_n(\beta)+\ln|\Lambda_n|,
\ee
where $\lex \cdot\rex_\beta$ denotes the expectation with respect to $P_\beta$. The thermodynamic limit of this expression is
\be
s_G(\beta)=\lim_{n\ra\infty} \frac{S_G(\beta)}{n}=\beta u_\beta -\varphi(\beta)+\ln|\Lambda|,
\ee
assuming that $u_\beta$ is the unique concentration point of $h_n$ with respect to $P_\beta$, that is, the unique equilibrium mean energy at inverse temperature $\beta$. This assumption is justified rigorously if $s(u)$ is strictly concave; see Example~\ref{exmec2}. In this case, it is known that $\varphi(\beta)$ is differentiable for all $\beta\in\reals$ and $\varphi'(\beta)=u_\beta$ by Legendre duality, so that
\be
s_G(\beta)=s(u_\beta)+\ln|\Lambda|.
\ee
Thus, in the thermodynamic limit, the Gibbs entropy $s_G(\beta)$ is equal (up to a constant) to the Boltzmann entropy $s(u)$ evaluated at the equilibrium mean energy value $u_\beta$. This holds again if $s(u)$ is strictly concave. If $s(u)$ is nonconcave or is concave but has an affine part, then $u_\beta$ need not be unique for a given $\beta$; see Example~\ref{exmec2}.
\end{example}

There are many more issues about the equivalence of the microcanonical and canonical ensembles that could be discussed; see, e.g., \cite{campa2008,campa2009}. One that deserves mention, but would take us too far to explain completely is that the two ensembles can be conceived as being equivalent or nonequivalent by comparing the equilibrium sets, $\mE^u$ and $\mE_\beta$, of each ensemble. This macrostate approach to the problem of ensemble equivalence was studied by Ellis, Haven and Turkington \cite{ellis2000} (see also \cite{eyink1993}), and has been illustrated so far for a model of 2D turbulence \cite{ellis2002}, as well as some mean-field spin models \cite{ellis2004,costeniuc2005a}, including a toy spin model \cite{touchette2008} based on the nonconvex rate function studied in Example~\ref{exmixedsum}. The essential result illustrated by these models is, in a simplified form, that the microcanonical and canonical ensembles are equivalent at the macrostate level if and only if they are equivalent at the thermodynamic level, that is, if and only if $s(u)$ is concave. Thus all models with a concave entropy $s(u)$ have equivalent microcanonical and canonical ensembles at the macrostate level. For a simple introduction to these results, see \cite{touchette2004b}; for the treatment of ensembles other than microcanonical and canonical, see \cite{ellis2000}.

The last example of this section explains how the concavity of $s(u)$ determines the behavior of the canonical equilibrium mean energy $u_\beta$ as a function of $\beta$. This example illustrates in the simplest way possible the theory of nonequivalent ensembles developed by Ellis, Haven and Turkington \cite{ellis2000}, and completes Example~\ref{exmec1}.

\begin{figure}
\centering
\includegraphics[scale=1.0]{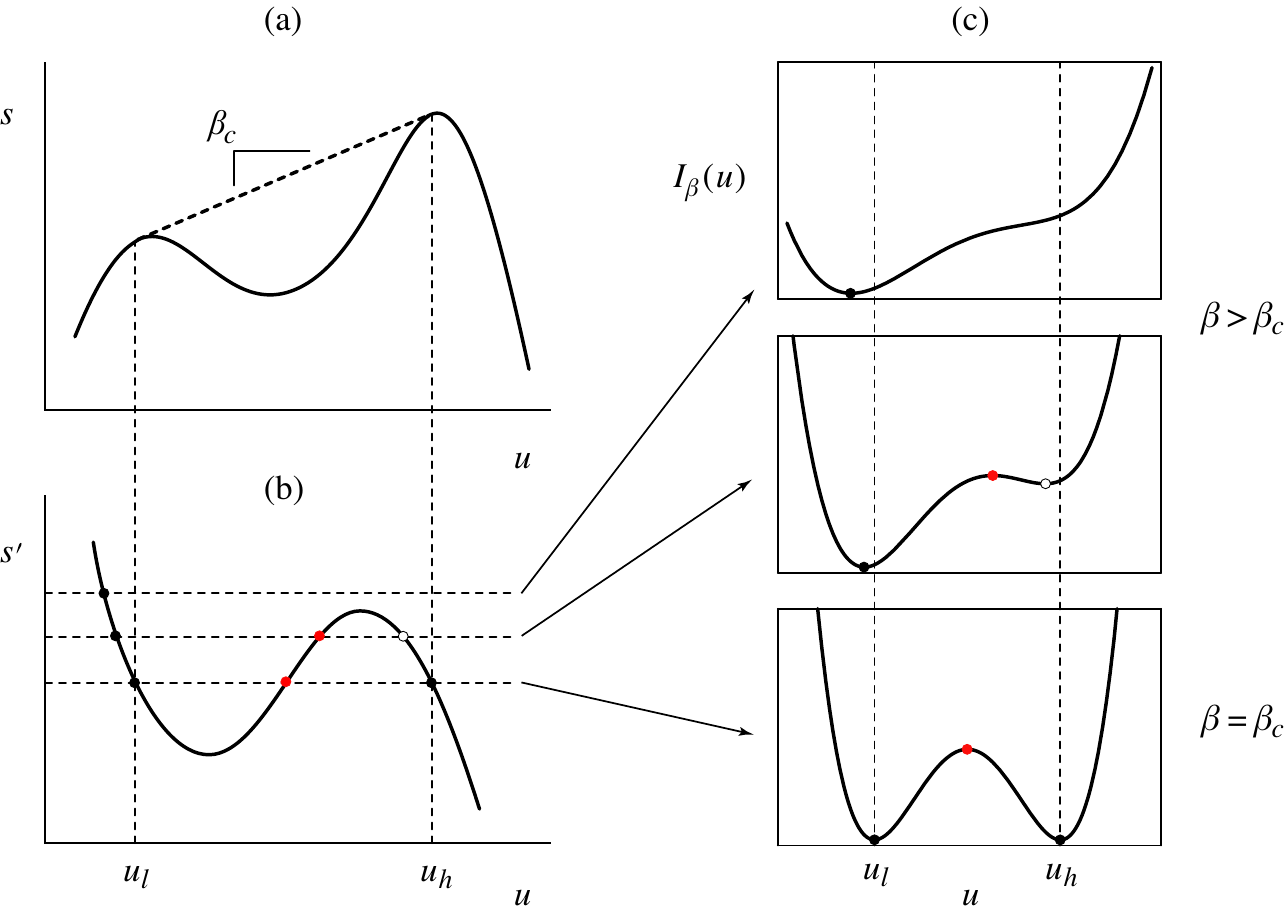}
\caption{(a) Generic nonconcave entropy function $s(u)$. (b) Derivative of $s(u)$. The nonconcavity of $s(u)$ translates into a non-monotonic derivative $s'(u)$. (c) Canonical rate function $I_\beta(u)$ for the mean energy, given in Eq.~(\ref{eqegme1}). The critical points of $I_\beta(u)$, for a given value of $\beta$, satisfy $s'(u)=\beta$. When $s(u)$ is nonconcave, $I_\beta(u)$ may have more than one critical point. The equilibrium mean energy $u_\beta$ in the canonical ensemble, which corresponds to the global minimum of $I_\beta(u)$ (black dot), is always located in the concave region of $s(u)$, i.e., in the region where $s(u)=s^{**}(u)$. The unstable (red dots) and metastable (open dot) critical points are located in the nonconcave region, i.e., in the region where $s(u)\neq s^{**}(u)$. At $\beta=\beta_c$, $I_\beta(u)$ has two global minima corresponding to $u_l$ and $u_h$ (phase coexistence).}
\label{figmens1}

\end{figure}

\begin{example}[Canonical equilibrium mean energy revisited]
\label{exmec2}
We have seen in Example~\ref{exmec1} that the equilibrium mean energy $u_\beta$ in the canonical ensemble is given by the zero(s) of the rate function $I_\beta(u)$ displayed in (\ref{eqegme1}). The behavior of $u_\beta$ as a function of $\beta$ is determined from this rate function by analyzing the concavity of $s(u)$. Three cases must be distinguished \cite{touchette2004b}. For simplicity, we assume that $s(u)$ is differentiable in all three cases.
\begin{itemize}
\item \emph{$s(u)$ is strictly concave}: In this case, $I_\beta(u)$ has a unique zero $u_\beta$ for all $\beta\in\reals$ such that $\varphi'(\beta)=u_\beta$ and $s'(u_\beta)=\beta$. Furthermore, the range of $u_\beta$ coincides with the domain of $s(u)$. These two results imply that there is a bijection between $u$ and $\beta$: to any $u$ in the domain of $s(u)$, there exists a unique $\beta$ such that $u=u_\beta$; to any $\beta\in\reals$, there exists a unique $u$ in the domain of $s(u)$ such that $u=u_\beta$. This bijection is an expression of the Legendre duality between $s(u)$ and $\varphi(\beta)$.

\item \emph{$s(u)$ is nonconcave}: In this case, $I_\beta(u)$ has more than one critical point satisfying $s'(u)=\beta$. The global minimum $u_\beta$ of $I_\beta(u)$ is one such critical point satisfying the additional condition $s(u_\beta)=s^{**}(u_\beta)$. If $s(u)\neq s^{**}(u)$, then $u\neq u_\beta$ for all $\beta\in\reals$. Therefore, all $u$ such that $s(u)\neq s^{**}(u)$ do not appear in the canonical ensemble as equilibrium values of $h_n$. This explains at the level of the mean energy why there is a first-order phase transition in the canonical ensemble when $s(u)$ is nonconcave. 
 
\item \emph{$s(u)$ is concave with an affine part of slope $\beta_c$}: In this case, $I_\beta(u)$ has a unique zero $u_\beta$ for all $\beta\neq\beta_c$. For $\beta=\beta_c$, $I_\beta(u)$ has a whole range of zeros corresponding to the range of affinity of $s(u)$. This range is a coexistence region for the mean energy $h_n$.
\end{itemize}
The case of a nonconcave $s(u)$ is illustrated in Fig.~\ref{figmens1}.
\end{example}

\subsection{Existence of the thermodynamic limit}

Proving that a large deviation principle holds for a macrostate is equivalent to proving the existence of a thermodynamic limit for an entropy function in the microcanonical ensemble, or a free energy function in the canonical ensemble. This equivalence is obvious in the microcanonical ensemble, for the statement $P(M_n\in \D m)\asymp \E^{n\rs(m)}\, \D m$ is equivalent to the existence of the limit
\be
\lim_{n\ra\infty} \frac{1}{n}\ln P(M_n\in \D m)=\rs(m).
\ee 
In particular, $P(h_n\in \D u)\asymp \E^{ns(u)}\, \D u$ if and only if the limit (\ref{eqnews1}) defining $s(u)$ exists. To establish the same correspondence in the canonical ensemble, note that the existence of the following macrostate free energy for $M_n$:
\be
\tilde{\varphi}(\eta)=\lim_{n\ra\infty} -\frac{1}{n}\ln \int_{\Lambda_n} \E^{-n\eta M_n(\om)}\, P(\D\om)
\ee
implies the existence of $\rs(m)$, since
\be
\rs(m)\leq \rs^{**}(m)=\inf_{\eta}\{\eta m -\tilde{\varphi}(\eta)\}.
\ee
The converse of this result also holds by Varadhan's Theorem: namely, $\tilde{\varphi}(\eta)$ exists if $\rs(m)$ exists.\footnote{Except possibly at boundary points; see Example~\ref{exnonsteep}.} As a particular case, we thus have that $s(u)$ exists if and only if $\varphi(\beta)$ exists; that is, the thermodynamic entropy exists if and only if the thermodynamic free energy exists \cite{lewis1988,lewis1995a}.

Our treatment of large deviation principles avoided, for the most part, any proofs of the thermodynamic limit. We either assumed the existence of large deviation principles in order to work out their consequences, or we established directly those large deviation principles by deriving their rate functions by contraction of other rate functions that are known to exist (e.g., the relative entropy). In the case of mean-field and long-range systems, for example, we are often led to prove that $s(u)$ exists simply by calculating this rate function with the general maximum entropy principle involving the macrostate entropy $\rs(m)$ and the energy representation function $\rh(m)$. Short-range models are more difficult to treat because, as mentioned, they necessitate the use of the empirical process, the Level-3 macrostate. The calculation of $s(u)$ for these models thus involves the contraction of an infinite-dimensional rate function---the relative entropy of the empirical process---down to the mean energy $h_n$. 

The existence of $s(u)$ may be established in a more general way by proving the existence of the limit defining this quantity for certain classes of Hamiltonians. This method was initiated by Ruelle \cite{ruelle1969,ruelle2004} and Lanford \cite{lanford1973} (see also Griffiths \cite{griffiths1965}), who proved that the limit defining $s(u)$ does exist for interactions that are \emph{stable}, in the sense that they do not lead to a collapse of the particles into a low energy state, and are \emph{tempered}, in the sense that they decay sufficiently quickly at large distances so as to limit surface or boundary effects in the thermodynamic limit. The method of proof relies on the notion of super-additivity, and establishes as an added result that $s(u)$ is concave, which implies, in turn, that the microcanonical and canonical ensembles are equivalent \cite{ruelle1969,galgani1971,gallavotti1999}. This result can be extended to macrostates other than $h_n$ to conclude that macrostates satisfying similar conditions of stability and temperedness have concave macrostate entropies. For an introduction to these results, the reader is referred to the work of Lanford \cite{lanford1973}, who treats a simpler class of short-range interactions and macrostates known as \emph{finite-range}.

Proofs of the existence of the canonical free energy $\varphi(\beta)$ have also been given, notably by  van Hove \cite{hove1949,hove1950}, Ruelle \cite{ruelle1963, ruelle1969}, Fisher \cite{fisher1964}, and Griffiths \cite{griffiths1964,griffiths1972}. The class of short-range interactions considered in this case is essentially the same as the one mentioned before, namely, stable are tempered. One interesting aspect of these proofs is that they also establish the equivalence of the microcanonical and canonical ensembles, among other ensembles. Therefore, one condition that appears to be necessary (although not sufficient) for having nonconcave entropies and nonequivalent ensembles in the thermodynamic limit is for the interaction in a system to be long-range. Gravitating particles, unscreened plasmas, and vortex models of 2D turbulence are examples of such long-range interaction systems; see \cite{draw2002} for others.

%%%%%%%%%%%%%%%%%%%%%%%%%%%%%%%%%%%%%%%%%%%%%%%
\section{Large deviations in nonequilibrium statistical mechanics}
\label{secnonequi}

We turn in this section to the study of large deviations arising in physical systems that dynamically evolve in time or that are maintained in out-of-equilibrium steady states by an external forcing. The methodology that will be followed for studying these nonequilibrium systems is more or less the one that we followed in the previous section. All that changes is the type of systems studied, and the fact that in nonequilibrium statistical mechanics the object of focus is most often not the Hamiltonian or the constraints imposed on a system, but the stochastic process (Markov chain, Langevin equation, master equation, etc.) used to model that system. 

The dynamical nature of nonequilibrium systems requires, of course, that we include time in the large deviation analysis, possibly as the extensive parameter controlling a large deviation principle (as is the case for the number of particles). Conceptually, this is a minor adjustment to take into account. A more fundamental difference between equilibrium and nonequilibrium is that there is no concept of statistical-mechanical ensemble for nonequilibrium systems, even those driven in out-of-equilibrium steady states \cite{derrida2007}. That is to say, when a system is out of equilibrium, we do not know in general what the underlying probability distribution of its states is (if such a distribution indeed exists). To find it, we must define the system precisely, calculate the probability distribution of its states from first principles, and proceed from there to derive large deviation principles for observables that are functions of the system's state. There is no general principle whereby one can calculate the distribution of the system's states from the sole knowledge of the system's invariants or external constraints imposed on the system. Such a general principle is precisely what a statistical-mechanical ensemble is, and what is missing from the theory of nonequilibrium systems. 

In spite of this, it is possible to formulate a number of general and interesting results for nonequilibrium systems, especially when these are modeled as Markov processes. The aim of this section is to give an overview of these results in the style of the previous section, with an emphasis on large deviations. We will see with these results that it is often possible to characterize the most probable states (trajectories or paths) of a nonequilibrium system as the minima of a rate function, and that these minima give rise to a variational principle that generalizes the maximum entropy or minimum free energy principles. The knowledge of this rate function also provides, as in the case of equilibrium systems, a complete description of the fluctuations of the system considered.

\subsection{Noise-perturbed dynamical systems}
\label{subsecfw}

The first class of large deviation results that we study concerns the fluctuations of deterministic dynamical systems perturbed by noise. The idea here is to consider a differential equation, which determines the motion of a dynamical system in time, and to perturb it with a Gaussian white noise of zero mean and small intensity (variance or power) $\vep\geq 0$. In the presence of noise ($\vep\neq 0$), the system's motion is random, but for a small noise, that random motion is expected to stay close to the unperturbed dynamics, and should converge, in the zero-noise limit $\vep\ra 0$, to the deterministic motion determined by the unperturbed differential equation. In terms of probabilities, this means that the probability distribution of the system's trajectories or \emph{paths} should concentrate, as $\vep\ra 0$, around the deterministic path of the unperturbed system. This concentration effect is akin to a Law of Large Numbers, so in the spirit of large deviation theory it is natural to inquire about the scaling of that concentration with $\vep$; that is, how is the probability decaying around its maximum as $\vep\ra 0$? The answer, as might be expected, is that the decay has the form of a large deviation principle. 

\subsubsection{Formulation of the large deviation principle}

The study of large deviations of random paths gives rise to a mathematical difficulty that we encountered before when we treated the continuous version of Sanov's Theorem: a trajectory is a function, which means that the probabilities that we must handle are probabilities over a function space. A rigorous mathematical treatment of large deviations exists in this setting (see, e.g.,~\cite{freidlin1984} or Chap.~5 of~\cite{dembo1998}), but for simplicity, and to give a clearer presentation of the ideas involved, we will follow the previous sections and deal with probabilities of random functions at a heuristic level. To simplify the presentation, we will also start our study with a simple, one-dimensional noise-perturbed system described by the following stochastic differential equation:
\be
\dot{X}_\vep(t)=b(X_\vep)+\sqrt{\vep}\eta(t),\quad X_\vep(0)=x_0.
\label{eqdyn1}
\ee
In this expression, $X_\vep\in\reals$, $b$ is a real function (sufficiently well-behaved), and $\eta(t)$ is a Gaussian white noise process characterized by its mean $\lex\eta(t)\rex=0$ and correlation function $\lex\eta(t)\eta(t')\rex=\delta(t-t')$. The real constant $\vep\geq 0$ is the large deviation parameter controlling the intensity of the noise. For the mathematically minded, Eq.~(\ref{eqdyn1}) should properly be interpreted as the It\^o form
\be
\D X_\vep(t)=b(X_\vep)\, \D t+\sqrt{\vep}\, \D W(t)
\label{eqdyn2}
\ee
involving the Brownian or Wiener motion $W(t)$ \cite{gardiner1985}. What we consider in (\ref{eqdyn1}) is the naive yet standard version of Eq.~(\ref{eqdyn2}), obtained by heuristically viewing Gaussian white noise as the time-derivative of Brownian motion.

The probability that we wish to investigate is the probability of a trajectory or path of the random dynamical system described by Eq.~(\ref{eqdyn1}), that is, the probability of a given realization $\{x(t)\}_{t=0}^\tau$ of that equation, extending from an initial time $t=0$ to some time $\tau>0$. Of course, one cannot speak of the probability of a \emph{single} trajectory, but only of a \emph{set} of trajectories, and this is where the difficulty mentioned above comes in. To avoid it, one may consider the probability that the system's trajectory lies in some cylinder or ``tube'' enclosing a given trajectory $\{x(t)\}_{t=0}^\tau$, or any other finite set of trajectories. 

This way of making sense of probabilities in trajectory space will not be followed here; instead, we will assume at a heuristic level that there is a probability density $P[x]$ over the different paths $\{x(t)\}_{t=0}^\tau$ of the system. Following the physics literature, we denote this density with square brackets to emphasize that it is a functional of the whole function $x(t)$. With this notation, we then write a large deviation principle for the random paths as $P_\vep[x]\asymp \E^{-a_\vep J[x]}$ to mean that $P[x]$ decays exponentially with \emph{speed} $a_\vep$ in such a way that $a_\vep\ra\infty$ as $\vep\ra 0$.\footnote{\label{notespeed}The term ``speed'' in this context has, of course, nothing to do with the time derivative of the position. The term is used in large deviation theory because $a_\vep$ determines how quickly $P_\vep$ decays to zero with $\vep$; see Appendices \ref{appldt} and \ref{appspeed}.} The rate function $J[x]$ is a functional of the paths. To be rigorous, we should write this large deviation principle as
\be
P\biggl(\sup_{0\leq t\leq\tau}| X_\vep(t)-x(t)|<\delta \biggr)\asymp \E^{-a_\vep J[x]},\quad \vep\ra 0
\ee
where $\delta$ is any small, positive constant. In this form, the rate function is then obtained by the limit
\be
\lim_{\vep\ra 0}\ \frac{1}{a_\vep}\ln P\biggl(\sup_{0\leq t\leq\tau}| X_\vep(t)-x(t)|<\delta \biggr)= -J[x].
\ee
The notation $P_\vep[x]\asymp \E^{-a_\vep J[x]}$ is obviously more economical.

The large deviation principle associated with the specific system described by  Eq.~(\ref{eqdyn1}) was derived rigorously by Freidlin and Wentzell \cite{wentzell1970,freidlin1984}, and through formal path integral methods by Graham and T\'el \cite{graham1973,graham1985}, as well as by Dykman and Krivoglaz \cite{dykman1979} among others.\footnote{Onsager and Machlup \cite{onsager1953} derived this result for linear equations as far back as 1953 (see also \cite{falkoff1956,falkoff1958,stratonovich1989}).} The result, in the path density notation, is
\be
P_\vep[x]\asymp \E^{-J[x]/\vep},\quad J[x]=\frac{1}{2}\int_0^\tau \left[\dot{x}-b(x)\right]^2\, \D t,
\label{eqldpfw1}
\ee
where $x(t)$ is any (absolutely) continuous path\footnote{\label{fnsmooth}That $x(t)$ should be a continuous path does not contradict the fact that the random paths of stochastic equations driven by Gaussian white noise are nondifferentiable with probability 1. Remember that $P_\epsilon[x]$ is a formal notation for the probability that a given random path lies inside an infinitesimal tube whose center follows a given smooth path $x(t)$. Thus, what we are interested in is the probability that a random path \emph{is close} to some smooth path $x(t)$, not the probability that a random path \emph{follows exactly} some smooth path $x(t)$ \cite{durr1978}.} satisfying the initial condition $x(0)=x_0$. The rate function $J[x]$ is sometimes called the \emph{action functional} \cite{freidlin1984} or \emph{entropy} of the path \cite{donsker1983}.\footnote{The form of $J[x]$ presented here is the form obtained in the It\^o interpretation of the stochastic equation. In the Stratonovich interpretation, there is an additional term involving the components of the vector $b(x)$, which vanishes in the zero-noise limit. The difference amounts to a Jacobian term in path integrals involving $P_\epsilon[x]$ \cite{horsthemke1975,hunt1981}.} Notice that $J[x]$ is positive and has a unique zero corresponding to the deterministic path $x^*(t)$ satisfying the unperturbed equation $\dot{x}^*=b(x^*)$. Therefore, $\|X_\vep(t)-x^*(t)\|\ra 0$ in probability as $\vep\ra 0$. The quadratic form of $J[x]$ stems from the Gaussian nature of the noise $\eta(t)$. For other types of noise, in particular correlated (colored) noise, $P_\vep[x]$ may still have a large deviation form, but with a rate function which is not quadratic or local in time. More details on these correlated large deviation principles can be found in \cite{bray1990,mckane1990,wio1989,dykman1994b,einchcomb1995} (see also \cite{moss19891,moss19892}).

\begin{figure*}[t]
\centering
\includegraphics[scale=1.0]{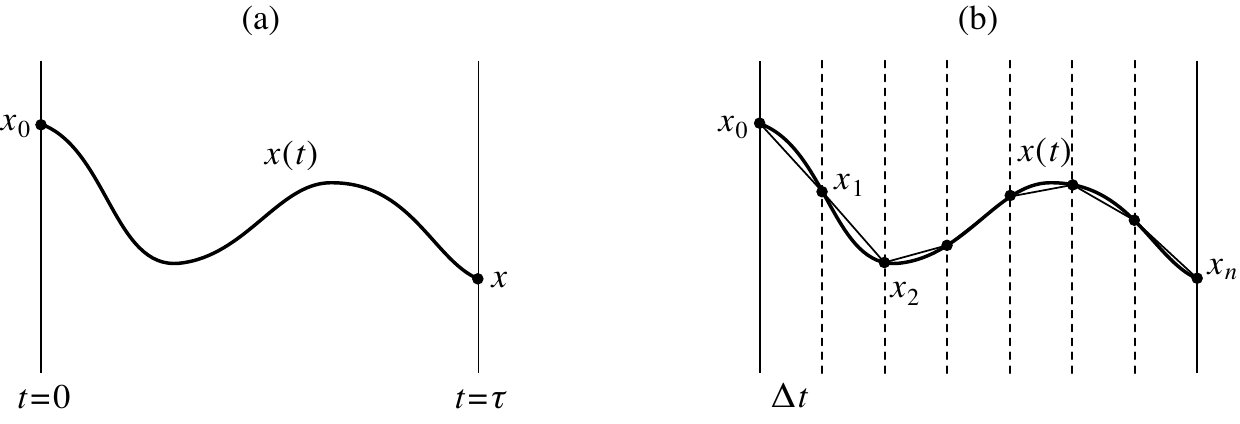}
\caption{(a) Trajectory starting at $x_0$ at time $t=0$ and ending at $x$ at time $t=\tau$. (b) Time-discretization of the continuous-time trajectory.}
\label{figpaths1}
\end{figure*}

\subsubsection{Proofs of the large deviation principle}

The large deviation result displayed in (\ref{eqldpfw1}) can be proved in many different ways. The simplest, perhaps, is to approximate the trajectories $\{x(t)\}_{t=0}^\tau$ in the spirit of path integral techniques by discrete-time trajectories $\{x_i\}_{i=1}^n$ involving $n$ points equally spaced between $t=0$ and $t=\tau$ at interval $\Delta t$; see Fig.~\ref{figpaths1}. This discretization or ``time-slicing'' procedure has the effect of transforming the Markov stochastic process of Eq.~(\ref{eqdyn1}) into a Markov chain, for which it is relatively easy to compute the probability density $p(x_1,x_2,\ldots,x_n)$ given the properties of the noise $\eta(t)$. The probability density $P_\vep[x]$ is then obtained from $p(x_1,x_2,\ldots,x_n)$ by taking the double limit $n\ra\infty$, $\Delta t\ra 0$. For more details, see Chap.~2 of \cite{kleinert2004}. 

More interesting from the point of view of large deviation theory is the fact that $P_\vep[x]$ can be obtained from the G\"artner-Ellis Theorem by calculating the functional Legendre-Fenchel transform of the scaled cumulant generating functional of $X_\vep(t)$, defined as
\be
\lambda[k]=\lim_{\vep\ra 0}\ \vep\ln \lex \E^{k\cdot X_\vep/\vep}\rex,
\ee
where
\be
\lex \E^{k\cdot X_\vep/\vep}\rex=\int\mathcal{D}[x]\, P_\vep[x]\, \exp\left(\frac{1}{\vep}{\int_0^\tau k(t)x(t)\, \D t}\right).
\ee
The path integral defining the expectation value above can be solved exactly for Gaussian white noise to obtain an explicit expression for $\lambda[k]$ \cite{wiegel1986,kleinert2004}. In this case, the measure $\mathcal{D}[x]$ on the space of trajectories is the Wiener measure. Alternatively, $\lambda[k]$ can be obtained in the context of the Donsker-Varadhan theory from the generator of the Markov process defined by Eq.~(\ref{eqdyn1}) (see, e.g., \cite{gartner1977,donsker1983,bucklew1990}). In both cases, $J[x]$ is then expressed as
\be
J[x]=\inf_{k}\{k\cdot x -\lambda[k]\}.
\ee
The same rate function can also be derived from the large deviation point of view by applying the contraction principle to the rate function governing the fluctuations of the scaled noise $\eta_\vep(t)=\sqrt{\vep} \eta(t)$ \cite{dembo1998}. We show in the next example how to obtain the latter rate function from the G\"artner-Ellis Theorem. The large deviation result obtained for $\eta_\vep(t)$ is known in the mathematical literature as \emph{Schilder's Theorem} \cite{schilder1966,dembo1998}. The calculation of $J[x]$ based on the contraction principle follows the presentation of that theorem.

\begin{example}[Schilder's Theorem]
The properties of a Gaussian white noise $\eta(t)$ with $\lex\eta(t)\rex=0$ for all $t$ and $\lex\eta(t)\eta(t')\rex=\delta(t-t')$ are completely determined by its characteristic function:
\be
G_\eta[k]=\lex \E^{ik\cdot\eta}\rex=\int \mathcal{D}[\eta]\, P[\eta]\, \E^{\I\int k(t)\eta(t)\, \D t}= \exp\left(-\frac{1}{2}\int k(t)^2\, \D t\right).
\ee
From this form of $G_\eta[k]$, the scaled cumulant generating functional of the scaled noise $\eta_\vep(t)=\sqrt{\vep}\eta(t)$ is easily found to be
\be
\lambda[k]=\lim_{\vep\ra 0}\ \vep \ln\lex \E^{k\cdot\eta_\vep/\vep}\rex=\frac{1}{2}\int k(t)^2\, \D t.
\ee
This result is the functional analog of the log-generating function of a Gaussian random variable with zero mean and unit variance; see Example~\ref{exgaussrev}. As in that example, $\lambda[k]$ is differentiable, but now in the functional sense. By applying the G\"artner-Ellis Theorem, we therefore conclude that $\eta_\vep(t)$ satisfies a large deviation principle in the limit $\vep\ra 0$ with a rate function given by the Legendre-Fenchel transform of $\lambda[k]$:
\be
P_\vep[\phi]\asymp \E^{-I[\phi]/\vep},\quad I[\phi]=\sup_{k}\{k\cdot\phi-\lambda[k]\}.
\ee
In this expression, $\phi(t)$ is a given trajectory or realization of the noise $\eta_\vep(t)$ starting at $\phi(0)=0$. As in the non-functional case, we can use the differentiability of $\lambda[k]$ to reduce the Legendre-Fenchel transform above to a Legendre transform, given by
\be
I[\phi]=k_\phi\cdot\phi-\lambda[k_\phi]=\int k_\phi(t)\phi(t)\, \D t-\lambda[k_\phi],
\ee
where $k_\phi$ is the functional root of 
\be
\frac{\delta\lambda[k]}{\delta k(t)}=\phi(t).
\ee
In the present case, $k_\phi(t)=\phi(t)$, so that
\be
I[\phi]=\phi\cdot\phi-\lambda[\phi]=\frac{1}{2}\int \phi(t)^2\, \D t.
\label{eqschilder1}
\ee
The expression of this rate function has an obvious similarity with the rate function of the sample mean of IID Gaussian random variables discussed in Example~\ref{exgaussrev}. 
\end{example}

We are now in a position to derive the rate function $J[x]$ of $X_\vep(t)$, shown in Eq.~(\ref{eqldpfw1}), from the rate function $I[\phi]$ of the scaled Gaussian noise $\eta_\vep(t)$. The main point to observe is that $X_\vep(t)$ is a contraction of $\eta_\vep(t)$, in the sense of the contraction principle (Sec.~5.6 of \cite{dembo1998}). This is obvious if we note that the stochastic differential equation (\ref{eqdyn1}) has for solution 
\be
x(t)=x_0+\int_0^t b(x(s))\, \D s+\int_0^t \phi(s)\, \D s,
\ee 
where $\phi(t)$ is, as before, a realization of $\eta_\vep(t)$. Let us denote this solution by the functional $f[\phi]=x$. Using the contraction principle, we then write
\be
J[x]=\inf_{\phi:f[\phi]=x} I[\phi]=\inf_{\phi:\phi=\dot{x}-b(x)}I[\phi]=I[\dot{x}-b(x)],
\ee
which is exactly the result of Eq.~(\ref{eqldpfw1}) given the expression of $I[\phi]$ found in Eq.~(\ref{eqschilder1}). 

For future use, we re-write $J[x]$ as
\be
J[x]=\int L(\dot{x},x)\, \D t,\quad L(\dot{x},x)=\frac{1}{2}\left[\dot{x}(t)-b(x(t))\right]^2.
\ee
The function $L(\dot{x},x)$ is called the \emph{Lagrangian} of the stochastic process $X_\vep(t)$. The next example gives the expression of $L(\dot x, x)$ and $J[x]$ for a more general class of stochastic differential equations involving a state-dependent diffusion term. This class is the one considered by Freidlin and Wentzell \cite{freidlin1984} (see also Sec.~5.6 of \cite{dembo1998} and \cite{roy1993,roy1995}).

\begin{example}[General stochastic differential equation]
Let $X_\vep(t)$ be a flow in $\reals^d$, $d\geq 1$, governed by the following (It\^o) stochastic differential equation:
\be
\D X_\vep(t)=b(X_\vep)\, \D t+\sqrt{\vep}\sigma(X_\vep)\, \D W(t),\quad X_\vep(0)=x_0,
\label{eqdyn3}
\ee
where $b$ is some function mapping $\reals^d$ to itself, $\sigma$ is a square, positive-definite matrix assumed to be nonsingular, and $W(t)$ is the usual Brownian motion. For this system, Freidlin and Wentzell \cite{freidlin1984} proved that $P_\vep[x]\asymp \E^{-J[x]/\vep}$ as $\vep\ra 0$ with rate function
\be
J[x]=\frac{1}{2}\int_0^\tau \left[\dot{x}(t)-b(x(t))\right]^TA^{-1}\left[\dot{x}(t)-b\left(x(t)\right)\right]\, \D t,
\ee
where $A=\sigma\sigma^T$ is the so-called \emph{diffusion matrix}. Two technical conditions complete this large deviation result. First, to ensure the existence and uniqueness of a solution for Eq.~(\ref{eqdyn3}), the \emph{drift} vector $b(x)$ must be Lipschitz continuous. Second, the realizations $x(t)$ of $X_\vep(t)$ for which the rate function $J[x]$ exists must verify the initial condition $x(0)=x_0$, in addition to having square integrable time-derivatives.
\end{example}
 
\subsubsection{Large deviations for derived quantities}

Once a large deviation principle has been proved for the path density $P_\vep[x]$, the way becomes wide open for deriving large deviation principles for all sorts of probabilities using the contraction principle. Of particular interest is the probability density 
\be
P_\vep(x,\tau|x_0)=\int_{x(0)=x_0}^{x(\tau)=x} \mathcal{D}[x]\, P_\vep[x].
\label{eqp1}
\ee
that the process $X_\vep(t)$ reaches a point $x$ at time $t=\tau$ given that it started at a point $x_0$ at time $t=0$; see Fig.~\ref{figpaths1}(a). Assuming that $P_\vep[x]$ satisfies a large deviation principle with rate function $J[x]$, it directly follows from the contraction principle that $P(x,\tau|x_0)$ also satisfies a large deviation principle of the form
\be
P_\vep(x,\tau|x_0)\asymp \E^{-V(x,\tau|x_0)/\vep}
\ee 
with rate function given by
\be
V(x,\tau|x_0)=\inf_{x(t):x(0)=x_0,x(\tau)=x} J[x].
\label{eqcp2}
\ee
This rate function is also called the \emph{quasi-potential}.

The large deviation approximation of $P(x,\tau|x_0)$ is often referred to as a \emph{WKB approximation}, following Wentzel,\footnote{Wentzel the physicist, not to be confused with Wentzell, the mathematician mentioned earlier.} Kramers and Brillouin, who developed a similar approximation in the context of quantum mechanics and differential equations \cite{kleinert2004}.\footnote{The WKB approximation is also referred to as the eikonal approximation.} The meaning of this approximation follows exactly the interpretation of the contraction principle, in that the dominant contribution to the probability of a fluctuation---here the observation of $x(\tau)=x$ starting from $x(0)=x_0$---is the probability of the most probable path leading to that fluctuation. This most probable or \emph{optimal} path $x^*(t)$, which is the path solving the variational problem (\ref{eqcp2}), can be determined by solving the \emph{Euler-Lagrange equation}
\be
\left.\frac{\delta J[x]}{\delta x(t)}\right|_{x^*(t)}=0,\quad x(0)=x_0,x(\tau)=x,
\ee
which has the well-known form
\be
\frac{\D}{\D t}\frac{\partial L}{\partial\dot{x}}-\frac{\partial L}{\partial x}=0,\quad x(0)=x_0,x(\tau)=x
\label{eqlag1}
\ee
in terms of the Lagrangian $L(\dot x,x)$. The optimal path\footnote{The optimal path is also called the \emph{maximum likelihood path} between two fixed endpoints, the \emph{phenomenological path} \cite{wiegel1986} or the \emph{instanton} \cite{dykman1994a,bray1989}.} can also be interpreted, by analogy with classical mechanics, as the solution of the \emph{Hamilton-Jacobi equations},
\be
\dot{x}=\frac{\partial H}{\partial p},\quad \dot{p}=-\frac{\partial H}{\partial x},\quad H(p,x)=\dot{x}p-L(\dot{x},x),
\label{eqham1}
\ee
which involve the \emph{Hamiltonian} $H(p,x)$ conjugated, in the Legendre-Fenchel sense, to the Lagrangian $L(\dot x,x)$. These observations are put to use in the next example to determine the WKB approximation of the stationary distribution of a simple but important Markov process.  

\begin{example}[Stationary distribution of the Ornstein-Uhlenbeck process] 
\label{exou1}
Consider the linear equation
\be
\dot{X}_\vep(t)=-\gamma X_\vep(t)+\sqrt{\vep}\eta(t),
\label{eqlang1}
\ee
and let $P_\vep(x)$ denote the stationary probability density of this process which solves the time-independent Fokker-Planck equation
\be
\gamma\frac{\partial }{\partial x}[xP_\vep(x)]+\frac{\vep}{2}\frac{\partial^2 P_\vep(x)}{\partial x^2}=0,\quad \vep>0.
\label{eqfp1}
\ee
We know that in the weak-noise limit $\vep\ra 0$, $P_\vep(x)$ obeys the WKB form $P_\vep(x)\asymp \E^{-V(x)/\vep}$. To find the expression of the quasi-potential $V(x)$, we follow Onsager and Machlup \cite{onsager1953} and solve the Euler-Lagrange equation (\ref{eqlag1}) with the terminal conditions $x(-\infty)=0$ and $x(\tau)=x$. The solution is $x^*(t)=x \E^{\gamma(t-\tau)}$. Inserting this back into $J[x]$ yields $V(x)=J[x^*]=\gamma x^2$. This result can also be obtained in a more direct way by expressing the force $b(x)$ as the derivative of a potential, i.e., $b(x)=-U'(x)$ with $U(x)=\gamma x^2/2$. In this case, it is known that $V(x)=2U(x)$ (see Sec.~4.3 of \cite{freidlin1984}).
\end{example}

The previous example can be generalized as follows. If the zero-noise limit of the stochastic differential equation given by Eq.~(\ref{eqdyn3}) has a unique attracting fixed point $x_s$, then the quasi-potential $V(x)$ associated with the stationary distribution of that equation is obtained in general by
\be
V(x)=\inf_{x(t):x(t_1)=x_s,x(t_2)=x} J[x],
\ee
where $t_1$ and $t_2$ are, respectively, the starting and ending times of the trajectory $x(t)$ (see Sec.~4.2 of \cite{freidlin1984}). In this variational formula, the two endpoints of the interval $[t_1,t_2]$ are not fixed, which means that they are variables of the variational problem. In many cases, however, it is possible to solve the infimum by letting $t_1\ra -\infty$ and by fixing $t_2$, as we have done in the previous example. For a discussion of this procedure, see Sec.~4.3 of \cite{freidlin1984}. Bertini \textit{et al.}~\cite{bertini2002} give an interesting derivation of the above formula by studying the rate function associated with the time-reverse image of the trajectories $x(t)$ determined by Eq.~({\ref{eqdyn3}). 

The example that follows presents another important problem for which the WKB approximation is useful, namely, that of estimating the average time of escape from an attractor.

\begin{figure*}[t]
\centering
\includegraphics[scale=1.0]{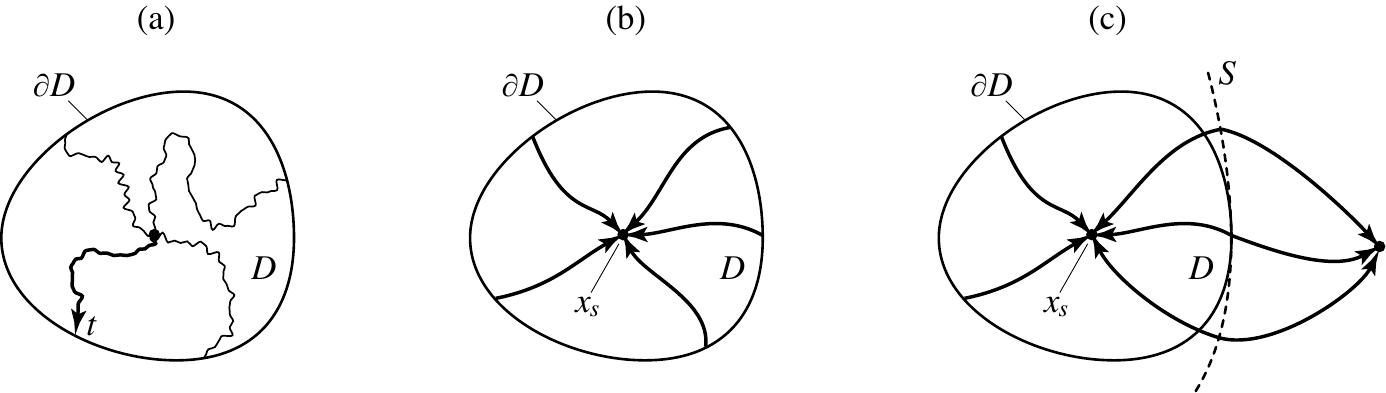}
\caption{(a) Random paths reaching the boundary $\partial D$ of a region $D$ in a time $t$. (b) Stable fixed point $x_s$ located in $D$. (c) Separatrix $S$ delimiting the basins of two attracting fixed points. Some paths on the boundary $\partial D$ lie on the separatrix and are attracted by one or the other fixed point depending on whether they start on the left or right of $S$.}
\label{figescdom1}
\end{figure*}

\begin{example}[Exit time from an attractor]
An attracting fixed point $x_s$ of a dynamical system does not remain attracting in the presence of noise: for $\vep\neq 0$, there is a non-zero probability, however small, that a trajectory starting in the vicinity of $x_s$ will be ``pushed'' by the noise out of some bounded region $D$ enclosing $x_s$; see Fig.~\ref{figescdom1}(a). The probability that such an escape occurs is very small, and decreases to zero as $\vep\ra 0$. Consequently, the random time needed for the system to reach the boundary $\partial D$ of $D$, which is defined as
\be
\tau_\vep=\inf\{t: x(t)\in \partial D\},
\ee
should increase in some probabilistic sense as $\vep\ra0$. This time $\tau_\vep$ is called the \emph{escape-} or \emph{exit-time} from $D$.

The calculation of $\tau_\vep$ is a classical problem in nonequilibrium statistical mechanics (see, e.g., \cite{gardiner1985,kampen1992}), and was solved on the mathematical front by Freidlin and Wentzell \cite{freidlin1984}, who treated it in the context of the general stochastic equation (\ref{eqdyn3}). Their main result assumes that the unperturbed dynamics associated with Eq.~(\ref{eqdyn3}) has a single attracting fixed point $x_s$ located in $D$, and that all the points on $\partial D$ are attracted to $x_s$, so that the case of a boundary $\partial D$ lying on a separatrix is excluded; see Fig.~\ref{figescdom1}(c). Under these assumptions, the following limit then holds:
\be
\lim_{\vep\ra 0} P\bigl(\E^{(V^*-\delta)/\vep}<\tau_\vep<\E^{(V^*+\delta)/\vep} \bigr)=1,
\label{eqtlim1}
\ee
where
\be
V^*=\inf_{x\in\partial D}\inf_{t\geq 0}\ V(x,t|x_s)
\label{eqv1}
\ee
and $\delta$ is any small positive constant. Moreover, 
\be
\lim_{\vep\ra 0}\ \vep\ln\lex\tau_\vep\rex=V^*.
\label{eqtelim1}
\ee
The first limit shown in (\ref{eqtlim1}) states that the most probable escape time scales as $\tau_\vep\asymp \E^{V^*/\vep}$ as $\vep\ra 0$. From this concentration result, the second limit follows.

The complete proof of these results is quite involved; see Sec.~4.2 of \cite{freidlin1984} or Sec.~5.7 of \cite{dembo1998}. However, there is a simple argument due to Kautz \cite{kautz1988} that can be used to understand the second result stating that $\lex \tau_\vep\rex\asymp \E^{V^*/\vep}$. The essential observation is that the average time $\lex \tau_\vep\rex$ of escape is roughly proportional to the escape rate $r_\vep$, which is itself proportional to the probability $P_\vep^\esc$ of escaping $D$. Thus $\lex\tau_\vep\rex\propto 1/P_\vep^\esc $, where
\be
P_{\vep}^\esc=\int_{\partial D}\D x\int_0^\infty \D t\, P_\vep(x,t|x_s).
\ee
Applying Laplace's approximation to this integral yields $P_{\vep}^\esc\asymp \E^{-V^*/\vep}$ with $V^*$ given by Eq.~(\ref{eqv1}), and therefore $\lex\tau_\vep\rex\asymp \E^{V^*/\vep}$, as Eq.~(\ref{eqtelim1}). The result of Freidlin and Wentzell is more precise, since it provides a Law of Large Numbers for $\tau_\vep$, not just an estimate for $\lex\tau_\vep\rex$.
\end{example}

The two previous examples are representative of the way large deviation techniques can be applied for calculating stationary probability densities $P_\vep(x)$ and fixed-time probability densities $P_\vep(x,t|x_0)$, as well as exit times and exit points. The second example, in particular, can be used to derive a whole class of Arrhenius-type results of the form $\tau_\vep\asymp \E^{V^*/\vep}$ for diffusion- or thermally-induced escape processes, including Kramers's classical result for the escape time of a Brownian particle trapped in a potential \cite{gardiner1985,kampen1992}. In the specific context of systems perturbed by thermal noise, the variational principle expressed by Eq.~(\ref{eqv1}) is often referred to as the \emph{principle of minimum available energy}, since $V^*$ can be shown to be proportional to the \emph{activation energy}, that is, the minimum energy required to induce the escape \cite{kautz1988,kautz1987}. An application of this principle for Josephson junctions is discussed by Kautz \cite{kautz1988,kautz1987}. 

For practical applications, it is important to note that Freidlin-Wentzell can be generalized to nonlinear systems having multiple attractors $A_i$, $i=1,2,\ldots$. For these, the escape time $\tau_{\vep,i}$ from a domain $D_i$ of attraction of $A_i$ is estimated as
\be
\tau_{\vep,i}\asymp \E^{V^*_i/\vep},\quad V^*_i=\inf_{x\in \partial D_i}\inf_{t\geq 0}\ V(x,t|x_{i}),
\ee
where $x_{i}$ is an initial point chosen inside $A_i$. For more than one attractor, the quasi-potential $V(x)$ characterizing the stationary distribution $P_\vep(x)$ over the whole state-space is also estimated as
\be
V(x)=\inf_{i} V_i(x),
\label{eqcp3}
\ee
where $V_i(x)$ is the quasi-potential of $P_\vep(x)$ restricted to the attractor $A_i$, i.e., the quasi-potential of a stationary probability density obtained by initiating paths inside $A_i$ \cite{freidlin1984,roy1995}. One important characteristic of many-attractor systems is that $V(x)$ is in general nonconvex, in addition to being nondifferentiable at points $x$ lying on a separatrix (see, e.g., \cite{freidlin1984,graham1986,graham1985,jauslin1987}). Mathematically, this arises because the infimum of Eq.~(\ref{eqcp3}) switches abruptly on a separatrix from one (generally smooth) quasi-potential $V_i(x)$ to another. A similar switching phenomenon was observed in the simpler context of sample means in Example~\ref{exmixedsum}.

\subsubsection{Experimental observations of large deviations}

Optimal paths and exit times are not just mathematical constructs---they can be, and have been, observed experimentally. The reader is refered to the extensive work of Dykman, Luchinsky, McClintock and collaborators \cite{dykman1992,dykman1996,luchinsky1997,luchinsky1997a} for a discussion of many properties of optimal paths observed in analog electronic circuits, including symmetry properties of these paths with respect to time inversion \cite{luchinsky1997}, and their singular patterns near coexisting attractors \cite{dykman1996}. All of these topics are reviewed in the excellent survey paper \cite{luchinsky1998}, which also discusses experimental measurements of exit times.

\subsection{Phenomenological models of fluctuations}
\label{subsecom}

Equilibrium statistical mechanics is a static theory of thermodynamics fluctuations: it provides a basis for calculating the probability of fluctuations of given macrostates, but says nothing about how these fluctuations arise in time. To describe the dynamics of these fluctuations, we must consider dynamical models of many-particle systems, and infer from these models the dynamical---and possibly stochastic---equations that govern the evolution of the macrostates that we are interested to study. Such a microstate-to-macrostate reduction of the dynamics of a many-body system is, as is well known, very difficult (if not impossible) to work out in practice, and so more modest approaches to this problem are usually sought. The most basic is the phenomenological approach, which consists in assuming that the time evolution of a macrostate, say $M_n$, follows a given stochastic dynamics of the form
\be
\dot M_n(t)=b(M_n)+\xi_n(t),
\label{eqpeq1}
\ee
where $b(M_n)$ is a force field, and $\xi_n(t)$ is a noise term that models the fluctuations of $M_n(t)$. The term ``phenomenological'' indicates that the dynamics of $M_n$ is postulated on the basis of a number of physical and mathematical principles, rather than being derived directly from an $n$-particle dynamics. Among these principles, we note the following:
\begin{enumerate}
\item The unperturbed dynamics $\dot m=b(m)$ should represent the macroscopic (most probable) evolution of $M_n(t)$;

\item The intensity of the noise $\xi_n(t)$ should vanish as $n\ra\infty$ to reflect the fact that the fluctuations of $M_n$ vanish in the thermodynamic limit;

\item Given that the fluctuations of $M_n$ arise from the cumulative and (we assume) short-time correlated interactions of $n$ particles, the noise $\xi_n(t)$ should be chosen to be a Gaussian white noise with zero mean;
 
\item The stationary probability distribution associated with Eq.~(\ref{eqpeq1}) should match the equilibrium probability distribution of $M_n$ determined by the statistical ensemble used to describe the $n$-particle system (at equilibrium).
\end{enumerate}

The second and third points imply that $\xi_n(t)$ should satisfy $\langle \xi_n(t)\rangle=0$ and $\langle \xi_n(t)\xi_n(t')\rangle=b_n \delta(t-t')$, with $b_n\ra 0$ as $n\ra\infty$. The precise dependence of $b_n$ on $n$ is determined self-consistently, following the last point, by matching the large-$n$ form of the stationary distribution of Eq.~(\ref{eqpeq1}) with the large deviation form of the equilibrium (ensemble) probability distribution of $M_n$. This is explained in the next example.

\begin{example}[Equilibrium fluctuations]
\label{exequifluct}
Consider a macrostate $M_n$ satisfying an equilibrium large deviation principle of the form
\be
p(M_n=m)\asymp \E^{-a_n I(m)},
\label{eqldpd1}
\ee
where $I(m)$ is any of the rate functions arising in the microcanonical or canonical ensemble, and $a_n$ is the speed of the large deviation principle.\footnote{Recall that in the microcanonical ensemble, $I(m)$ is interpreted as an entropy function, whereas in the canonical ensemble, $I(m)$ is interpreted as a free energy function; see Sec.~\ref{secequi}.} If $I(m)$ has a unique global minimum, then we know from Example~\ref{exou1} that the stochastic dynamics
\be
\dot M_n(t)=-\frac{1}{2}I'(M_n)+\xi_n(t)
\label{eqratef1}
\ee
has a stationary density given by $p(M_n=m)\asymp \E^{-I(m)/b_n}$ for small $b_n$. By matching this asymptotic with the large deviation principle of (\ref{eqldpd1}), we then obtain $b_n=a_n^{-1}$. Thus, if the speed $a_n$ of the large deviation principle is the number $n$ of particles, as is typically the case, then $b_n=n^{-1}$. Near phase transitions, the speed of a large deviation principle may change (see Examples~\ref{ex2disingm} and \ref{ex2disingc}), and this should be reflected in $b_n$.
\end{example}

Models of fluctuation dynamics based on the phenomenological model of Eq.~(\ref{eqpeq1}) or the more specific equation found in (\ref{eqratef1}), based on the rate function $I(m)$, are used to answer a variety of questions, such as: 
\begin{itemize}
\item What is the most probable \emph{fluctuation path} $\{m(t)\}_{t=0}^\tau$ connecting over a time $\tau$ the equilibrium or stationary value $m^*$ of $M_n$ to some other value $m\neq m^*$? 

\item What is the most probable \emph{decay path} connecting the nonequilibrium state $M_n(0)=m$ to the equilibrium state $M_n(\tau)=m^*$? 

\item Is there a relationship between a given fluctuation path and its corresponding decay path? For instance, are decay paths the time-reverse image of fluctuation paths? 

\item What is the typical or expected time of return to equilibrium? That is, what is the typical or expected time $\tau$ for which $M_n(\tau)=m^*$ given that $M_n(0)=m\neq m^*$? 

\item If $I(m)$ has local minima in addition to global minima, what is the typical time of decay from a local minimum to a global minimum? In other words, what is the typical decay time from a metastable state?
\end{itemize}

It should be clear from our experience of the last subsection that all of these questions can be answered within the framework of the Freidlin-Wentzell theory of differential equations perturbed by noise. In the thermodynamic limit, fluctuation and decay paths are optimal paths, and can be determined as such by the variational principle of Eq.~(\ref{eqcp2}), the Lagrangian equation (\ref{eqlag1}) or its Hamiltonian counterpart, Eq.~(\ref{eqham1}). These equations also hold the key for comparing the properties of decay and fluctuation paths. As for the calculation of decay times from nonequilibrium states, including metastable states, it closely follows the calculation of exit times that we have discussed in the previous subsection (see also \cite{olivieri2003,olivieri2005}).

Other results about nonequilibrium fluctuations can be translated in much the same way within the framework of the Freidlin-Wentzell theory. The minimum dissipation principle of Onsager and Machlup \cite{onsager1953}, for example, which states that the fluctuation and decay paths of equilibrium systems minimize some dissipation function, can be re-interpreted in terms of the variational principle of Eq.~(\ref{eqcp2}), which determines the optimal paths of noise-perturbed systems. The next example is intended to clarify this point by explicitly translating the theory of Onsager and Machlup into the language of large deviations. For simplicity, we consider the dynamics of a single, one-dimensional macrostate.
 
\begin{example}[Linear fluctuation theory]
\label{exom1}
The linear theory of equilibrium fluctuations proposed by Onsager and Machlup \cite{onsager1953} (see also \cite{falkoff1956,falkoff1958}) is based on what is essentially a linear version of Eq.~(\ref{eqratef1}), obtained by assuming that $I(m)$ has a unique and locally quadratic minimum at $m^*=0$, and that $M_n$ fluctuates close to this equilibrium value. By approximating $I(m)$ to second order around its minimum
\be
I(m)\approx am^2,\quad a=\frac{I''(0)}{2}>0,
\ee
we thus write
\be
\dot M_n(t)=-aM_n(t)+\xi_n(t).
\label{eqlin1}
\ee
The scaling of $\xi_n(t)$ with $n$ is not specified by Onsager and Machlup \cite{onsager1953}, but it is obvious from their analysis that, if Eq.~(\ref{eqlin1}) is to have a macroscopic limit, then the variance of the noise should scale inversely with the number $n$ of particles, as explained in Example~\ref{exequifluct}. In this case, we can write the path probability density $P_n[m]$ of $M_n(t)$ as
\be
P_n[m]\asymp \E^{-nJ[m]},\quad J[m]=\frac{1}{2}\int_0^\tau [\dot m(t)+am(t)]^2\, \D t
\ee
in the limit of large $n$, which is more or less what Onsager and Machlup obtain in \cite{onsager1953}. The Lagrangian of this rate function can be re-written as
\be
L(\dot m,m)=\Phi(\dot m)+\Psi(m)+\frac{\dot I(m)}{2}
\ee
by defining what Onsager and Machlup call the \emph{dissipation functions} $\Phi(\dot m)=\dot m ^2/2$ and $\Psi(m)=a^2m^2/2$. With this form of $L(\dot m,m)$, a fluctuation path is then characterized as a path that globally minimizes
\be
\int_0^\tau [2\Phi(\dot m)+2\Psi(m)+\dot I(m)]\, \D t = I(m(\tau))-I(m(0))+2\int_0^\tau [\Phi(\dot m)+\Psi(m)]\, \D t
\label{eqmdp1}
\ee
subject to the terminal conditions $m(0)=0$ and $m(\tau)=m\neq 0$. This variational principle is equivalent to the general variational principle of Eq.~(\ref{eqcp2}), and is what Onsager and Machlup refer to as the \emph{minimum dissipation principle} \cite{onsager1953}. The decay path bringing an initial fluctuation $m(0)=m\neq 0$ back to the equilibrium point $m^*=0$ also satisfies this principle, but with the terminal conditions exchanged, i.e., with $m(0)=m$ and $m(\tau)=0$. 

From the symmetry of the associated Lagrange equation, it can be shown that the decay path is the time-reverse image of the corresponding fluctuation path. This holds, in general, whenever the dynamics of $M_n(t)$ is derived from a quasi-potential $I(m)$, that is, when $m^*$ is an \emph{equilibrium state} in the thermodynamic sense. When the dynamics of $M_n(t)$ involves external forces or non-conservative forces (in more than one dimension), the forward and backward optimal paths need not be the time-reverse of one another; see \cite{luchinsky1998} for examples. In this case, $m^*$ is called a stationary state rather than an equilibrium state.
\end{example}

The linear model of Onsager and Machlup serves as a template for constructing and studying other models of fluctuation dynamics, including models of nonequilibrium steady states (see, e.g., \cite{hasegawa1976,taniguchi2007,taniguchi2008a}), and for ultimately building a general theory of nonequilibrium processes (see, e.g., \cite{graham1981,eyink1990,paniconi1997,oono1998,suarez1995,bertini2002}). In going beyond this model, one may replace the linear force $b(m)=-am$ by nonlinear forces, consider noise processes with a non-zero mean or noise processes that are correlated in time, in addition to studying several (possibly coupled) macrostates rather than just one as we did in the previous example. One may also model the fluctuations of a field $\rho(x,t)$ using a general equation of the form
\be
\partial_t \rho(x,t)=D(\rho(x,t))+\xi(x,t),
\ee
where $D$ is some operator acting on $\rho(x,t)$, and $\xi(x,t)$ is a space-time noise process. Stochastic field equations of this form are known as \emph{hydrodynamic equations}, and are used to model turbulent fluids \cite{falkovich2001,chetrite2007,gourcy2007}, as well as the macroscopic dynamics of particles evolving and interacting on lattices \cite{eyink1990,spohn1991,kipnis1999}; see Sec.~\ref{secintpart}. Note that for a field $\rho(x,t)$, the analog of the path probability density $P[m]$ of $M_n(t)$ is the functional probability density $P[\rho]=P(\{\rho(x,t)\}_{t=0}^\tau)$, which gives the probability density that $\rho(x,t)$ follows a given ``trajectory'' or history $\{\rho(x,t)\}_{t=0}^\tau$ in some function space.

In the next subsection, we will apply methods inspired from the results of Onsager and Machlup to study the fluctuations of physical observables defined as time-averages over the paths of stochastic systems. 

\subsection{Additive processes and fluctuation relations}
\label{subsecfr}

The large deviation results that we have surveyed so far were mostly concerned with the trajectories or paths of stochastic processes, and the probability density of these paths. Here we shall be concerned with random variables defined on these paths as additive functionals of the form
\be
A_\tau[x]=\frac{1}{\tau}\int_0^\tau f(x(t))\, \D t,
\label{eqaf1}
\ee 
where $f$ is a smooth function mapping the state $x(t)$ of some stochastic process $X(t)$ to $\reals^d$, $d\geq 1$. The random variable $A[x]$ is called the \emph{time-average} of $f(x(t))$ over the time interval $[0,\tau]$. The usual problem that we are concerned with is to investigate whether, for a given stochastic process $X(t)$, $A_\tau$ satisfies a large deviation principle and, if so, to determine its rate function. 

\subsubsection{General results}

As in the case of sample means of random variables, a large deviation principle can be derived for $A_\tau$, at least in principle, via the G\"artner-Ellis Theorem. The scaled cumulant generation function in this case is
\be
\lambda(k)=\lim_{\tau\ra\infty} \frac{1}{\tau}\ln\lex \E^{\tau k\cdot A_\tau}\rex,\quad k\in\reals^d,
\ee
where
\be
\lex \E^{\tau k\cdot A_\tau}\rex=\int \E^{\tau k\cdot a}P(A_\tau\in \D a)=\int \mathcal{D}[x]\, P[x]\, \E^{\tau k\cdot A_\tau[x]}
\ee
and $P[x]$ is, as before, the probability density over the paths $\{ x(t)\}_{t=0}^\tau$ extending from $t=0$ to $t=\tau$. Provided that $\lambda(k)$ exists and is differentiable, we then have 
\be
P(A_\tau\in \D a)\asymp \E^{-\tau I(a)}\D a,\quad I(a)=\sup_k\{k\cdot a-\lambda(k)\}.
\ee

If $X(t)$ is an ergodic Markov process, the large deviations of $A_\tau$ can be determined, also in principle, using the Donsker-Varadhan theory of Markov additive processes. In this case, $\lambda(k)$ is evaluated as the logarithm of the largest eigenvalue of the operator $L_k=L+k\cdot f$, $L$ being the generator of the stochastic process $X(t)$ \cite{donsker1975,donsker1975a,donsker1976}; see also Sec.~V.A of \cite{bucklew1990} and \cite{majumdar2002}.\footnote{There is a mathematical difficulty that will not be discussed here, namely, that the largest eigenvalue of $L_k$ has to be isolated in order for the large deviation principle to hold.} This result is the continuous-time generalization of the result of Sec.~\ref{subsecmp} stating that, for an ergodic Markov chain, $\lambda(k)$ is given by the logarithm of the largest eigenvalue of the ``tilted'' transition matrix $\Pi_k$. An important example of additive random variables, which has been extensively studied by Donsker and Varadhan \cite{donsker1975,donsker1975a,donsker1976}, is presented next.

\begin{example}[Occupation measure]
Let $1_A(x)$ denote the indicator function for the set $A$ which equals $1$ if $x\in A$ and $0$ otherwise. The time-average of this function, given by
\be
M_\tau(A)=\frac{1}{\tau}\int_0^\tau 1_A(x(t))\, \D t,
\ee
gives the fraction of the time $\tau$ that the path $\{ x(t)\}_{t=0}^\tau$ spends in $A$, and plays, as such, the role of the empirical vector for continuous-time dynamics. To make the connection more obvious, take $A$ to be an infinitesimal interval $[x,x+\D x]$ anchored at the point $x$. Then $M_\tau(\D x)=M_\tau([x,x+\D x])$ ``counts'' the number of times $x(t)$ goes inside $[x,x+\D x]$. The density version of $M_\tau(\D x)$, defined as
\be
L_\tau(x)=\frac{1}{\tau}\int_0^\tau \delta(x(t)-x)\, \D t,
\ee 
``counts'', similarly as for the empirical density defined in Eq.~(\ref{eqed1}), the number of times that $x(t)$ hits a given point $x$ as opposed to an interval of points.\footnote{$L_\tau(x)$ is a density, so the number of times that $x(t)$ hits the interval $[x,x+\D x]$ is actually $L_\tau(x)\,\D x$.}

For many stochastic processes, $L_\tau$ is observed to converge in probability to a given stationary density in the long-time limit $\tau\ra\infty$. The fluctuations of $L_\tau$ around this concentration point can be characterized by a rate function, which can formally be expressed via the G\"artner-Ellis Theorem as 
\be
I[\mu]=\sup_{k}\{\mu\cdot k-\lambda[k]\},
\label{eqgemp1}
\ee
where
\be
\mu\cdot k=\int_{\reals^d} \mu(x) k(x)\, \D x
\ee 
and 
\be
\lambda[k]=\lim_{\tau\ra\infty}\frac{1}{\tau}\ln\lex\exp\left\{\int_0^\tau k(x(t))\, \D t\right\}\rex.
\ee
This result can be found in G\"artner \cite{gartner1977}. More explicit expressions for $I[\mu]$ can be obtained from the general Legendre-Fenchel transform shown in Eq.~(\ref{eqgemp1}) by considering specific random processes $X(t)$. In the particular case of an ergodic Markov process with generator $G$, Donsker and Varadhan obtained
\be
I[\mu]=-\inf_{u>0} \lex\frac{Gu}{u}\rex_\mu=-\inf_{u>0}\int \mu(x) \frac{(Gu)(x)}{u(x)}\, \D x.
\ee
as part of their general theory of large deviations of Markov processes~\cite{donsker1975,donsker1975a, donsker1976} (see also Sec.~V.B of \cite{bucklew1990}). This rate function is the continuous-time analog of the rate function presented in Example~\ref{exsanovmc}. As in that example, the  minimum and zero of $I[\mu]$ is the stationary probability density $\rho^*$ of the ergodic Markov process generated by $G$. The rate function $I[\mu]$ characterizes the fluctuations of $L_\tau$ around that concentration point.
\end{example}

The next example shows how a large deviation principle can be derived for $A_\tau$ when the stochastic process $X(t)$ falls in the Freidlin-Wentzell framework of stochastic differential equations perturbed by Gaussian noise. The large deviation principle that one obtains in this case applies in the limit of vanishing noise, which is different from the $\tau\ra\infty$ limit that we have just considered.

\begin{example}
Consider a general Markov process $X_\vep(t)$ arising, as in Sec.~\ref{subsecfw}, as the solution of a dynamical system perturbed by a Gaussian white noise of strength $\vep$, and let $A_\tau[x]$ be a time average defined over $X_\vep(t)$. From the Freidlin-Wentzell theory, we know that the large deviation principle $P[x]\asymp \E^{-J[x]/\vep}$ applies in the small noise limit $\vep\ra 0$, with rate functional $J[x]$ given by Eq.~(\ref{eqldpfw1}). Since $A_\tau$ is a functional of $X_\vep(t)$, the contraction principle immediately implies that $A_\tau$ also satisfies a large deviation principle in the limit $\vep\ra 0$, with rate function $I(a)$ given by the contraction of $J[x]$:
\be
I(a)=\inf_{x(t):A_\tau[x]=a}\ J[x].
\label{eqvarnoneq1}
\ee
As always, we can use Lagrange's multiplier method to transform this constrained maximization into an unconstrained optimization problem. This was done in Example~\ref{exsmst1} for the contraction of Sanov's Theorem, as well as in Example~\ref{exjaynes1}, which illustrated the maximum entropy principle. The unconstrained functional that we have to optimize in the present case is $K[x]=J[x]-\beta A_\tau[x]$, and involves the Lagrange multiplier $\beta$ which takes care of the constraint $A_\tau[x]=a$. 
\end{example}

Although the large deviation principle obtained in the previous example applies, strictly speaking, in the limit $\vep\ra 0$, it can often be transformed into a large deviation principle for $A_\tau$ in the limit $\tau\ra\infty$ by studying the extensivity of $I(a)$ with $\tau$ (see, e.g., \cite{paniconi1997,taniguchi2007}). The large deviation principle that one obtains in this case applies in the dual limit $\tau\ra\infty$ and $\vep\ra 0$, which means that it is only an approximation of the large deviation principle that governs the fluctuations of $A_\tau$ in the limit $\tau\ra\infty$ for an \emph{arbitrary} noise power $\vep$, i.e., without the limit $\vep\ra\infty$. In technical terms, this means that the knowledge of the optimal path solving the variational principle (\ref{eqvarnoneq1}) is in general not sufficient to derive the long-time large deviations of $A_\tau$ for any noise power. The only exception to this statement, noted by Onsager and Machlup \cite{onsager1953}, are linear stochastic differential equations, i.e., linear Langevin equations. For these, the evaluation of a path integral by its most probable path actually gives the exact value of the path integral for all $\vep>0$, up to a normalization constant, which can usually be omitted for the purpose of deriving large deviation results.

\subsubsection{Fluctuation relations}

The next example is concerned with the large deviations of an additive process often studied in nonequilibrium statistical mechanics. This example is also our point of departure for studying large deviations of \emph{nonequilibrium observables}, that is, random variables defined in the context of nonequilibrium systems, and for introducing an important class of results known as \emph{fluctuation relations} or \emph{fluctuation theorems}.

\begin{example}[Work fluctuations for a Brownian particle \cite{zon2003a}] 
\label{exbrown1}
Consider a Brownian particle immersed in a fluid, and subjected to the ``pulling'' force of a harmonic potential moving at constant velocity $v_p$. The dynamics of the particle is modeled, in the overdamped limit, by the Langevin equation 
\be
\dot x(t)=-[x(t)-v_pt]+\zeta(t),
\ee
where $x(t)$ denotes the position of the particle at time $t$, with $x(0)=0$, and $\zeta(t)$ is a Gaussian white noise characterized by $\langle\zeta(t)\rangle=0$ for all $t$ and $\langle \zeta(t)\zeta(t')\rangle=2\delta(t-t')$.\footnote{Dimensional units are used here; see \cite{zon2003a} for the full, physical version of this equation.} The \emph{work per unit time} or \emph{intensive work} $W_\tau$ done by the pulling force $F(t)=-[x(t)-v_p t]$ over an interval of time $[0,\tau]$ has for expression:
\be
W_\tau=\frac{1}{\tau}\int_0^\tau v_p F(t)\, \D t=-\frac{v_p}{\tau}\int_0^\tau [x(t)-v_pt]\, \D t.
\ee
The large deviation principle governing the fluctuations of this additive process is found, following the G\"artner-Ellis Theorem, by calculating the scaled cumulant generating function $\lambda(k)$ of $W_\tau$. This calculation can be performed using various methods (e.g., characteristic functions \cite{zon2003a}, differential equation techniques \cite{touchette2007}, path integrals \cite{taniguchi2007}, etc.), which all lead to 
\be
\lambda(k)=ck+ck^2=ck(1+k),
\ee
where $c=(v_p)^2$. Since this function is quadratic, the rate function of $W_\tau$ given by the Legendre-Fenchel transform of $\lambda(k)$ must also be quadratic, which implies that the fluctuations of $W_\tau$ are Gaussian. To be more precise, let $p(W_\tau=w)$ denote the probability density of $W_\tau$. Then
\be
p(W_\tau=w)\asymp \E^{-\tau I(w)},\quad I(w)=\inf_k\{kw-\lambda(k)\}=\frac{(w-c)^2}{4c}.
\ee
The main conclusion that we draw from this result is that positive amounts of work done by the pulling force on the Brownian particle are exponentially more probable than negative amounts of equal magnitude, since $c>0$ for $v_p\neq 0$. To make this more obvious, consider the probability ratio
\be
R_\tau(w)=\frac{p(W_\tau=w)}{p(W_\tau=-w)}.
\label{eqratio1}
\ee
Given the quadratic form of $I(w)$, it is easy to see that
\be
R_\tau(w)\asymp \E^{\tau [I(-w)-I(w)]}= \E^{\tau w}.
\label{eqratio2}
\ee
Accordingly, the probability that $W_\tau=w>0$ is, in the large time limit, exponential larger than the probability that $W_\tau=-w$.
\end{example}

The study of the probability ratio $R_\tau(w)$ for physical observables other than the work $W_\tau$ defined above has become an active topic of study in nonequilibrium statistical mechanics; see \cite{maes2006,harris2007,kurchan2007,marconi2008} for theoretical surveys of this topic, and \cite{ritort2003,evans2002} for more experimental surveys. The importance of $R_\tau(w)$ is justified by two observations. The first is that $R_\tau(w)$ provides a measure of how ``out of equilibrium'' a system is, since it yields information about the positive-negative asymmetry of nonequilibrium fluctuations that arises in general because of the irreversibility of the fluctuations paths \cite{bochkov1981,maes2003a}. The second observation is that the precise exponential form of $R_\tau(w)$ displayed in (\ref{eqratio2}) appears to be a general law characterizing the fluctuations of several different nonequilibrium observables, not just the work $W_\tau$ considered above. A precise formulation of this law, now commonly referred to as the \emph{fluctuation theorem} \cite{gallavotti1995,gallavotti1995a}, can be given as follows. Let $A_\tau$ denote a nonequilibrium observable integrated over a time interval $\tau$. For simplicity, let us assume that $A_\tau$ is a real random variable, and that $p(A_\tau=a)$ is non-zero for all $a\in\reals$. Then $A_\tau$ is said to satisfy the fluctuation theorem if
\be
R_\tau(a)=\frac{p(A_\tau=a)}{p(A_\tau=-a)}\asymp \E^{\tau c a},
\label{eqratio3}
\ee
in the limit of large $\tau$, with $c$ a constant independent of $a$ and $\tau$. Equivalently, $A_\tau$ satisfies the fluctuation theorem if
\be
\varrho(a)=\lim_{\tau\ra\infty}\frac{1}{\tau}\ln R_\tau(a)=ca.
\label{eqratio4}
\ee
The expression ``fluctuation theorem'' should be used, strictly speaking, when the asymptotic result of (\ref{eqratio3}) or (\ref{eqratio4}) is proved for a specific nonequilibrium observable. When this result is experimentally or numerically verified rather than being proved, the expression ``fluctuation relation'' is more appropriate.

The first observation of a fluctuation relation was reported by Evans, Cohen and Morriss \cite{evans1993}, who numerically studied the fluctuations of sheared fluids. Based on these results, Gallavotti and Cohen \cite{gallavotti1995,gallavotti1995a} then proved a fluctuation theorem for the entropy rate of chaotic, deterministic systems, which was later extended to general Markov processes by Kurchan \cite{kurchan1998}, Lebowitz and Spohn \cite{lebowitz1999}, and Maes \cite{maes1999}. These results form the basis of several experimental studies of fluctuation relations arising in the context of particles immersed in fluids \cite{wang2002,andrieux2007a}, electrical circuits \cite{zon2004a,garnier2005,andrieux2007a}, granular media \cite{aumaitre2001,feitosa2004,puglisi2005,visco2005,visco2006}, turbulent fluids \cite{ciliberto1998,ciliberto2004}, and the effusion of ideal gases \cite{cleuren2006}, among other systems. The next example gives the essence of the fluctuation theorem for the entropy rate of Markov processes. This example is based on the results of Lebowitz and Spohn \cite{lebowitz1999}, and borrows some notations from Gaspard \cite{gaspard2004} (see also \cite{chetrite2007a,chetrite2008}). For a discussion of the entropy production rate based on the Donsker-Varadhan rate function of the empirical measure, the reader is referred to \cite{maes2007}.

\begin{example}[Entropy production]
\label{exentprod1}
Let $\sig=\sig_1,\sig_2,\ldots,\sig_n$ be the trajectory of a discrete-time ergodic Markov chain starting in the state $\sig_1$ at time $1$ and ending with the state $\sig_n$ at time $n$. Denote by $\sig^R$ the \emph{time-reversed} version of $\sig$ obtained by reversing the order in which the states $\sig_1,\sig_2,\ldots,\sig_n$ are visited in time, that is, $\sig^R=\sig_n,\sig_{n-1},\ldots,\sig_1$. If the Markov chain is \emph{reversible}, that is, if $P(\sig)=P(\sig^R)$ for all $\sig$, then the \emph{entropy production rate} of the Markov chain, defined as 
\be
W_n(\sig)=\frac{1}{n}\ln\frac{P(\sig)}{P(\sig^R)},
\ee
equals zero for all $\sig$. Accordingly, to study the irreversibility of the Markov chain, we may study how $W_n$ fluctuates around its mean $\langle W_n\rangle$, as well as how the mean differs from zero.

For an ergodic Markov chain, the Asymptotic Equipartition Theorem mentioned in Sec.~\ref{subsecsp} directly implies \cite{jiang2003,gaspard2004}
\be
\lim_{n\ra\infty} \langle W_n\rangle=h^R-h,
\ee
where $h$ is the \emph{forward} entropy rate defined in Sec.~\ref{subsecsp}, and $h^R$ is the \emph{backward entropy} rate, defined as
\be
h^R=\lim_{n\ra\infty} -\frac{1}{n}\sum_{\sig} P(\sig) \ln P(\sig^R).
\ee
To find the rate function governing the fluctuations of $W_n$ around its mean, note that
\be
\lex \E^{nkW_n}\rex=\sum_{\sig} P(\sig) \frac{P(\sig)^k}{P(\sig^R)^k}=\sum_{\sig} P(\sig^R) \frac{P(\sig)^{k+1}}{P(\sig^R)^{k+1}}.
\ee
Summing over the time-reversed trajectories $\sig^R$ instead of $\sig$ leads to
\be
\lex \E^{nkW_n}\rex=\sum_{\sig^R} P(\sig^R) \frac{P(\sig)^{k+1}}{P(\sig^R)^{k+1}}=\sum_{\sig^R} P(\sig) \frac{P(\sig^R)^{k+1}}{P(\sig)^{k+1}}.
\ee
Thus $\langle \E^{nkW_n}\rangle=\langle \E^{n(-1-k)W_n}\rangle$ and
\be
\lambda(k)=\lim_{n\ra\infty} \frac{1}{n}\ln\lex \E^{nkW_n}\rex=\lambda(-1-k)
\label{eqsym1}
\ee 
for all $k\in\reals$.\footnote{The convergence of $\lambda(k)$ for all $k\in\reals$ follows from the assumption that the space of $\sig$ is finite; see Sec.~\ref{subsecmp}.} From the G\"artner-Ellis Theorem, we therefore obtain
\be
I(w)=\sup_{k}\{kw-\lambda(k)\}=\sup_{k}\{kw-\lambda(-1-k)\}=I(-w)-w
\label{eqftldt1}
\ee
and, consequently, $\varrho(w)=I(-w)-I(w)=w$. This shows, as announced, that $W_n$ satisfies the fluctuation theorem.
\end{example}

The fluctuation theorem for $W_n$ can be generalized to continuous-state and continuous-time Markov processes by adapting the definition of the entropy production rate to these processes, and by studying the time reversibility of their paths or, equivalently, the time reversibility of the master equation governing the evolution of probabilities defined on these paths (see, e.g., \cite{bochkov1981,lebowitz1999, kurchan1998,maes2000,maes2003b,seifert2005,imparato2006a}). In the case of ergodic, continuous-time Markov processes, the symmetry of $\lambda(k)$ expressed in Eq.~(\ref{eqsym1}) can be related to a time-inversion symmetry of the processes' generator \cite{kurchan1998,lebowitz1999}. This time-inversion symmetry can also be used to prove fluctuation theorems for other nonequilibrium observables related to the entropy production rate, such as the work \cite{bochkov1981,kurchan1998}. 

\subsubsection{Fluctuation relations and large deviations}

Fluctuation relations and fluctuation theorems are intimately linked to large deviation principles, as is obvious from the theory and examples studied so far. In a sense, one implies the other. This equivalence was observed by Gallavotti and Cohen in their original derivation of the fluctuation theorem \cite{gallavotti1995,gallavotti1995a} (see also \cite{gallavotti1998a,gallavotti2008}), and can be made more explicit by examining the chain of equalities displayed in (\ref{eqftldt1}). To put these equalities in a more general light, let us consider a general nonequilibrium observable $A_\tau$ integrated over a time $\tau$, and let $\lambda(k)$ be its scaled cumulant generating function. By re-stating the result of (\ref{eqftldt1}) for $A_\tau$, it is first obvious that, if $\lambda(k)$ satisfies the conditions of the G\"artner-Ellis Theorem (differentiability and steepness), in addition to the symmetry property $\lambda(k)=\lambda(-k-c)$ for all $k\in\reals$ and $c$ a real constant, then $\varrho(a)=ca$. By inverting the Legendre-Fenchel transform involved in (\ref{eqftldt1}), we also obtain the converse result, namely that, if $A_\tau$ satisfies a large deviation principle with rate function $I(a)$, and $\varrho(a)=ca$ for some real constant $c$, then $\lambda(k)=\lambda(-k-c)$. By combining these two results, we thus see that the symmetry $\lambda(k)=\lambda(-c-k)$ is essentially equivalent to having a fluctuation theorem for $A_\tau$. 

When applying these results, it is important to note that the symmetry property $\lambda(k)=\lambda(-k-c)$ can be satisfied even if the generating function of $A_\tau$ satisfies the same property but only approximately in the limit of large $\tau$. In Example~\ref{exentprod1}, it so happens that this property is satisfied exactly by the generating function. Moreover, for processes having a countably-infinite or continuous state space, $\lambda(k)$ does not exist in general for all $k\in\reals$, but only for a convex subset of $\reals$ (see Examples~\ref{exskew} and \ref{exnonsteep}). In this case, results similar to those above apply but in a pointwise sense. That is, if $\lambda(k)$ is differentiable at $k$  and satisfies the symmetry $\lambda(k)=\lambda(-k-c)$ for the same value $k$, then $\varrho(a)=ca$ for $a$ such that $a=\lambda'(k)$.\footnote{This follows by applying the local Legendre transform of Eq.~(\ref{eqllt1}), which we have discussed in the context of nondifferentiable points of $\lambda(k)$ and nonconvex rate functions; see Secs.~\ref{subseciid} and \ref{subsecncrf}.} To formulate a converse to this result, we can follow the same arguments as above to prove that, if $\lambda(k)\neq \lambda(-k-c)$ for at least one value $k$, then $\varrho(a)$ is not proportional to $a$ for at least one value $a$. Therefore, if $\lambda(k)$ satisfies the symmetry property only for a subset of $\reals$, then $\varrho(a)=ca$ only for a subset of the values of $A_\tau$. In this case, we say that $A_\tau$ satisfies an \emph{extended} fluctuation theorem.

To make sure that extended fluctuation theorems are not confused with the fluctuations theorems defined at the start of this section, it is common to refer to the latter ones as \emph{conventional} fluctuation theorems \cite{zon2003,zon2004,marconi2008}. Thus an observable $A_\tau$ is said to satisfy a \emph{conventional} fluctuation theorem if its fluctuation function $\varrho(a)$, defined by the limit of (\ref{eqratio4}), is linear in $a$, i.e., if $\varrho(a)=ca$. If $\varrho(a)$ is a nonlinear function of $a$, then $A_\tau$ is said to satisfy an \emph{extended} fluctuation theorem. The reader will judge by him- or herself whether these definitions are useful. In the end, it should be clear that the rate function $I(a)$ completely characterizes the fluctuations of $A_\tau$ in the long time limit, so that one might question the need to attach the terms ``conventional'' or ``extended'' to $I(a)$. Indeed, one might even question the need to define $\varrho(a)$ when one has $I(a)$.

One example of nonequilibrium observables that illustrates the notion of extended fluctuation theorem is the heat per unit time dissipated by the dragged particle of Example~\ref{exbrown1}. For this observable, van Zon and Cohen \cite{zon2003,zon2004} have shown that the symmetry on $\lambda(k)$ holds only on a bounded interval due to the fact that $\lambda(k)$ does not converge for all $k\in\reals$. Violations of the fluctuation theorem have also been shown to arise from the choice of initial conditions \cite{farago2002,puglisi2006a,visco2006a}, the unboundness of the observable considered \cite{bonetto2006}, or the restriction of the domain of $\lambda(k)$ \cite{harris2006,rakos2008}. These violations are all included in large deviation theory insofar as they reflect special properties of rate functions and their associated scaled cumulant generating functions. Indeed, many of the limiting cases of rate functions that we have discussed in Sec.~\ref{secmathapp} do arise in the context of nonequilibrium fluctuations, and lead to violations and possible extensions of the fluctuation theorem. The extended fluctuation theorem of van Zon and Cohen \cite{zon2003,zon2004}, for instance, is closely related to affine rate functions, studied in Examples~\ref{exbp1} and \ref{exnonsteep}. A model for which $\lambda(k)$ is found to be nondifferentiable is discussed in \cite{rakos2008}. Finally, a model of nonequilibrium fluctuations having a zero rate function is discussed in \cite{touchette2007}. This model is a Markov equivalent of the sample mean of IID symmetric L\'evy random variables that was considered in Example~\ref{exsymlevy}.

To close our discussion of fluctuation relations and fluctuation theorems, note that a fluctuation theorem may hold approximately for the small values of $A_\tau$ even if its associated $\lambda(k)$ does not satisfy the symmetry $\lambda(k)=\lambda(-k-c)$. This follows by noting that if a large deviation principle holds for $A_\tau$ with a rate function $I(a)$ which is differentiable at $a=0$, then $\varrho(a)\approx -2I'(0)a$ to first order in $a$ \cite{aumaitre2001,farago2002}. If $I(a)$ has a parabolic minimum $a^*$, we can also write $\varrho(a)\approx2 I''(a^*)a^*a$ to second order in $a-a^*$. In both cases, $\varrho(a)$ is linear in $a$, which is the defining property of conventional fluctuation relations.

\subsection{Interacting particle models}
\label{secintpart}

Markovian models of interacting particles, such as the exclusion process, the zero-range process, and their many variants (see~\cite{spohn1991,liggett2004}), have been, and still are, extensively studied from the point of view of large deviations. The interest for these models comes from the fact that their macroscopic or \emph{hydrodynamic} behavior can be determined from their ``microscopic'' dynamics, sometimes in an exact way. Moreover, the typicality of the hydrodynamic behavior can be studied by deriving large deviation principles which characterize the probability of observing deviations in time from the hydrodynamic evolution \cite{kipnis1999}. The interpretation of these large deviation principles follows the Freidlin-Wentzell theory, in that a deterministic dynamical behavior---here the hydrodynamic behavior---arises as the global minimum and zero of a given (functional) rate function. From this point of view, the hydrodynamic equations, which are the equations of motion describing the hydrodynamic behavior, can be characterized as the solutions of a variational principle similar to the minimum dissipation principle of Onsager \cite{bertini2004}.

Two excellent review papers \cite{derrida2007,bertini2007} have appeared recently on interacting particle models and their large deviations, so we will not review this subject in detail here. The next example illustrates in the simplest way possible the gist of the results that are typically obtained when studying these models. The example follows the work of Kipnis, Olla and Varadhan \cite{kipnis1989}, who were the first to apply large deviation theory for studying the hydrodynamic limit of interacting particle models. 

\begin{figure*}[t]
\centering
\includegraphics[scale=1.0]{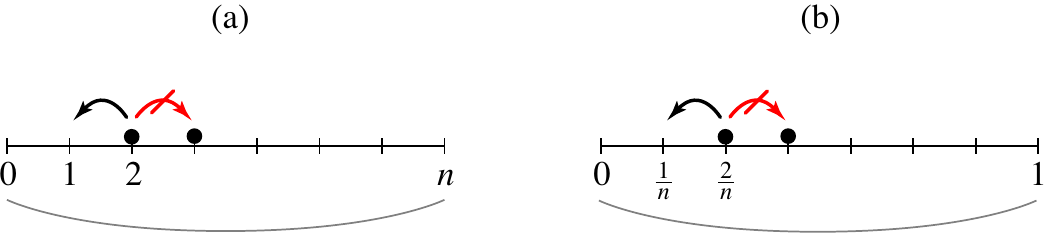}
\caption{(a) Exclusion process on the lattice $\Zn$ and (b) rescaled lattice $\Zn/n$. A particle can jump to an empty site (black arrow) but not to an occupied site (red arrow). The thin line at the bottom indicates the periodic boundary condition $\eta(0)=\eta(1)$.}
\label{figexclproc1}
\end{figure*}

\begin{example}[Simple symmetric exclusion process]
Consider a system of $k$ particles moving on the lattice $\Zn$ of integers ranging from $0$ to $n$, $n>k$; see Fig.~\ref{figexclproc1}(a). The rules that determine the evolution of the particles are assumed to be the following:
\begin{itemize}
\item A particle at site $i$ waits for a random exponential time with mean $1$, then selects one of its neighbors $j$ at random;
\item The particle at $i$ jumps to $j$ if $j$ is unoccupied; if $j$ is occupied, then the particle stays at $i$ and goes to a waiting period again before choosing another neighbor to jump to (exclusion principle).
\end{itemize}

We denote by $\eta_t(i)$ the occupation of the ``site'' $i\in\Zn$ at time $t$, and by $\eta_t=(\eta_t(0),\eta_t(1),\ldots,\eta_t(n-1))$ the whole configuration or \emph{microstate} of the system. Because of the exclusion principle, $\eta_t(i)\in\{0,1\}$. Moreover, we impose boundary conditions on the lattice by identifying the first and last site.   
 
The generator of the Markovian process defined by the rules above can be written explicitly by noting that there can be a jump from $i$ to $j$ only if $\eta(i)=1$ and $\eta(j)=0$. Therefore,
\be
(Lf)(\eta)=\frac{1}{2}\sum_{|i-j|=1} \eta(i)[1-\eta(j)][f(\eta^{i,j})-f(\eta)], 
\ee
where $f$ is any function of $\eta$, and $\eta^{i,j}$ is the configuration obtained after one jump, that is, the configuration obtained by exchanging the occupied state at $i$ with the unoccupied state at $j$:
\be
\eta^{i,j}(k)=\left\{
\begin{array}{ll}
\eta(i) & \textrm{if } k=j\\
\eta(j) & \textrm{if } k=i\\
\eta(k) & \textrm{otherwise}.
\end{array}
\right.
\ee
To obtain a hydrodynamic description of this dynamics, we rescale the lattice spacing by a factor $1/n$, as shown in Fig.~\ref{figexclproc1}(b), and take the limit $n\ra\infty$ with $r=k/n$, the density of particles, fixed. Furthermore, we speed-up the time $t$ by a factor $n^2$ to overcome the fact that the diffusion dynamics of the particle system ``slows'' down as $n\ra\infty$. In this limit, it can be proved that the empirical density of the rescaled dynamics, defined by
\be
\pi^n_t(x)=\frac{1}{n}\sum_{i\in\Zn} \eta_{n^2t}(i)\, \delta(x-i/n),
\ee
where $x$ is a point of the unit circle $C$, weakly converges in probability to a field $\rho_t(x)$ which evolves on $C$ according to the diffusion equation
\be
\partial_t \rho_t(x)=\partial_{xx} \rho_t(x).
\ee 
It can also be proved that the fluctuations of $\pi^n_t(x)$ around the deterministic field $\rho_t(x)$ follows a large deviation principle, expressed heuristically as
\be
P_n[\pi_t^n=\pi_t]=P_n(\{\pi_t^n(x)=\pi_t(x)\}_{t=0}^\tau)\asymp \E^{-nI[\pi_t]}.
\label{eqldpexcl1}
\ee
The interpretation of this expression follows the interpretation of the density $P_\vep[x]$ considered earlier: $P_n[\pi_t^n=\pi]$ is the probability density for the evolution of a field in time, so that the rate function shown in (\ref{eqldpexcl1}) is a space and time functional of that field. The expression of this rate function is relatively complicated compared to all the rate functions studied in this review. It involves two parts: a static part, which measures the ``cost'' of the initial field $\pi_0^n$, and a dynamic part, which measures the cost of the deviation of $\pi_t^n$ from $\rho_t$; see \cite{kipnis1989} for the full expression of the rate function, and \cite{eyink1990} for a discussion of its physical interpretation.
\end{example}

The previous example can be generalized in many different ways. One can consider asymmetric exclusion processes for which the diffusion is enhanced in one direction (see, e.g., \cite{derrida1998,derrida1998a}), or extended exclusion processes for which jumps to sites other than first neighbors are allowed. One can also consider models that allow more than one particle at each site, such as the zero-range process (see, e.g., \cite{benois1995,landim1993}), or models with particle reservoirs that add and remove particles at given rates. Moreover, one can choose not to impose the exclusion rule, in which case the particles jump independently of one another \cite{kipnis1990,kipnis1995}. 

For many of these models, large deviation principles have been derived at the level of the empirical density or density field \cite{landim1991,derrida2001,derrida2002,derrida2002a,derrida2003}, as well as at the level of the current \cite{bodineau2004,bodineau2005,bodineau2006}, which measures the average number of particles moving on the lattice. The rate functions associated with these observables show many interesting properties. In the case of the totally asymmetric exclusion process, for instance, the rate function of the density field is nonconvex \cite{derrida2002a,derrida2003}. This provides a functional analog of nonconvex rate functions. The reader will find many details about these large deviation results in the two review papers mentioned earlier. The first one, written by Derrida \cite{derrida2007}, is useful for gaining a feeling of the mathematics involved in the derivation of large deviation results for interacting particle models. The review written by Bertini \textit{et al.}~\cite{bertini2007}, on the other hand, is useful for gaining an overview of the different models that have been studied, and of the theory that describes the fluctuations of these models at the macroscopic level. For a study of interacting particle systems based on the Donsker-Varadhan theory, see \cite{bodineau2004a}.

To close this short discussion of large deviations in interacting particle models, let us mention that Derrida and Bodineau \cite{bodineau2004} have formulated a useful calculation tool for obtaining the rate function of the current in interacting particle models, which they dubbed the \emph{additivity principle}. This principle is close in spirit to the Freidlin-Wentzell theory (see Sec.~\ref{subsecfw}), and appears to be based, as for that theory, on a Markov property of fluctuations. For a presentation of this principle and its applications, see Derrida~\cite{derrida2007}.

%%%%%%%%%%%%%%%%%%%%%%%%%%%%%%%%%%%%%%%%%%%%%%%
\section{Other applications}
\label{secother}

The results, techniques, and examples compiled in the previous sections make for a more or less complete toolbox that can be used to study other applications of large deviations in statistical mechanics. We conclude this review by mentioning four more important applications related to multifractals, chaotic systems, disordered systems, and quantum systems. Our discussion of these applications is far from exhaustive; our aim is merely to mention them, and to point out a few useful references for those who want to learn more about them.

\subsection{Multifractals}

The subject of multifractal analysis was developed independently of large deviation theory, and is typically not presented from the point of view of this theory (see, e.g., \cite{paladin1987,mccauley1990,beck1993,falconer1997}). The two subjects, however, have much in common. In fact, one could say that multifractal analysis is a large deviation theory of self-similar measures, or a large deviation theory of the measure equivalent of self-processes, studied in Sec.~\ref{subsecsp}. A presentation of multifractal analysis in these terms is given in \cite{zohar1999,veneziano2002}, as well as in the book of Harte \cite{harte2001}. 

The idea that multifractal analysis is related to large deviation theory, or is an application of large deviation theory, becomes more obvious by noting the following:
\begin{itemize}
\item The two basic quantities commonly employed to characterize multifractals---the so-called \emph{multifractal spectrum} and \emph{structure function}---are the analogs of an entropy and a free energy function, respectively;
\item The scaling limit underlying the multifractal spectrum and the structure function has the form of a large deviation limit;
\item The multifractal spectrum and structure function are related by Legendre transforms. 
\end{itemize}

The last point is the perhaps the most revealing: the fact that two functions are found to be related by a Legendre (or Legendre-Fenchel) transform is often the sign that a large deviation principle underlies these functions. This is the case for the entropy and the free energy of equilibrium statistical mechanics, as we have seen in Sec.~\ref{secequi}, and this is the case, too, for the multifractal spectrum and the structure function. 

By re-interpreting in this way multifractal analysis in terms of large deviations, we do more than just translating a theory in terms of another---we gain a rigorous formulation of multifractals, as well as a guide for deriving new results about multifractals. One case in point concerns nonconvex rate functions. It had been known for some time that the structure function of multifractal analysis, which is the analog of the function $\varphi(\beta)$ or $\lambda(k)$ studied here, can be nondifferentiable, and that the nondifferentiable points of this function signal the appearance of a multifractal analog of first-order phase transitions (see \cite{beck1993} and references cited therein). Some confusion reigned as to how the multifractal spectrum had to be calculated in this case. Many authors assumed that the multifractal spectrum is always the Legendre-Fenchel transform of the structure function, and so concluded that the spectrum must be affine if the structure function is nondifferentiable \cite{tominaga1990,hata1989}. The correct answer given by large deviation theory is more involved: the multifractal spectrum can be concave or nonconcave, in the same way that an entropy can be concave or nonconcave. If it is concave, then it can be calculated as the Legendre-Fenchel transform of the structure function, otherwise, it cannot. A recent discussion of this point can be found in \cite{touchette2006a}; see also \cite{riedi1995,testud2005,testud2006} for mathematical examples of multifractals having nonconcave spectra.

\subsection{Thermodynamic formalism of chaotic systems}

The Freidlin-Wentzell theory of differential equations perturbed by noise has its analog for discrete-time dynamical maps, which was developed by Kifer \cite{kifer1988,kifer1990}. One interesting aspect of dynamical systems, be they represented by flows or maps, is that they often give rise to large deviation principles without a perturbing noise. In many cases, the chaoticity and mixing properties of a deterministic system are indeed such that they induce a seemingly stochastic behavior of that system, which induces, in turn, a stochastic behavior of observables of that system. The study of this phenomenon is the subject of the theory of chaotic systems and ergodic theory (see, e.g.,~\cite{alekseev1981,eckmann1985,lasota1994,gaspard1998}), and the study of large deviations in the context of these theories is the subject of the so-called \emph{thermodynamic formalism} developed by Ruelle \cite{ruelle1989,ruelle2004} and Sinai \cite{sinai1972,sinai1994}. For an introduction to this formalism, see \cite{beck1993,keller1998}.

As in the case of multifractals, the thermodynamic formalism was developed independently of large deviation theory. But it is also clear with hindsight that this formalism can be re-interpreted or recast in the language of large deviations. The basis of this interpretation can be summarized with the following basic observations:
\begin{itemize}
\item The so-called topological pressure, which plays a central role in the thermodynamic formalism, is a scaled cumulant generating function;

\item The entropy function of an observable, as defined in the thermodynamic formalism, is a rate function;

\item The topological pressure and entropy are related by Legendre transforms when the entropy is concave;

\item An equilibrium state in the thermodynamic formalism has the same large deviation interpretation as an equilibrium state in equilibrium statistical mechanics: both are the solution of a variational principle which can be derived from the contraction principle.
\end{itemize}

The reader is referred to the review paper of Oono \cite{oono1989} for an explanation of some of these points; see also \cite{kai1980,mori1989,takahashi1984}. A number of results that establish a direct connection between dynamical systems and large deviation theory can be found in \cite{young1990,lopes1990,waddington1996,pollicott1998,keller1998,young2003}. For a derivation of fluctuation theorems in the context of chaotic maps, see \cite{maes2003}. Finally, for a large deviation study of multiplicative processes in deterministic systems having many degrees of freedom, see \cite{tailleur2007} and references therein.

At the time of writing this review, a complete presentation of the thermodynamic formalism that refers explicitly to large deviation theory has yet to appear. The fact that chaotic maps can be thought of as continuous-state Markov chains with transition matrix given by the Frobenius operator appears to be a good starting point for establishing a direct link between the thermodynamic formalism and large deviation theory (see, e.g., \cite{oono1978, oono1980}).
 
\subsection{Disordered systems}

The application of large deviation techniques for studying disordered systems focuses in the literature on two different models: random walks in random environments (see, e.g.,~\cite{gantert1999,comets2000,varadhan2003a,zeitouni2006}) and spin glasses (see, e.g.,~\cite{dorlas2001,dorlas2002,talagrand2007}). Large deviation principles can be derived, for both applications, at the \emph{quenched} level, i.e., for a fixed realization of the random disorder, or at the \emph{annealed} level, which involves an average over the disorder. An interesting question in the context of random walks in random environments is whether a large deviation arises out of an atypical state of the walk or out of the atypicality of a specific random environment. A similar question arises for spin glasses in the form of, is the equilibrium state of a spin glass obtained for a specific random interaction typical in the ensemble of all interactions? The book of den Hollander \cite{hollander2000} and the recent review paper by Zeitouni \cite{zeitouni2006} offer two good entry points to the first question; see \cite{talagrand2003,bovier2006} for a mathematical discussion of spin glasses.

From the large deviation point of view, the difference between disordered and regular systems is that generating functions defined in the context of the former systems have an extra dependence on a ``disorder'' variable, which implies that these generating functions are random variables themselves. Therefore, in addition to studying the ``quenched'' large deviations associated with a given ``random'' generating function (i.e., a generating function arising for a given realization of the disorder), one can study the large deviations of the generating function itself, in order to determine the most probable value of the generating function. This concentration value of the generating function often simplifies the study of ``annealed'' large deviations, which are obtained from generating functions averaged over the disorder. For a discussion of spin glasses which follows this point of view, see the recent book of M\'ezard and Montanari \cite{mezard2009}.

\subsection{Quantum large deviations}

Quantum systems have entered the large deviation scene relatively recently compared to classical systems: end of 1980s compared to early 1970s. Applications of large deviations for studying boson gases are described in \cite{cegla1988,berg1988,dorlas2005}; quantum gases are considered in \cite{lenci1999,lebowitz2000,gallavotti2002}, while quantum spin systems are considered in \cite{hiai2007,lenci2005,netocny2004}. For an application of Varadhan's Theorem for a class of mean-field quantum models, see \cite{petz1989}.

A quantum version of Sanov's Theorem is presented in \cite{bjelakovia2005}. As for classical version of that theorem, the quantum version plays an important role in the theory of estimation and in information theory, as generalized to the quantum world \cite{keyl2006,audenaert2007,ahlswede2003}. Finally, a quantum adaptation of the Freidlin-Wentzell theory of dynamical systems perturbed by noise can be found in \cite{blanchard1985}.

%%%%%%%%%%%%%%%%%%%%%%%%%%%%%%%%%%%%%%%%%%%%%%%
\appendix
%%%%%%%%%%%%%%%%%%%%%%%%%%%%%%%%%%%%%%%%%%%%%%%
\section{Summary of main mathematical concepts and results}
\label{appsum}

\begin{itemize}
\item \textbf{Large deviation principle (Sec.~\ref{secldt}):} Let $\{A_n\}$ be a sequence of random variables indexed by the positive integer $n$, and let $P(A_n\in \D a)=P(A_n\in [a,a+\D a])$ denote the probability measure associated with these random variables. We say that $A_n$ or $P(A_n\in\D a)$ satisfies a \emph{large deviation principle} if the limit
\be
I(a)=\lim_{n\ra\infty} -\frac{1}{n}\ln P(A_n\in\D a)
\ee
exists (see Appendix~\ref{appldt} for a more precise definition). The function $I(a)$ defined by this limit is called the \emph{rate function}; the parameter $n$ of decay is called in large deviation theory the \emph{speed}.\footnote{See footnote~\ref{notespeed}.}

\item \textbf{Asymptotic notation:} The existence of a large deviation principle for $A_n$ means concretely that the dominant behavior of $P(A_n\in\D a)$ is a decaying exponential with $n$, with rate exponent $I(a)$. We summarize this property by writing
\be
P(A_n\in\D a)\asymp \E^{-n I(a)}\, \D a.
\ee 
The infinitesimal element $\D a$ is important in this expression; if we work with the probability density $p(A_n=a)$ instead of the probability measure $P(A_n\in\D a)$, then the large deviation principle is expressed simply as $p(A_n=a)\asymp\E^{-nI(a)}$. 

\item \textbf{Generating function:} The \emph{generating function} of $A_n$ is defined as
\be
W_n(k)=\langle\E^{nkA_n}\rangle=\int \E^{nka}\, P(A_n\in\D a),\quad k\in\reals.
\ee
In terms of the density $p(A_n)$, we have instead
\be
W_n(k)=\int \E^{nka}\, p(A_n=a)\, \D a.
\ee
In both expressions, the integral is over the domain of $A_n$.

\item \textbf{Scaled cumulant generating function:} The function $\lambda(k)$ defined by the limit
\be
\lambda(k)=\lim_{n\ra\infty}\frac{1}{n}\ln W_n(k)
\ee
is called the \emph{scaled cumulant generating function} of $A_n$. It is also called the \emph{log-generating function} or \emph{free energy function} of $A_n$. The existence of this limit is equivalent to writing $W_n(k)\asymp\E^{n\lambda(k)}$.

\item \textbf{Legendre-Fenchel transform:} The \emph{Legendre-Fenchel transform} of a function $f(x)$ is defined by 
\be
g(k)=\sup_x \{kx-f(x)\}.
\ee
This transform is often written in convex analysis in the compact form $g=f^*$. This transform is also sometimes written with an infimum instead of a supremum. With the supremum, the Legendre-Fenchel transform reduces to the standard \emph{Legendre transform} when $f(x)$ is strictly convex and differentiable, for then
\be
g(k)=kx(k)-f(x(k)),
\ee
where $x(k)$ is the unique root of $f'(x)=k$. The Legendre-Fenchel transform involving the infimum reduces to the standard Legendre transform when $f(x)$ is strictly concave and differentiable. The formula of the Legendre transform, in this case, is the same as above.

\item \textbf{G\"artner-Ellis Theorem:} If $\lambda(k)$ is differentiable, then $A_n$ satisfies a large deviation principle with rate function $I(a)$ given by the Legendre-Fenchel transform of $\lambda(k)$:
\be
I(a)=\sup_{k}\{ka-\lambda(k)\}.
\ee
This Legendre-Fenchel transform is expressed in convex analysis by the shorthand notation $I=\lambda^*$. (See Sec.~\ref{secldt} for a more precise statement of this theorem.)

\item \textbf{Varadhan's Theorem:} If $A_n$ satisfies a large deviation principle with rate function $I(a)$, then its scaled cumulant generating function $\lambda(k)$ is the Legendre-Fenchel transform of $I(a)$:
\be
\lambda(k)=\sup_k \{ka-I(a)\}.
\ee
In shorthand notation, this is expressed as $\lambda=I^*$. (See Sec.~\ref{secldt} for a more precise statement of this theorem.)

\item \textbf{Convex versus nonconvex rate functions (Sec.~\ref{secmathapp}):} If $I(a)$ is convex, then $I=\lambda^*$. If $I(a)$ is nonconvex, then $I\neq \lambda^*$. As a corollary, rate functions that are nonconvex cannot be calculated via the G\"artner-Ellis Theorem because rate functions obtained from this theorem are always convex (strictly convex, in fact). 

\item \textbf{Properties of $\lambda(k)$ (Sec.~\ref{secldt}):}
\begin{enumerate}
\item $\lambda(0)=0$. This follows from the normalization of probabilities.
\item $\lambda'(0)=\displaystyle\lim_{n\ra\infty} \langle A_n\rangle$. This property is related to the Law of Large Numbers.
\item $\lambda''(0)=\displaystyle\lim_{n\ra\infty} n\, \var (A_n)$. This property is related to the Central Limit Theorem.
\item $\lambda(k)$ is convex. This implies, among other things, that $\lambda(k)$ can be nondifferentiable only at isolated points.
\item $\lambda(k)$ is differentiable if $I(a)$ is strictly convex, i.e., convex with no linear parts.
\item $\lambda(k)$ has at least one nondifferentiable point if $I(a)$ is nonconvex or has linear parts. 
\item Suppose that $\lambda(k)$ is differentiable. Then the value $k$ such that $\lambda'(k)=a$ has the property that $k=I'(a)$. This is the statement of the Legendre duality between $\lambda$ and $I$, which can be expressed in words by saying that the slopes of $\lambda$ correspond to the abscissas of $I$, while the slopes of $I$ correspond to the abscissas of $\lambda$. (This property can be generalized to a nondifferentiable $\lambda(k)$ and a nonconvex $I(a)$ with the concept of supporting lines \cite{rockafellar1970}.)
\end{enumerate}

\item \textbf{Contraction principle (Sec.~\ref{secldt}):} Consider two sequences of random variables $\{A_n\}$ and $\{B_n\}$ such that $A_n=f(B_n)$, and assume that $B_n$ obeys a large deviation principle with rate function $I_B$. Then $A_n$ obeys a large deviation principle with rate function $I_A$ given by
\be
I_A(a)=\inf_{a:f(b)=a} I_B(b).
\ee

\item \textbf{Connection with physics:} Entropies are rate functions; free energies are scaled cumulant generating functions.
\end{itemize}

\section{Rigorous formulation of the large deviation principle}
\label{appldt}

This appendix is an attempt at explaining the rigorous formulation of the large deviation principle for the benefit of physicists not versed in topology and measure theory. The formulation presented here is inspired from the work of Ellis \cite{ellis1984, ellis1995, ellis1999,ellis2006}, which is itself inspired from Varadhan \cite{varadhan1966}. For background material on topology and measure theory, the reader should consult Appendices B and D of \cite{dembo1998}. 

The rigorous definition of the large deviation principle is based on four basic ingredients:

\begin{itemize}
\item A sequence of probability spaces $\{(\Lambda_n,\mathcal{F}_n,P_n),n\in\mathbb{N}\}$ consisting of a probability measure $P_n$ defined on the set $\mathcal{F}_n$ of all (Borel) sets of the ``event'' set $\Lambda_n$;

\item A sequence of random variables $\{Y_n,n\in\mathbb{N}\}$ mapping $\Lambda_n$ into a complete, separable metric space $\mathcal{X}$, also known as a \emph{Polish} space;

\item A sequence $\{a_n:n\in\mathbb{N}\}$ of positive constants such that $a_n\ra \infty$ as $n\ra\infty$;

\item A lower semi-continuous function $I(x)$ mapping $\mathcal{X}$ into $[0,\infty]$.
\end{itemize}

From the point of view of statistical mechanics, $\Lambda_n$ can be thought of as the space of the microstates of an $n$-particle system. The set $\mathcal{F}_n$ is the set of all possible events (sets) on $\Lambda_n$, whereas $P_n$ is a probability measure on $\mathcal{F}_n$. The fact that we are dealing with a ``sequence'' of probability spaces is there, of course, because we are interested in studying the behavior of $P_n$ in the limit $n\ra\infty$, which we call the thermodynamic limit. In the same vein, $Y_n$ should be thought of as a macrostate, and $\mathcal{X}$ as the macrostate space. One can think of $Y_n$, for example, as the mean magnetization of a simple spin model, in which case $\mathcal{X}=[-1,1]$. In all the applications covered in this review, $\mathcal{X}$ is a subset of $\mathbb{R}^d$, and so we need not bother with the fact that $\mathcal{X}$ is a ``complete, separable metric'' space. This requirement is a technicality used by mathematicians to make the theory of large deviations as general as possible. 

The random variable for which we are interested to formulate a large deviation principle is $Y_n$. The probability measure $P_n$ defined at the level of $\Lambda_n$ is extended to $Y_n$ via
\be
P_n(Y_n\in B)=\int_{\{\om\in\Lambda_n:Y_n(\om)\in B \}} P_n(\D\om),
\ee
where $B$ is any subset of $\mathcal{X}$. Given this probability, we say that the sequence $\{Y_n,n\in\mathbb{N}\}$ satisfies a \emph{large deviation principle} on $\mathcal{X}$ with \emph{rate function} $I$ and \emph{speed} $a_n$ if for any \emph{closed} set $C$, 
\be
\limsup_{n\ra\infty} \frac{1}{a_n} \ln P_n(Y_n\in C) \leq - \inf_{y\in C} I(y),
\label{eqldpub1}
\ee
and for any \emph{open} set $O$,
\be
\liminf_{n\ra\infty} \frac{1}{a_n} \ln P_n(Y_n\in O) \geq - \inf_{y\in O} I(y).
\label{eqldplb1}
\ee 
The lower semi-continuity of $I$ guarantees that this function achieves its minimum on any closed sets (a lower semi-continuous function has closed level sets; see Chap.~5 of~\cite{tiel1984}).

The two limits (\ref{eqldpub1}) and (\ref{eqldplb1}) give a rigorous meaning to the two bounds mentioned in our formal discussion of the large deviation principle; see Sec.~\ref{secldt}. To understand why the first limit involves closed sets and the second open sets, we need to invoke the notion of weak convergence. The idea, as partly explained in Sec.~\ref{secldt}, is that we wish to approximate a measure $\mu_n$ by a limit measure $\mu$ such that
\be
\lim_{n\ra\infty} \int_\mathcal{X} f(y)\mu_n(\D y)=\int_\mathcal{X} f(y)\mu(\D y)
\label{eqwc1}
\ee
for all bounded and continuous functions $f:\mathcal{X}\ra\reals$. Though less transparent, an equivalent and often more practical way of expressing the weak convergence of $\mu_n$ to $\mu$ is provided by the so-called \emph{Portmanteau Theorem} (see Sec.~D.2 of \cite{dembo1998}), which states that the limit (\ref{eqwc1}) is equivalent to
\be
\limsup_{n\ra\infty} \mu_n(C)\leq \mu(C)
\ee
for all closed subsets $C$ of $\mathcal{X}$, and
\be
\liminf_{n\ra\infty} \mu_n(O)\geq \mu(O)
\ee
for all open subsets $O$ of $\mathcal{X}$. These two limits correspond to the two limits shown in (\ref{eqldpub1}) and (\ref{eqldplb1}), with $\mu_n$ equal to $a_n^{-1}\ln P_n$ to account for the scaling $P_n \asymp \E^{-a_n I}$, $I\geq 0$. 

The heuristic form of the large deviation principle that we use as the basis of this review is a simplification of the rigorous formulation, in that we assume, as in Sec.~\ref{secldt}, that the large deviation upper and lower bounds, defined by (\ref{eqldpub1}) and (\ref{eqldplb1}) respectively, are the same. This is a strong simplification, which happens to be verified only for so-called \emph{$I$-continuity sets}, that is, sets $A$ such that
\be
\inf_{y\in\bar{A}}I(y)=\inf_{y\in A^\circ}I(y),
\ee
where $\bar{A}$ and $A^\circ$ denote, respectively, the closure and relative interior of $A$; see Sec.~3 of \cite{ellis1995} or \cite{ellis2006} for more details. In treating large deviations, we also take the simplifying step of considering  probabilities of the form $P(Y_n\in dy)$, where $dy=[y,y+dy]$ with a bit of abuse of notation, in which case 
\be
\lim_{n\ra\infty} \frac{1}{a_n} \ln P(Y_n\in dy)= - \inf_{x\in [y,y+dy]} I(x)=-I(y).
\ee
Finally, in most examples covered in this review, the speed $a_n$ is equal to $n$. In statistical mechanics, the proportionality of $a_n$ with $n$ is an expression of the concept of extensivity.

%%%%%%%%%%%%%%%%%%%%%%%%%%%%%%%%%%%%%%%%%%%%%%%
\section{Derivations of the G\"artner-Ellis Theorem}
\label{appge}

We give here two derivations of the G\"artner-Ellis Theorem for random variables taking values in $\reals$. The first derivation is inspired from the work of Daniels \cite{daniels1954} on saddle-point approximations in statistics, and is presented to reveal the link that exists between the large deviation principle, the saddle-point approximation, and Laplace's approximation.\footnote{The so-called Darwin-Fowler method \cite{fowler1966} used in statistical mechanics is yet another example of saddle-point or Laplace approximation applied to discrete generating functions.} The second derivation is based on a clever change of measure which goes back to Cram\'er \cite{cramer1938}, and which is commonly used nowadays to prove large deviation principles. None of the derivations are rigorous.

\subsection{Saddle-point approximation}

Consider a random variable $S_n(\om)$ which is a function of a sequence $\om=(\om_1,\om_2,\ldots,\om_n)$ of $n$ random variables. The random variable $S_n$ need not be a sample mean, but it is useful to think of it as being one. For simplicity, assume that the $\om_i$'s are also real random variables, so that $\om\in\reals^n$. Denoting by $p(\om)$ the probability density of $\om$, we write the probability density of $S_n$ as
\be
p(S_n=s)=\int_{\{\om\in\reals^n:S_n(\om)=s\}} p(\om)\, \D\om=\int_{\reals^n} \delta(S_n(\om)-s)\, p(\om)\, \D\om =\lex \delta(S_n-s)\rex,
\ee
just as in Eq.~(\ref{eqprob1}) of Sec.~\ref{secexamples}. Using the Laplace transform representation of Dirac's delta function,
\be
\delta(s)=\frac{1}{2\pi \I}\int_{a-\I\infty}^{a+\I\infty} \E^{\zeta s}\, \D\zeta,\quad a\in\reals,
\ee
we then write
\be
p(S_n=s)=\frac{1}{2\pi\I}\int_{a-\I\infty}^{a+\I\infty} \D\zeta \int_{\reals^n} \D\om\, p(\om)\, \E^{\zeta [S_n(\om)-s]}=\frac{1}{2\pi\I}\int_{a-\I\infty}^{a+\I\infty} \D\zeta\, \E^{-\zeta s} \int_{\reals^n} \D\om\, p(\om)\, \E^{\zeta S_n(\om)}.
\label{eqrepi1}
\ee
The integral of the Laplace transform is performed along the so-called \emph{Bromwich contour}, which runs parallel to the imaginary axis from $\zeta=a-\I\infty$ to $\zeta=a+\I\infty$, $a\in\reals$. 

At this point, we anticipate the scaling of the large deviation principle by performing the change of variable $\zeta\ra n\zeta$, and note that if
\be
\lambda(\zeta)=\lim_{n\ra\infty} \frac{1}{n}\ln \lex \E^{n\zeta S_n}\rex
\ee
exists, then
\be
p(S_n=s)\asymp \int_{a-\I\infty}^{a+\I\infty} \D(\zeta)\, \E^{-n[\zeta s-\lambda(\zeta)]}
\ee
with sub-exponential corrections in $n$. By deforming the contour so that it goes through the saddle-point $\zeta^*$ of $\zeta s-\lambda(\zeta)$, and by considering only the exponential contribution to the integral coming from the saddle-point, we then write
\be
p(S_n=s)\asymp \int_{\zeta^*-\I\infty}^{\zeta^*+\I\infty} \D(-\I\zeta)\, \E^{-n[\zeta s-\lambda(\zeta)]}\asymp \E^{-n[\zeta^* s-\lambda(\zeta^*)]}.
\ee
The last approximation is the saddle-point approximation (see Chap.~6 of~\cite{bender1978}). This result is completed by noting that the saddle-point $\zeta^*$ must be real, since $p(S_n=s)$ is real. Moreover, if we assume that $\lambda(\zeta)$ is analytic, then $\zeta^*$ is the unique minimum of $\zeta s-\lambda(\zeta)$ satisfying $\lambda'(\zeta^*)=s$ along the Bromwich contour. The analyticity of $\lambda(k)$ also implies, by the Cauchy-Riemann equations, that the point $\zeta^*$, which is a minimum of $\zeta s-\lambda(\zeta)$ along the Bromwich contour, is a maximum of $\zeta s-\lambda(\zeta)$ for $\zeta$ real. Therefore, we can write
\be
\lim_{n\ra\infty} -\frac{1}{n}\ln p(S_n=s)=\sup_{k\in\reals} \{ks-\lambda(k)\},.
\ee
This concludes our first derivation of the G\"artner-Ellis Theorem. For a discussion of cases for which $\lambda(k)$ is not analytic, see Sec.~\ref{secmathapp}.

\subsection{Exponential change of measure}

We consider the same random variable $S_n(\om)$ as in the previous derivation, but now we focus on the probability measure $P(\D\om)$ instead of the probability density $p(\om)$. We also introduce the following modification or ``perturbation'' of $P(\D\om)$:
\be
P_k(\D\om)=\frac{\E^{nkS_n(\om)}}{\lex \E^{nk S_n}\rex} P(\D\om),
\ee
which involves the parameter $k\in\reals$. This probability has the same form as the probability $P_\beta(\D\om)$ defining the canonical ensemble. In large deviation theory, $P_k(\D\om)$ is called the \emph{tilted measure}, and the family of such measures indexed by $k$ is often called the \emph{exponential family} \cite{ellis1985}.\footnote{In statistics and actuarial mathematics, $P_k$ is also known as the \emph{associated law} or \emph{Esscher transform} of $P$ \cite{feller1970}.} 

Starting from the definition of $P_k(\D\om)$, one can prove the following properties (see \cite{gartner1977,ellis1984,ellis1985}):

\begin{description}
\item[Property 1:] If
\be
\lambda(k)=\lim_{n\ra\infty}\frac{1}{n}\ln \lex \E^{nkS_n}\rex
\ee
exists, then
\be
P_k(\D\om)\asymp \E^{n[kS_n(\om)-\lambda(k)]} P(\D\om).
\ee
The so-called Radon-Nikodym derivative of $P_k(\D\om)$ with respect to $P(\D\om)$ is thus written as
\be
\frac{\D P_k(\om)}{\D P(\om)}=\frac{P_k(\D\om)}{P(\D\om)}\asymp \E^{n[kS_n(\om)-\lambda(k)]}.
\ee
\end{description}

\begin{description}
\item[Property 2:] If $\lambda(k)$ is differentiable at $k$, then
\begin{eqnarray}
\lim_{n\ra\infty} \lex S_n\rex_k &=& \lim_{n\ra\infty} \int_{\reals^n} S_n(\om)\, P_k(\D\om)\nonumber\\
&=&\lim_{n\ra\infty}\frac{1}{\lex \E^{nkS_n}\rex}\int_{\reals^n} S_n(\om)\, \E^{nkS_n(\om)}\, P(\D\om)\nonumber\\
&=&\lambda'(k).
\end{eqnarray}
\end{description}

\begin{description}
\item[Property 3:] The value $s_k=\lambda'(k)$ is the concentration point (viz., typical value) of $S_n$ with respect to $P_k(\D\om)$, that is,
\be
\lim_{n\ra\infty} P_k(S_n\in [s_k,s_k+\D s])=1.
\ee
This limit expresses a Law of Large Numbers for $S_n$ with respect to $P_k(\D\om)$.\footnote{Recall that the concentration point of $S_n$ with respect to $P(\D\om)$ is $s_0=\lambda'(0)$; see Sec.~\ref{propk0}.}
\end{description}

From these properties, we obtain a large deviation principle for $P(S_n\in \D s)$ as follows. Starting with
\be
P(S_n\in \D s)=\int_{\{\om\in\reals^n: S_n(\om)\in \D s\}} P(\D\om) =\int_{\{\om\in\reals^n: S_n(\om)\in \D s\}} \frac{P(\D\om)}{P_k(\D\om)}\, P_k(\D\om),
\ee
we use the first property to obtain
\begin{eqnarray}
P(S_n\in \D s) &\asymp& \int_{\{\om\in\reals^n: S_n(\om)\in \D s\}} \E^{-n[kS_n(\om)-\lambda(k)]}\, P_k(\D\om)\nonumber\\
&=&\E^{-n[ks-\lambda(k)]}\int_{\{\om\in\reals^n: S_n(\om)\in \D s\}} P_k(\D\om),
\end{eqnarray}
which implies
\be
P(S_n\in \D s)\asymp \E^{-n[ks-\lambda(k)]}\, P_k(S_n \in \D s).
\ee
Next we choose $k$ such that $\lambda'(k)=s$. According to the second and third properties, we must have
\be
\lim_{n\ra\infty} P_k(S_n\in [s,s+\D s])=1
\ee
or, equivalently, $P_k(S_n\in \D s)\asymp \E^{n0}\, \D s$ using the asymptotic notation, so that
\be
P(S_n\in \D s)\asymp \E^{-n[ks-\lambda(k)]}\, \D s.
\ee
Therefore, $P(S_n\in \D s)\asymp \E^{-nI(s)}$, where
\be
I(s)=ks-\lambda(k),\quad \lambda'(k)=s. 
\ee
We recognize in the last expression the Legendre transform of $\lambda(k)$.

This derivation can be adapted to other random variables and processes, and is useful in practice for deriving large deviation principles, as the Radon-Nikodym derivative can often be calculated explicitly. In the case of Markov processes, for example, $\D P_k/\D P$ is given by Girsanov's formula \cite{varadhan2003}. Other perturbations of $P$, apart from the exponential one, can also be used. The general idea at play is to change the measure (or process) $P$ into a measure $P'$, so as to make an unlikely event under $P$ a typical event under $P'$, and to use the relationship between $P$ and $P'$ to infer the probability of that event under $P$.

%%%%%%%%%%%%%%%%%%%%%%%%%%%%%%%%%%%%%%%%%%%%%%%
\section{Large deviation results for different speeds}
\label{appspeed}

The G\"artner-Ellis Theorem is stated and used throughout this review mostly for large deviation principles having a linear speed $a_n=n$. The following is the general version of that theorem which applies to any speed $a_n$ such that $a_n\ra\infty$ as $n\ra\infty$ \cite{ellis1985}. Consider a random variable $W_n$ such that 
\be
\lambda(k)=\lim_{n\ra\infty}\frac{1}{a_n}\ln\lex\E^{a_n k W_n}\rex
\ee
exists and is differentiable. Then $P(W_n\in\D w)\asymp\E^{-a_n I(w)}\, \D w$, where $I(w)$ is, as before, the Legendre-Fenchel transform of $\lambda(k)$. 

The version of Varadhan's Theorem that applies to general speeds is the following \cite{dembo1998}. Let $W_n$ be a random variable satisfying a large deviation principle with speed $a_n$ and rate function $I(w)$, and let $f$ be a bounded function of $W_n$. Then
\be
\lambda(f)=\lim_{n\ra\infty}\frac{1}{a_n}\ln\lex\E^{a_nf(W_n)}\rex=\sup_w\{f(w)-I(w)\}.
\ee
For the (unbounded) linear function $f(W_n)=kW_n$, it can also be proved, with an additional mild assumption on $W_n$ (see, e.g., Theorem 5.1 of \cite{ellis1995} or Theorem 4.3.1 of \cite{dembo1998}), that
\be
\lambda(k)=\lim_{n\ra\infty}\frac{1}{a_n}\ln\lex\E^{a_nkW_n}\rex=\sup_w\{kw-I(w)\}.
\ee

%%%%%%%%%%%%%%%%%%%%%%%%%%%%%%%%%%%%%%%%%%%%%%%
\begin{acknowledgments}
The writing of this review has benefited from the help of many people who provided ideas, criticisms, guidance, encouragements, as well as useful opportunities to lecture about large deviations. Among these, I would like to thank Fulvio Baldovin, Julien Barr\'e, Freddy Bouchet, Eddie G. D. Cohen, Claude Cr\'{e}peau, Thierry Dauxois, Bernard Derrida, Richard S. Ellis, Vito Latora, Michael C. Mackey, Stefano Ruffo, Attilio Stella, Julien Tailleur, Tooru Taniguchi, Bruce Turkington, the Touchette-Ostiguy family, and my colleagues of the statistical mechanics study group at QMUL. I also thank Rosemary J. Harris, Michael Kastner, and Michael K.-H. Kiessling for reading the manuscript. A special thank is also due to Ana Belinda Pe\~nalver Pe\~na for her more than needed support. 

My intermittent work on this review has been supported over the last few years by NSERC (Canada), the Royal Society of London (Canada-UK Millennium Fellowship), and RCUK (Interdisciplinary Academic Fellowship). 
\end{acknowledgments}

%%%%%%%%%%%%%%%%%%%%%%%%%%%%%%%%%%%%%%%%%%%%%%%
\bibliographystyle{plain}
\bibliography{masterbibmin}

\end{document}